\documentclass[lettersize,journal]{IEEEtran}
\usepackage{amsmath,amsfonts}
\usepackage{algorithmic}
\usepackage{algorithm}
\usepackage{array}
\usepackage[caption=false,font=normalsize,labelfont=sf,textfont=sf]{subfig}
\usepackage{textcomp}
\usepackage{stfloats}
\usepackage{url}
\usepackage{verbatim}
\usepackage{graphicx}
\usepackage{cite}
\usepackage{tabularray}
\usepackage{url}
\usepackage{graphicx}
\usepackage{outlines}
\usepackage{xcolor}

\usepackage{tikz}
\newcommand*\emptycirc[1][1ex]{\tikz\draw (0,0) circle (#1);} 
\newcommand*\halfcirc[1][1ex]{%
  \begin{tikzpicture}
  \draw[fill] (0,0)-- (90:#1) arc (90:270:#1) -- cycle ;
  \draw (0,0) circle (#1);
  \end{tikzpicture}}
\newcommand*\fullcirc[1][1ex]{\tikz\fill (0,0) circle (#1);} 

\begin{document}

\title{Modern DDoS Threats and Countermeasures: Insights into Emerging Attacks and Detection Strategies}

\author{Jincheng Wang, Le Yu, John C.S. Lui, and Xiapu Luo \thanks{Jincheng Wang and Xiapu Luo are from The Hong Kong Polytechnic University. Le yu is from Nanjing University of Posts and Telecommunications. John C.S. Lui is from The Chinese University of Hong Kong.}}


\maketitle

\begin{abstract}
Distributed Denial of Service (DDoS) attacks persist as significant threats to online services and infrastructure, evolving rapidly in sophistication and eluding traditional detection mechanisms.
This evolution demands a comprehensive examination of current trends in DDoS attacks and the efficacy of modern detection strategies.
This paper offers an comprehensive survey of emerging DDoS attacks and detection strategies over the past decade.
We delve into the diversification of attack targets, extending beyond conventional web services to include newer network protocols and systems, and the adoption of advanced adversarial tactics. Additionally, we review current detection techniques, highlighting essential features that modern systems must integrate to effectively neutralize these evolving threats.
Given the technological demands of contemporary network systems, such as high-volume and in-line packet processing capabilities, we also explore how innovative hardware technologies like programmable switches can significantly enhance the development and deployment of robust DDoS detection systems.
We conclude by identifying open problems and proposing future directions for DDoS research.
In particular, our survey sheds light on the investigation of DDoS attack surfaces for emerging systems, protocols, and adversarial strategies.
Moreover, we outlines critical open questions in the development of effective detection systems, e.g., the creation of defense mechanisms independent of control planes.
\end{abstract}


\section{Introduction}\label{sec:introduction}

Distributed Denial of Service (DDoS) attacks have persistently been one of the most prevalent threats to the stability and availability of online services and infrastructures.
According to CloudFlare's report~\cite{cloudflare-ddos-report}, a 117\% year-over-year increase in DDoS attacks has been observed.
Famous DDoS attacks like Mirai~\cite{antonakakis2017understanding} and Github paralyze~\cite{chadd2018ddos} demonstrates the severity of these attacks, where significant traffic volumes flood the critical network infrastructures or services.
The proliferation of Internet of Things (IoT) devices, combined with the increased availability of DDoS-as-a-Service platforms, further lowered the barrier to launching sophisticated attacks that can overwhelm even the most robust defensive mechanisms.

The landscape of DDoS attacks has not only expanded in scale but also advanced in complexity, demonstrating a two-pronged evolution.
The first aspect of this evolution is the diversification of targets and techniques.
Attackers are no longer limited to traditional transport-layer protocols. 
They are increasingly exploiting a variety of application-layer protocols for their attacks.
This trend also extends to the targets themselves, with sophisticated systems such as blockchain technologies and cellular network infrastructures coming under siege.
The second facet of this evolution pertains to the subtlety of the attacks.
There is a noticeable increase in the stealthiness with which these actions are executed.
Adversaries are crafting more sophisticated strategies designed to circumvent not only the current commercial detection systems but also the most advanced detection algorithms.
This advancement is indicative of a continuous arms race between attackers seeking invisibility and defenders aiming to maintain visibility and control.

Simultaneously, the advancement of sophisticated DDoS detection methodologies is advancing at a rapid pace.
These approaches meticulously model behavioral patterns and traffic characteristics to delineate the boundary between legitimate users and malevolent attackers.
Furthermore, the emergence of state-of-the-art network hardware, such as programmable switches, signifies a burgeoning domain ripe with opportunities to fortify network resilience.
Initiatives to amalgamate these cutting-edge technologies into comprehensive detection frameworks are currently in progress, marking a significant stride toward more robust defense mechanisms.

The preceding discussion underscores the necessity for a systematic literature review to meticulously examine the progression of DDoS attack trends.
Additionally, a critical evaluation of nascent detection technologies is crucial to yield insights that are fundamental to the architecture of a contemporary detection system.
Regrettably, existing surveys~\cite{agrawal2019defense,kumari2023comprehensive,chaudhary2023ddos,yan2015software,li2023comprehensive,zhang2024revealing} typically focus on specific scenarios (e.g., cloud computing~\cite{agrawal2019defense} and IoT~\cite{kumari2023comprehensive}).
This narrow focus often results in a lack of a comprehensive perspective that encompasses the full spectrum of DDoS attack characteristics and trends.
Consequently, these studies frequently overlook crucial aspects necessary for the development of modern DDoS detection systems, such as attack-agnostic detection capabilities and cross-domain data sharing.
Additionally, the potential benefits of emerging advanced network hardware in enhancing DDoS detection are rarely discussed.
As a result, we identify three critical questions that need to be explored to advance the understanding of DDoS attacks and their detection.
\begin{enumerate}
    \item What are the prevailing trends in emerging DDoS attacks, and what insights can be gleaned to inform the vulnerability analysis of nascent network protocols and systems (Section~\ref{sec:ddos-attack})?
    \item Given the diversity and increasing stealthiness of DDoS attacks, what guiding principles should inform the construction of modern detection systems (Section~\ref{ddos-detection})?
    \item With the rollout of advanced network hardware, such as programmable switches, which feature can be leveraged to augment DDoS detection (Section~\ref{sec:ddos-detection-system-deployment})?
\end{enumerate}

This survey is committed to tackling the questions previously delineated.
To answer the first question, we carefully unravel the progression of DDoS threats, providing actionable insights for identifying vulnerabilities within the DDoS landscape.
Instead of targeting single scenario, we studied emerging DDoS vulnerabilities lurking in nine popular communication protocols and eight advanced systems, revealing novel protocol features (e.g., DNS recursive resolution) and system weakness (e.g., vulnerable resource sharing mechanism) for attackers to orchestrate DDoS attacks.
Moreover, our study reveals three emerging types of adversarial DDoS tactics, which can efficiently bypass commercial and state-of-the-art DDoS detection techniques.
Our investigation shines a light on the frailties that adversaries target in emerging protocols and systems, while also assisting in the forecast of impending attack methodologies.

To answer the second and the third question, we present a detailed analysis of current detection systems to bridge the gap between present-day challenges and the novel solutions taking shape in the field of DDoS protection.
We also examine contemporary research that leverages emerging network primitives (SDN and programmable switches) to enhance DDoS detection, and summarize their benefits.
Finally, we highlight the unresolved challenges in the analysis of DDoS vulnerabilities and the development of contemporary detection systems.
We also spotlight promising methodologies and suggest future avenues for research to tackle these unresolved issues.

\begin{figure*}
    \centering
    \includegraphics[scale=.3]{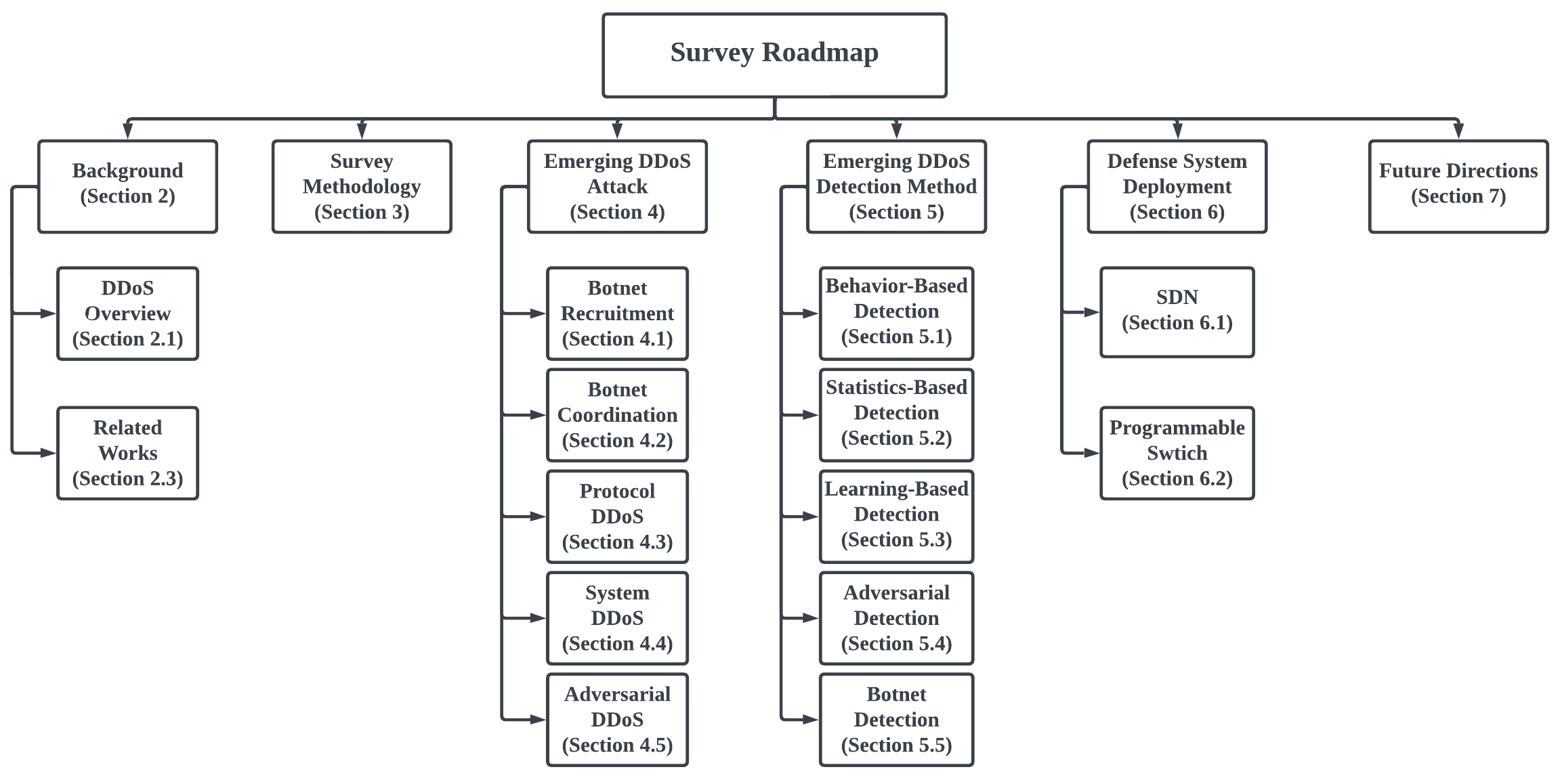}
    \caption{Survey roadmap}
    \label{fig:survey-taxonomy}
\end{figure*}

The roadmap of the survey is shown in Figure~\ref{fig:survey-taxonomy}.
In Section 2, we present an overview of DDoS attacks, highlighting significant historical cases, and delving into the existing research in the field.
Section 3 outlines the methodology behind our survey.
Section 4 examines the sophisticated evolution of botnet recruitment and coordination techniques, as well as the progression of DDoS attacks that exploit new network protocols, systems, and incorporate adversarial tactics.
Advancing into Section 5, we explore a variety of proposed methods for DDoS attack detection, organizing them by their heuristic approaches and techniques employed.
The conversation further evolves in Section 6, where we assess the innovative deployment methods for DDoS defense systems, made possible by cutting-edge network primitives such as programmable switches.
Lastly, Section 7 contemplates the future of DDoS attack vulnerability analysis and modern detection systems, providing a roadmap for ongoing research in the domain.
\section{Background and Related Work}

\subsection{DDoS Overview}
DDoS attacks represent a formidable threat in the cyber landscape, designed to overwhelm and incapacitate an online service, rendering it inaccessible to its intended users~\cite{mirkovic2004taxonomy}.
These attacks are achieved by bombarding the target network or system with malicious traffic that far exceeds its processing capacity or triggers specific protocol/system vulnerabilities, leading to a breakdown in its services and functionalities.
This disruption can have significant repercussions, ranging from financial losses and damage to reputation~\cite{sachdeva2010ddos}, to broader impacts on internet infrastructure and service availability~\cite{sommese2022investigating,abhishta2019measuring}.
The anatomy of a DDoS attack is typically structured around a meticulously planned and executed workflow, generally encompassing three critical phases: Bot recruitment, malicious traffic generation, and traffic orchestration.

The bot recruitment stage lays the groundwork for the attack by establishing a network of compromised devices, known as a botnet~\cite{wang2018delving,rodriguez2013survey}.
Cyber attackers infiltrate these devices through malware or exploiting vulnerabilities to gain control over them without the owners' knowledge.
Each compromised device (called "bot") is then poised to contribute to the deluge of traffic directed at the target.
The size of the botnet can be a determining factor in the potential impact of the ensuing attack, with larger botnets capable of generating more traffic and causing more significant disruption.

Once an attacker has established a botnet, the next step is to coordinate the production of malicious traffic~\cite{wang2018data,de2018ddos}.
This traffic is not legitimate user data; instead, it is designed to mimic or disrupt normal traffic, thereby creating an overload condition.
Methods of generating this traffic can range from simple, such as flooding the target with superfluous requests, to complex, involving crafted packets that exploit specific vulnerabilities or weaknesses in the target's infrastructure.
The sophistication of this phase can vary, but the goal remains consistent: Generate enough traffic to exceed the target's handling capacity.

The final phase involves the strategic management and direction of the generated traffic towards the target system.
This step is akin to conducting an orchestra, with the attacker ensuring that the compromised devices in the botnet act in unison to deliver the attack traffic in a coordinated and timely manner~\cite{obaidat2023creating}.
Effective orchestration can amplify the impact of the attack, as it seeks to exploit choke points in the network or times of peak user activity to maximize disruption.
Moreover, the pattern and volume of the traffic generated by each bot can also be adjusted to evade detection and mitigation efforts~\cite{wang2018data}, making the attack more difficult to counter and resolve.

\subsection{Related Work}\label{subsec:related-work}
Several survey papers have been published on DDoS attacks and detection.
The comparative analysis of the relevant survey papers is given in Table~\ref{tab:comparison-existing-survey}.
Specifically, the comparison is performed on the scope of DDoS attack and detection.
For DDoS attack, plenty of works focus on protocol exploitation.
These works~\cite{kumari2023comprehensive,li2023comprehensive,yan2015software} surveyed traditional protocol vulnerabilities, e.g., ICMP, TCP, and HTTP.
However, vulnerabilities for emerging protocols (e.g., HTTP/2 and IoT-specific protocols) are rarely discussed.
Moreover, existing surveys usually focus on specific systems.
For instance, Agrawal et al.~\cite{agrawal2019defense} present a survey that explores DDoS attacks within the context of cloud computing, and Kumari et al.~\cite{kumari2023comprehensive,chaudhary2023ddos} focuses on the IoT ecosystem instead.
The narrow scope of protocols and systems under investigation hinders the investigation of DDoS attack trends like low attack cost and common vulnerability patterns.
Additionally, the discussion of botnets and adversarial attack tactics is limited, while they become increasingly important in modern DDoS attacks.
In this survey, we studied DDoS attacks on nine network protocols and eight systems, from which we gained insight into the DDoS attack trends and the common vulnerability pattern.
We also surveyed related works which focus on botnet recruitment and coordination, and summarize the common exploits (e.g., weak authentication).
Finally, we summarized three types of adversarial tactics, revealing the emerging trend for adversarial DDoS. 
\begin{table*}
\centering
\caption{Comparisons with existing surveys on DDoS attacks and detection. The empty, half, and full circles mean ``Not mentioned", ``Partially mentioned", and ``Mentioned", respectively.}
\label{tab:comparison-existing-survey}
\scalebox{.8}{
\begin{tblr}{
  cells = {c},
  cell{1}{1} = {r=2}{},
  cell{1}{2} = {c=4}{},
  cell{1}{6} = {c=4}{},
  vlines,
  hline{1,3-4} = {-}{},
  hline{2} = {2-9}{},
}
\textbf{Authors} & \textbf{DDoS Attack}       &                          &                 &                                       & \textbf{DDoS Detection\textbf{}}                      &                                            &                                                         &                                          \\
                 & \textbf{\textbf{Protocol}} & \textbf{\textbf{System}} & \textbf{Botnet} & {\textbf{Adversarial}\\\textbf{DDoS}} & {\textbf{\textbf{Behavior}}\\\textbf{\textbf{Based}}} & {\textbf{Adversarial}\\\textbf{Detection}} & {\textbf{\textbf{Botnet}}\\\textbf{\textbf{Detection}}} & {\textbf{Innovative}\\\textbf{Hardware}} \\
Agrawal et al.~\cite{agrawal2019defense}                 &      \emptycirc                      &   \halfcirc                       &    \emptycirc             &   \emptycirc                                    &        \fullcirc                                               &     \emptycirc                                       &               \emptycirc                                          & \emptycirc \\
Li et al.~\cite{li2023comprehensive}                 &      \fullcirc                      &   \emptycirc                       &    \halfcirc             &   \emptycirc                                    &        \emptycirc                                               &     \emptycirc                                       &               \emptycirc                                          & \fullcirc \\
\hline
Kumari et al.~\cite{kumari2023comprehensive,chaudhary2023ddos}                 &      \fullcirc                      &   \halfcirc                       &    \fullcirc             &   \emptycirc                                    &        \emptycirc                                               &     \emptycirc                                       &               \emptycirc                                          & \emptycirc \\
\hline
Yan et al.\cite{yan2015software}                 &      \fullcirc                      &   \halfcirc                       &    \emptycirc             &   \emptycirc                                    &        \emptycirc                                               &     \emptycirc                                       &               \emptycirc                                          & \halfcirc \\
\hline
Zhang et al.\cite{zhang2024revealing}                 &      \halfcirc                      &   \emptycirc                       &    \emptycirc             &   \emptycirc                                    &        \halfcirc                                               &     \emptycirc                                       &               \emptycirc                                          & \halfcirc \\
\hline
Praseed et al.\cite{praseed2018ddos}                 &      \fullcirc                      &   \halfcirc                       &    \emptycirc             &   \emptycirc                                    &        \fullcirc                                               &     \emptycirc                                       &               \emptycirc                                          & \emptycirc \\
\hline
Our work                 &      \fullcirc                      &   \fullcirc                       &    \fullcirc             &   \fullcirc                                    &        \fullcirc                                               &     \fullcirc                                       &               \fullcirc                                          & \fullcirc \\
\hline
\end{tblr}
}
\end{table*}

For DDoS detection, a subset of works studied detection strategies based on the attack behavior~\cite{agrawal2019defense,praseed2018ddos,zhang2024revealing}.
For instance, Agrawal and Praseed et al.~\cite{agrawal2019defense,praseed2018ddos} discuss detection strategies for volumetric, low-rate, amplification attacks, e.t.c.
Zhang et al.~\cite{zhang2024revealing} mainly focus on the volumetric attack instead.
Besides the behavior-based detection strategy, our work includes a more comprehensive detection taxonomy, including behavior-based, statistics-based, learning-based, adversarial-based, and botnet detection methods.
Notably, the adversarial-based detection methods are rarely discussed in existing surveys, while our work covers it.
Some works also discuss innovative use of emerging hardware technologies for DDoS attack defense~\cite{yan2015software,li2023comprehensive}.
In particular, Yan et al.\cite{yan2015software} explore the implications of software-defined networking (SDN) in this domain, and Li et al.\cite{li2023comprehensive} consider the role of programmable switches.
These discussions focus on how such hardware can lower the costs associated with deploying DDoS attacks.
However, they fail to fully explore the unique capabilities of these technologies, such as line-speed packet processing, which could significantly enhance the efficiency and adaptability of DDoS attack detection systems.
\section{Survey Methodology}
We start by introducing how we collect the papers from the literature and filter out most relevant papers. 
We aim to collect well-researched papers that span the last decade and are from the literature of DDoS attack and detection.
Specifically, we first leverage advanced searches to collect a number of papers from the conferences and transactions that are sponsored by IEEE, USENIX, ACM, and Elsevier.
We search papers with keywords "DDoS" and "distributed denial of service".
Moreover, we restrict the type to be the research article.
In this context, we acquire 3,348, 5,408, 666, and 31 papers from IEEE Xplore, ACM library, Elsevier ScienceDirect, and USENIX, respectively.
Then we filter out papers based on the ranking of their publication venue, retaining only papers from highly-ranked conferences and transactions to ensure quality.
In particular, we selects top-tier venues from Google Scholar Metrics, Conference Ranks, Core Conference Rankings, and China Computer Federation.
Specifically, we focused on subcategory of system, network, and security on these ranking sources, and pulled the top-20 ranking lists.
As a result, we totally selected 87 venues from these ranking sources, and examples of the selected venues are CCS, S\&P, USENIX Security, and TDSC.

Finally, considering that some papers do not focus on DDoS attack/detection but just occasionally mention the word "DDoS" somewhere in the paper, we extend the keyword dictionary by including "attack" and "detection".
For each paper, we further calculate a relevance degree by counting the frequency of keywords in the extended keyword dictionary, and sort these papers in descending order of their relevance degrees.
As a result, papers with low degrees or even zero degrees (e.g., false positives that are wrongly returned by the sponsor's search engine) are discarded.
Each preserved paper describes a concrete attack or detection technique about distributed denial of service.
Eventually, we select 184 papers for deep examination. 

\section{DDoS Attacks}\label{sec:ddos-attack}
In this section, we present a comprehensive review of the latest developments in DDoS attacks.
A successful DDoS attack include botnet formation, exploiting target selection, and malicious traffic generation.
Specifically, botnet formation involves botnet recruitment and coordination.
Attackers first create a network of compromised computers, known as a botnet, by exploiting vulnerabilities in devices to install malware.
Then the attacker uses command and control (C\&C) servers to manage the botnet for synchronized coordination of the attack.
With the botnet, the next step is to select exploiting targets.
Attackers may identify vulnerabilities in network protocols, such as HTTP, DNS, or TCP/IP, to exploit during the attack.
Specific features and weaknesses in the target system (e.g., content caching) can also be leveraged.
Finally, the attacker decides on the type of malicious traffic to generate (e.g., SYN request).
Attackers also design the traffic pattern to maximize disruption, potentially using slow and low attacks to evade detection or high-volume bursts to overwhelm the target quickly.
Adversarial tactics can be enforced during the attack (e.g., encrypted traffic), which bypass traditional detection efforts.

\begin{figure*}
    \centering
    \includegraphics[scale=.3]{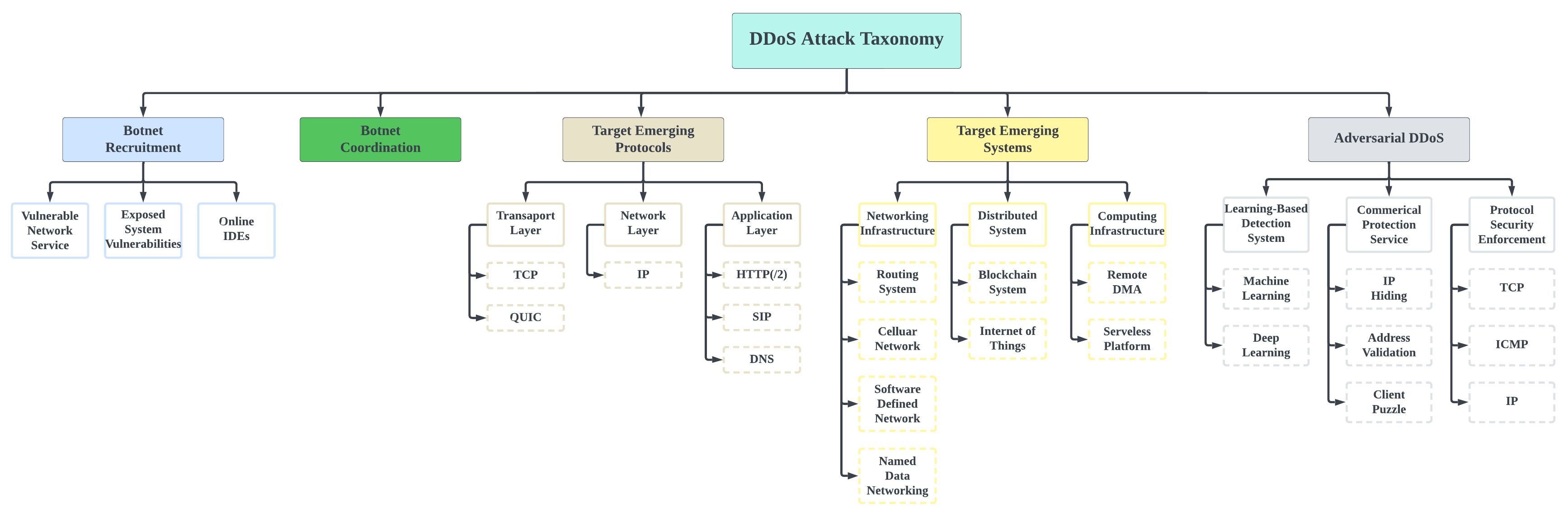}
    \caption{Taxonomy of DDoS attacks.}
    \label{fig:ddos-attack-taxonomy}
\end{figure*}

Following the DDoS attack workflow, our exploration begins with an examination of sophisticated methodologies for botnet recruitment and coordination, which serve as the primary mechanism for attackers to orchestrate DDoS campaigns (see Section~\ref{subsec:botnet-setup} and Section~\ref{subsec:botnet-coordination} for details).
We proceed to classify the spectrum of current DDoS threats from three innovative angles:
Firstly, attacks that exploit emerging network protocols are addressed in Section~\ref{subsec:ddos-targeting-protocol};
secondly, we discuss attacks that specifically target new and evolving systems in Section~\ref{subsec:ddos-targeting-system};
and thirdly, we delve into adversarial strategies designed to evade detection mechanisms, outlined in Section~\ref{subsec:adversarial-ddos-bypassing-detection}.
A visual representation of the DDoS attack taxonomy and a summarizing overview can be found in Figure~\ref{fig:ddos-attack-taxonomy}.

\subsection{Botnet Recruitment}\label{subsec:botnet-setup}
To mount a formidable DDoS attack, attackers must first construct a botnet by amassing a collection of compromised devices known as bots. This section outlines three prevalent strategies for bot recruitment.

\textbf{Network services with weak authentication.}
The pivotal role of vulnerable network services as the primary channels for malware distribution has been consistently underscored by recent research.
Pa et al.~\cite{pa2016iotpot} innovatively created IoTPOT, a honeypot emulating the Telnet vulnerabilities specific to IoT devices.
This honeypot’s effectiveness was proven by the 481,521 malware download attempts it recorded, vividly illustrating the allure of weakly secured network services to potential attackers.
The study detailed the malware propagation process, beginning with attackers exploiting a list of common Telnet credentials to infiltrate devices.
Once access was gained, the attackers proceeded to download a malicious binary, setting the stage for monetization of the breach.
The compromised devices were predominantly used for DDoS attacks, demonstrating the severe consequences of inadequate authentication measures.
Supporting this, Choi et al.~\cite{choi2022understanding} provided an analysis of malware's traffic patterns, noting a substantial sharing of target IPs across varying malware sources—a testament to the widespread exploitation of network service vulnerabilities.
Their findings indicated a preferential exploitation of ports 80 (HTTP service) and 22 (TCP service), although attackers also strategically targeted lesser-known ports such as 111 and 123.

\textbf{Exposed system vulnerabilities.}
The strategic exploitation of device vulnerabilities for the dissemination of DDoS malware is a widely recognized tactic among attackers.
Al et al.~\cite{al2023bin} conducted a comprehensive analysis of 11,893 malware binaries, revealing that a significant number—2,629 binaries—specifically targeted known vulnerabilities documented in databases like CVE and NVD.
This starkly highlights how reported vulnerabilities can serve as a beacon for attackers to compromise devices and assimilate them into botnets.
The research further categorized the vulnerabilities, with remote code execution vulnerabilities being the most prevalent, followed by command injection.
This prioritization indicates a tactical selection by attackers, opting for vulnerabilities that provide the most control over compromised devices.
In their targeting strategy, attackers exhibited a preference for unpatched vulnerabilities or those with complex patching processes, revealing a calculated exploitation of slower mitigation responses.
Additionally, Al et al. pointed out that the availability of public proof-of-concept (PoC) exploits considerably increased the risk of a vulnerability being targeted.
These PoC exploits act as a double-edged sword: while they serve to inform security practitioners about potential vulnerabilities, they also provide attackers with a roadmap for exploitation.
Contrary to what might be expected, the study observed that the severity of a vulnerability did not necessarily correlate with its exploitation frequency. This insight suggests that attackers are opportunistic, focusing less on the potential impact of a vulnerability and more on the ease of exploitation and the effectiveness in spreading DDoS malware.

\textbf{Online integrated development environments (IDEs)}.
Srinivasa et al.~\cite{srinivasa2022bad} have unearthed a concerning trend wherein online Integrated Development Environments (IDEs) emerge as unsuspecting tools in the hands of attackers to propagate DDoS malware.
Their study reveals that the inherent vulnerabilities within unregulated online IDEs can be systematically exploited, turning the IDE servers into unwilling participants in DDoS attacks as part of a botnet.
The crux of the issue lies in the permissiveness of these online IDEs, which, with their unrestricted imports and non-sandboxed operational environments, allow attackers to effortlessly introduce malicious libraries.
In addition to the ease of infiltration, the potential for unbounded resource consumption within these platforms presents an opportunity for attackers to orchestrate attacks of substantial scale.
The research highlighted the alarming prevalence of such vulnerabilities, with 719 out of 2,269 online Python IDEs identified as uncontrolled.
The capability of these compromised online IDEs to generate massive amounts of network traffic serves as a testament to their potency in DDoS attacks.
The study indicates that a coordinated utilization of just 32 of these vulnerable IDEs is enough to unleash an average of 6 million requests per minute, capable of overwhelming systems and bringing down critical online services.

\subsection{Botnet Coordination}\label{subsec:botnet-coordination}
Once the bots are recruited, the attackers coordinate them for the targets.
While the size of the botnet and the target may vary, existing works point out some common coordination strategies.
Wang et al.~\cite{wang2018data} investigate 50,704 different Internet DDoS attacks across the globe in a seven-month period, and study how attackers scheduled their controlled bots.
The result shows that attackers deliberately schedule their controlled bots in a dynamic fashion.
For example, attackers do not uniformly sample bots across different countries. Instead, the distribution shifts over time, and such shifts can be statistically modeled.
Moreover, the author identifies similar shifting patterns over different botnets, e.g., Dirtjumper and Pandora.
This implies that different botnets may collaborate or share resources when launching DDoS, and a single bot may participate in attacks launched by different botnets.

Attackers may also validate the effectiveness of their coordination strategies by simulation.
For example, Obaidat et al.~\cite{obaidat2023creating} provide a simulation framework DDOSim.
It allows attacker to build attacker/victim nodes by loading docker containers with their malicious binaries and software.
Network stacks, malware, attack scripts, and softwarized defense systems can be contained in the container to support diverse evaluation tasks.
Moreover, attackers can customize the simulated network topology and configuration, such that he can evaluate the attack impacts under different network conditions.
Finally, with quantitative measurement of various metrics, e.g., server network throughput, DDOSim achieves real-time monitoring and evaluation of attack/defense progression.

\subsection{DDoS Targeting Protocols}\label{subsec:ddos-targeting-protocol}
As network protocols evolve to offer improved speed, reliability, and security, malicious actors adapt their strategies accordingly.
DDoS attacks are now progressively aimed at these advanced protocols, taking advantage of vulnerabilities that have yet to be addressed with security patches or protocol enhancements.
At the same time, the advanced features of these protocols (e.g., multiplexing) are being exploited to design advanced DDoS attacks.
This section will explore the protocols that are susceptible to such attacks, identify their specific vulnerabilities, and describe the methods by which attackers exploit these weaknesses to carry out DDoS campaigns.

\subsubsection{Transport-Layer Protocol}
\textbf{Transmission Control Protocol (TCP)}.
TCP serves as the backbone of the internet, facilitating reliable communication across its vast network.
However, the ubiquity of TCP also makes it a prime target for security exploits, with its inherent vulnerabilities becoming a focal point for researchers.
The most famous attacks targeting TCP are SYN flooding and Shrew DDoS attacks.
Specifically, SYN flooding attacks leverage the TCP three-way handshake by sending excessive SYN requests, which the server responds to with SYN/ACK packets, awaiting completion of the connection that never occurs~\cite{wang2002detecting}.
This results in an accumulation of half-open connections in the server's backlog queue—a memory structure with limited capacity.
Once this queue is full, legitimate connection requests are rejected, leading to a denial of service.
This attack method exploits the finite size of the backlog queue and highlights a critical vulnerability within the TCP session management. 

Luo et al.~\cite{luo2014mathematical} conducted a formal analysis of the Shrew DDoS attack, which manipulates the TCP's retransmission timeout (RTO) mechanism.
By sending high-rate packet bursts at a frequency that aligns with the RTO intervals, the Shrew attack induces repeated timeouts in legitimate TCP connections.
This results in severe congestion at the network bottleneck upon each recovery attempt, causing the throughput of legitimate traffic to plummet, potentially to near-zero levels.
Luo et al. developed a mathematical model to assess how varying attack patterns and network conditions influence the success of this attack, delineating the least resources required for a successful Shrew attack and its potential maximum impact.

Expanding upon this concept, Tang et al.~\cite{tang2013modeling} generalized the principles behind the Shrew attack to formulate a model for low-rate denial-of-service attacks, also known as Reduction of Quality (RoQ) attacks.
These attacks exploit the feedback control mechanisms within network protocols, e.g., TCP's dynamic congestion window adjustment, forcing the victim's system into a suboptimal state controlled by the attacker.
This strategy effectively diverges the system from its intended operational state.

It’s important to note that low-rate denial-of-service attacks are not exclusive to the TCP protocol.
Schuchard et al.~\cite{schuchard2010losing} extended the attack vector to the Border Gateway Protocol (BGP), introducing the Coordinated Cross Plane Session Termination (CXPST) attack.
This method disrupts BGP sessions by causing intermittent link congestion, leading to repeated session disconnections and reconnections between victim routers.
In the wireless domain, Chen et al.~\cite{chen2008feasibility} demonstrated the feasibility of such attacks on 802.11 networks by periodically interfering with TCP acknowledgment (ACK) packets, forcing the sender into unnecessary retransmissions and throttling the transmission rate.
Lastly, He et al.~\cite{he2009reduction} applied the low-rate DDoS concept to peer-to-peer (P2P) protocols, showing that attackers can destabilize these networks through timed patterns of joining and leaving, knocking the system off its stable equilibrium.

\textbf{Quick UDP Internet Connection (QUIC).}
QUIC, a modern transport protocol analogous to UDP, was created by Google and has been standardized by the Internet Engineering Task Force (IETF).
This protocol is designed to enhance transport layer security and privacy while simultaneously reducing connection establishment latency.
Despite these improvements, research has uncovered that QUIC can be susceptible to Distributed Denial of Service (DDoS) attacks.

Nawrocki et al.~\cite{nawrocki2021quicsand} explore two primary attack strategies that threaten the integrity of QUIC: state-overflow and reflective amplification attacks.
A state-overflow attack involves an attacker impersonating a QUIC client to inundate the server with a deluge of connection states.
The attacker initiates handshakes repeatedly, compelling the server to consume resources to track each supposed connection by issuing a unique Source Connection ID (SCID) and its corresponding Transport Layer Security (TLS) certificate.
These fraudulent requests impose a heavy cryptographic load and exhaust server resources dedicated to managing connection states.
The attacker exacerbates this situation by using spoofed IP addresses and port numbers, inflating the server's state management workload to the point where it may become incapable of serving legitimate requests, leading to service outages.

Reflective amplification attacks exploit the UDP foundation of QUIC, enabling IP spoofing.
In this scenario, attackers control numerous bots that send QUIC Initial packets with falsified source IP addresses—typically that of the intended victim—to a QUIC server.
The server then replies with QUIC Initial messages that include TLS handshakes, which, by including server certificates, are significantly larger than the incoming requests.
The QUIC standard restricts servers from sending more than three times the data they receive prior to client verification.
Attackers circumvent this by padding Initial packets with superfluous bytes, inflating the response size.
This tactic is deceptive, as it resembles a recommended practice for streamlining the handshake process, where large initial packets enable the server to transmit certificates in a single message, thus reducing delays.
Consequently, the victim's network is bombarded with an overwhelming response from the server.

\subsubsection{Network-Layer Protocol}
\textbf{Internet Protocol (IP).}
The integrity of the Internet Protocol (IP) is critical for the stable operation of networked systems.
However, research has brought to light substantial vulnerabilities within the IP protocol that can be weaponized to execute denial of service attacks.
A primary method of exploitation is the IP fragmentation attack, which targets inherent weaknesses in the IP fragmentation process.
Attackers can send meticulously crafted packets with overlapping fragment offsets, which confound the target system's ability to correctly reassemble the fragments.
This confusion can lead to improper packet reassembly or even buffer overflows, potentially resulting in crashes and service interruptions.

Gilad et al.~\cite{gilad2011fragmentation} have demonstrated that even attackers who lack a direct path to the communication stream can predict the IP identification values that packets will use.
These values are crucial for the reassembly process, as they indicate which fragments belong to which packets.
Attackers can then send fraudulent fragments bearing these anticipated identifiers to a victim.
When the legitimate fragments are received, they are incorrectly reassembled with the attacker's fragments, causing packet corruption and loss.
In the realm of IPv6, Atlasis et al.~\cite{atlasis2012attacking} reveal that the exploitation potential is more severe.
Unlike IPv4, IPv6 introduces a more complex system of extension headers.
Attackers can exploit this by deliberately crafting packets that split critical TCP/UDP header information across multiple fragments and bypass firewall detection.
The initial malicious fragment which contains the IPv6 header and possibly a fragment of the TCP/UDP header can pass through the firewall, since it typically inspects only the IPv6 header of the traffic. Subsequent fragments, which may contain the rest of the TCP/UDP headers, can slip through the firewall unchecked since many security systems do not perform full reassembly of packet fragments before inspection.

\subsubsection{Application-Layer Protocol}
\textbf{HTTP(/2)}.
DDoS attacks have evolved significantly with the advancement of web protocols.
Attacks that exploit the HTTP protocol, particularly HTTP/1.x and HTTP/2, have become a pressing concern.
Dantas et al.~\cite{dantas2014selective} have identified three distinct DDoS attack strategies that exploit HTTP/1.x: HTTP GET, HTTP PRAGMA, and HTTP POST.
Slow Write attacks, which make use of the HTTP GET and POST requests, operate by sending request fragments to a server at a deliberately slow pace.
This was notably employed in the Slowloris attack, which emerged after the 2009 Iranian Presidential elections.
By transmitting tiny fragments of a request and pausing until just before the server's timeout interval expires, attackers can keep connections open indefinitely.
This forces servers to maintain these malicious connections, eventually exhausting their capacity to accept legitimate requests.
The efficacy of these attacks is further enhanced by exploiting the HTTP PRAGMA header.
By including this header in requests, attackers can reset the server's timeout timer, allowing the malicious connection to persist even longer.
This tactic effectively monopolizes server resources, leaving fewer sockets available for genuine users.

The research by Beckett et al.\cite{beckett2017http} and Praseed et al.\cite{praseed2019multiplexed} reveals the susceptibility of HTTP/2 to DDoS attacks.
While HTTP/2 introduced advanced features like multiplexing (the ability to send multiple requests/responses in a single connection) and server push (the server pre-emptively sends resources to the client), they can be weaponized by attackers.
Beckett et al.~\cite{beckett2017http} demonstrate that HTTP/2's multiplexing capability amplifies the impact of HTTP floods.
Attackers can bundle numerous requests into a single packet, leading to a flood that is significantly more potent than one using HTTP/1.1.
The study showed that with the same rate of packet transmission, an attack could be magnified by up to 95 times when compared to HTTP/1.1.
Building on this, Praseed et al.~\cite{praseed2019multiplexed} introduce an enhanced multiplexing attack that selects high-workload requests to maximize the target server's CPU usage while remaining under the radar.
The server push feature exacerbates the situation; it prompts the server to handle not only the direct requests but also the associated inline requests, such as resources linked to a web page.
This can lead to the server's CPU usage spiking to 80\% with as few as four attacking bots, illustrating the severe impact of such attacks.

\textbf{SIP}.
The Session Initiation Protocol (SIP) is a cornerstone of Voice over Internet Protocol (VoIP) technologies, facilitating a wide array of communication services.
As an application-layer protocol, SIP relies heavily on the functionality of intermediate proxy servers to manage the signaling and control of voice sessions.
Despite its widespread adoption, SIP's reliance on these servers and its session management mechanisms introduce vulnerabilities ripe for exploitation.
Sisalem et al.~\cite{sisalem2006denial} and Tang et al.~\cite{tang2014sip} have conducted extensive research into the vulnerabilities inherent in SIP, identifying multiple avenues through which attackers can launch Denial of Service (DoS) attacks.
Their work categorizes three primary types of SIP flooding attacks that affect the stability and availability of SIP proxy servers—INVITE flooding, BYE flooding, and Multi-Attribute flooding.

The INVITE flooding attack is a method by which an attacker overwhelms the SIP proxy with an excessive number of INVITE requests.
These requests aim to initiate new SIP sessions, and the proxy server, in attempting to maintain state information for each session, eventually depletes its memory resources.
This form of attack targets the fundamental role of the proxy in establishing communication sessions, thereby crippling its ability to service legitimate users.
BYE flooding takes a different approach.
In this scenario, the attacker sends a large volume of spoofed BYE messages, which are protocol methods designed to terminate existing SIP sessions.
By generating these messages with brute-forced user addresses, the attacker can trick the proxy into prematurely ending a substantial number of active VoIP calls from benign users, causing widespread disruption.
Finally, the Multi-Attribute flooding attack combines various forms of SIP flooding, such as INVITE and BYE flooding, to create a more complex and damaging assault.
By varying the attack vectors, attackers can inflict compounded harm on the proxy server while simultaneously evading detection systems that typically rely on analyzing the proportion of different SIP methods used in the traffic flow.

\textbf{Domain Name System (DNS).}
The Domain Name System (DNS) is a critical component of the Internet's infrastructure, underpinning the resolution of domain names into IP addresses.
Its significance is paralleled by its attractiveness as a target for Distributed Denial of Service (DDoS) attacks, particularly those aiming for amplification.
DNS servers, especially those that are publicly accessible, are essential for handling name resolution requests from clients.
These servers are capable of querying multiple DNS zones—each containing a set of DNS records—in a single request.
This capability makes them prime targets for amplification attacks because they often return responses significantly larger than the incoming requests.
Moreover, since DNS protocol relies on UDP, which does not require a connection and is susceptible to IP address spoofing, it allows attackers to alter the source address of DNS queries, making the responses go to the victim's IP (i.e., reflection).

The DNS amplification attack, as described by Kim et al.~\cite{kim2017preventing}, exploits these characteristics.
Attackers forge the source IP in a DNS request to match that of their intended victim.
The DNS server, unaware of the spoofing, sends a response, which can be many times larger than the request, to the victim's IP address, thereby flooding it with unsolicited traffic.
The attack is highly efficient and difficult to trace for two reasons.
(1) Economical: It requires minimal effort from the attacker, who need only generate small query packets to elicit large responses, resulting in a significant amplification of traffic.
(2) The attack traffic appears to originate from legitimate DNS servers, not the attacker, making it challenging to identify the true source through traffic analysis.
To identify candidate DNS resolvers, Yazdani et al.~\cite{yazdani2022mirrors} explore the misuse of cloud-based DNS infrastructures.
Their findings indicate that a substantial number—around 12\%—of the 3 million DNS resolvers analyzed are hosted within cloud networks.
These cloud-based resolvers can be powerful instruments for attackers seeking to amplify their attacks. Furthermore, some cloud providers' lack of destination-side address validation exacerbates the vulnerability of their DNS resolvers to external attacks.

Griffioen et al.~\cite{griffioen2021scan} examined the procedures of amplification DDoS attacks through the deployment of 549 honeypots across five public cloud platforms.
The researchers implemented traffic shaping techniques to ensure that these honeypots did not contribute to the attacks while enabling the monitoring and analysis of the attacks' characteristics.
Throughout the duration of the study, approximately 13,000 attacks were recorded, leading to several noteworthy discoveries.
Firstly, the study found that attackers engage in preliminary testing of servers to evaluate their potential use in amplification attacks.
This testing involves sending bursts of requests to assess whether the servers' responses are consistent with expected protocol behaviors.
Moreover, the data revealed that attackers keep track of servers that have previously demonstrated a high amplification factor.
Evidence from honeypot records indicated that attackers would revisit IP addresses of servers that had been effective amplifiers in the past, even if those servers had since ceased responding.
Lastly, the research highlighted a tactical approach employed by sophisticated attackers: Pulsing their traffic instead of sending a constant stream.
This strategy is indicative of an effort to optimize the cost-efficiency of their attacks.

Note that in addition to DNS, a multitude of network protocols has been exploited to facilitate amplification attacks. In 2020, the Federal Bureau of Investigation (FBI) issued an alert regarding the exploitation of three specific network protocols: Apple Remote Management Services (ARMS), Web Services Dynamic Discovery (WS-DD), and the Constrained Application Protocol (CoAP)~\cite{fbi-amplification-attack}. These protocols were found to be vulnerable to misuse for amplifying malicious traffic in Distributed Denial of Service (DDoS) attacks.
In the following year, SECURELIST expanded on this list by identifying three additional protocols that had been abused: the Microsoft Remote Desktop Protocol (RDP), the Chameleon Protocol for Virtual Private Networks (VPNs), and the Datagram Transport Layer Security Protocol (DTLS)~\cite{securelist-amplification-attack}.
The discovery of such a diverse array of exploitable protocols for amplification underscores the attractiveness of these methods to attackers.
The potential for significant damage and the relative ease of orchestrating these attacks make them a persistent threat in the cyber landscape.

Beyond amplification DDoS, Yin et al.~\cite{yin2023waterpurifier} introduce the concept of the DNS water torture attack.
This method exploits the recursive nature of DNS resolution, where queries are forwarded between servers, rather than leveraging the size disparity between request and response.
In this attack, a botnet inundates a target domain, such as example.com, with requests for non-existent subdomains.
Due to the recursive lookup required to resolve these fabricated subdomains, the authoritative server for example.com becomes overwhelmed.
Consequently, legitimate DNS queries for the domain fail, leading to service disruption for the targeted domain.

Pan et al.~\cite{pan2024loopy} reveal that with spoofed packets, attackers can create loops between two servers.
Severely, attackers can create infinite communication loops between two servers with even single packet.
The root cause is the flaw design about error message handling.
An error message as input can create an error message as output for two DNS systems.
As a result, with a spoofed error message,  two DNS systems will keep
sending error messages back and forth indefinitely.
Such vulnerabilities can be easily exploited to create DDoS attacks.
For example, an attacker can create many loops with other loop
servers, all of which concentrate on a single target loop server.
As a result, the target server either exhausts its host
bandwidth or computational resources.

\textbf{IoT protocols.}
The Modbus protocol is designed to facilitate communication among IoT devices within industrial control systems, such as electricity and gas supply networks.
Due to their critical importance, these systems often become targets for attackers seeking to hijack and disable them, with the exploitation of the Modbus protocol being a primary focus.
Mohammed et al.~\cite{mohammed2023detection} introduced a novel field flooding attack that leverages the structure of Modbus packets to execute a DoS attack.
The adversary can craft malicious packets by modifying the ModbusTCP packet header, and these modifications aim to enlarge the allocated memory for the malicious packets.
For instance, the adversary can modify the length field in the write packet header with extremely large valuess.
Consequently, this triggers the target control units to allocate large memory and create an overflow of the memory bank, causing them to crash.

Besides IoT protocols used in the industrial system, researchers show that protocols for smart home devices are also vulnerable.
Wang et al.~\cite{wang2022zigbee} exposes a specific DoS vulnerability lurking in the device rejoin procedure of the Zigbee protocol.
The study demonstrates how an attacker can exploit this vulnerability by coordinating compromised Zigbee devices to send falsified rejoin requests to target routers.
A flaw in the aging-out process of the rejoin procedure is revealed, allowing these unauthorized connections to overwhelm the router’s capacity, with the router erroneously maintaining these connections.
This in turn prevents legitimate Zigbee devices from joining the network, effectively resulting in a denial of service for the intended users.

\subsection{DDoS Targeting Systems}\label{subsec:ddos-targeting-system}
The advancement of technology brings with it innovative systems aimed at enhancing efficiency and user experience.
However, alongside these developments, a worrying trend emerges in the landscape of cyber threats.
Recent patterns in denial-of-service attacks reveal a targeted interest in these modern systems.
Malicious actors are keenly searching for and exploiting vulnerabilities in these cutting-edge frameworks, particularly those still in early stages of deployment.
Within this section, we will delve into the systems impacted by these threats, examine their specific vulnerabilities, and explore the methods attackers employ to leverage these weaknesses, resulting in denial-of-service incidents.

\subsubsection{Networking Infrastructure}

\textbf{Routing system.}
The routing system, an intricate web of routers and connecting links, is pivotal in directing network traffic.
Its seamless operation is essential for maintaining the integrity and availability of network services.
However, recent research illustrates that this system is not impervious to attack; malefactors have developed methods to exploit it, potentially causing widespread denial of service that could cripple a district or even bring a nation's digital infrastructure to a standstill.

Studer et al.~\cite{studer2009coremelt} have uncovered novel attack strategies that specifically target and overwhelm crucial network links, severing connections to the intended victim host.
For instance, the Coremelt attack operates by utilizing a network of compromised machines that exchange high volumes of data, thereby inundating and incapacitating a vital link within the network.
This results in a denial of service for all servers dependent on the affected link.
The insidious nature of this attack lies in the fact that the compromised machines, being the recipients of the flooding traffic, enable the attacker to circumvent traditional filtering-based DoS defenses that are typically employed to protect the server.

The Coremelt attack presupposes a substantial botnet under the attacker's control, with these bots strategically positioned both upstream and downstream of the target link.
Recognizing the limitation of this assumption, further research by Kang et al. introduced the Crossfire attack~\cite{kang2013crossfire}, which aims to mitigate the dependency on a vast botnet.
This method involves coordinating the bots to send traffic to a series of decoy servers, strategically situated downstream of the critical link.
The malicious traffic, destined for the decoy servers, must traverse the targeted link, resulting in its congestion. Consequently, all servers within the region that rely on this link would suffer from a denial of service.
The Crossfire attack represents an evolution in denial of service techniques by reducing the reliance on the number and distribution of bots, and instead focusing on the strategic generation of traffic to exploit the routing system's vulnerabilities.
Through the aggregation of traffic at critical junctures, attackers can induce a significant impact with fewer resources, posing a grave threat to the robustness of modern network infrastructure.

\textbf{Cellular network.}
The cellular network plays a crucial role in mobile communication, with the Long Term Evolution (LTE) standard—developed by the 3rd Generation Partnership Project (3GPP)—serving as the backbone for current and emerging cellular technologies, including 4G and 5G.
Despite its advancements, LTE is vulnerable to Distributed Denial of Service (DDoS) attacks that pose significant challenges to network stability and user security.

Attackers, aiming to disrupt the LTE network, amass mobile malware to create a formidable botnet.
By exploiting LTE's architecture—which distinctly separates the control plane (responsible for signaling) from the data plane (responsible for user data)—these attackers can specifically target the control plane with a deluge of signaling traffic.
Research has identified critical vulnerabilities in LTE procedures, such as the user attach and handover processes, where attackers can induce signaling storms with minimal effort, creating an amplification effect that leads to service disruptions~\cite{henrydoss2014critical,silva2020repel}.
Another aspect of DDoS susceptibility in mobile communications is network slicing, a key technology in 5G networks that enables differentiated services through a shared infrastructure.
Literature suggests that due to the inherent design of physical resource sharing in network slicing, an attack on a single service can have cascading effects, disrupting multiple services across different slices and magnifying the attack's impact~\cite{olimid20205g,sattar2019towards,javadpour2023reinforcement}.

The emergency service supported by the cellular network can also be exploited by attackers.
In particular, the 911 emergency service system, a critical component of public safety, is not immune to these threats.
Mirsky et al. highlighted the vulnerability of 911 services to DDoS attacks perpetrated through mobile phone botnets~\cite{mirsky2020ddos}.
An attacker orchestrates this by infecting smartphones with malware to form a botnet, which is then directed to place continuous emergency calls using randomized IMSIs.
The Federal Communications Commission's (FCC) mandate to route unidentified emergency calls without blocking creates an exploitable loophole.
The botnet, leveraging this policy and the randomized IMSIs, can evade detection by the cellular network and flood the 911 service infrastructure, resulting in a critical service outage.

\textbf{Software-Defined Network (SDN).}
SDN has revolutionized network architecture by decoupling the control plane from the data plane, thereby introducing greater flexibility and programmability.
However, this paradigm shift also presents novel vulnerabilities, particularly to denial of service attacks.
Among these, DDoS poses a critical threat due to its capability to leverage multiple launch points and its potential to inflict severe service disruptions.

A pivotal study by Shin et al.~\cite{shin2013avant} elucidates the vulnerability inherent in the separation of the control and data planes, particularly to what is termed a control plane saturation attack.
In an SDN environment, when a switch encounters a packet from an unrecognized flow, it refers the packet to the centralized controller for further instructions.
This controller is integral to the SDN's operation as it manages flow requests and configures the network dynamically. However, it is also a singular point of failure.
An adversary can exploit this by coordinating a multitude of compromised devices, or bots, to generate an overwhelming number of unique flow requests.
This orchestrated effort can saturate the control plane, effectively paralyzing the network's ability to manage legitimate traffic.

In addition to the control plane, the data plane is also susceptible to a similar form of exploitation.
As demonstrated in another work by Shin et al.~\cite{shin2013attacking}, attackers are capable of initiating a data plane saturation attack by inundating the network with a vast array of unique flows.
This barrage of flow requests leads to the generation of numerous redundant flow rules, which the data plane must process and store.
The data plane, encumbered by this deluge of spurious rules, becomes less efficient or even incapable of handling legitimate network flows, severely degrading network performance.

Cao et al.~\cite{cao2019crosspath} have further identified that attackers could exploit the shared links between control and data traffic paths, thereby disrupting the SDN control channel.
The proposed CrossPath Attack involves first probing the SDN with data traffic bursts to identify shared links by observing control message delays.
Once identified, the attacker can employ a low-rate, TCP-targeted DoS attack to create data traffic pulses, inducing congestion on these critical links and impairing control message transmission.

\textbf{Named Data Networking (NDN}).
The Named Data Networking (NDN) paradigm represents a promising shift in network infrastructure, focusing on content-centric operations rather than the traditional location-centric approach characteristic of the IP protocol.
Unlike the IP architecture which relies on specific location addresses, NDN operates on a named resource basis, allowing users to request content by name without requiring knowledge of its physical location.
For instance, a news article from CNN could be requested with the name `/ndn/cnn/news/2012May20', which NDN routers can process to retrieve the content directly, bypassing the need to locate the CNN server.
While NDN's inherent features, such as in-network caching and the symmetry of interest and content paths, offer a degree of resistance against conventional DDoS attacks like bandwidth depletion and reflection attacks, they are not a panacea.
Research has shown that modified traditional DDoS attacks can still effectively exploit these features and compromise NDN's operations.

Interest Flooding~\cite{gasti2013and,mannes2019naming} is an attack that overwhelms NDN routers by exploiting their caching capability for unsatisfied Interest requests.
Attackers can coordinate botnets to generate excessive Interest requests, saturating the router's cache and obstructing the processing of legitimate interests.
Additionally, Content Poisoning attacks aim to corrupt the content caches within benign routers, obstructing the caching of legitimate content.
This attack involves using bots to issue a multitude of interest requests, followed by a compromised host responding with poisoned content, leading to the proliferation of tainted content across the network's cache.

\subsubsection{Distributed System}

\textbf{Internet of Things.}
In the burgeoning landscape of interconnected devices, the susceptibility of smart home devices to cyber-attacks poses a significant threat.
The study by Tushir et al.~\cite{tushir2020quantitative} quantitatively assesses the impact of DDoS on these devices.
The findings highlight a considerable variation in the minimum attack rate required to disrupt different smart home devices and cause power outrage.
The research underscores a critical vulnerability inherent in the dependency of these devices on WiFi connections; specifically, the process of group key updating in WiFi, which is shown to exacerbate the risk of DDoS attacks by precipitating faster disconnections of devices.
Furthermore, the study delineates several key factors that influence the energy consumption of victim devices during an attack, including the utilized communication protocols, the rate and size of the attack payloads, and the state of the device ports.

In the realm of smart grid systems, vulnerabilities to DDoS attacks have been identified~\cite{vukovic2014security}.
The research focuses on the distributed state estimation module, which is integral to the operation and supervision of the power system.
An attacker gaining control of a central control center can manipulate the state data communicated between this center and its adjacent centers.
This manipulation can compromise the reliability of the data used by neighboring control centers, which is crucial for their operational decision-making.
Consequently, the dissemination of false data can incapacitate these systems, leading to a denial of service and potentially catastrophic failures in power system management.

\textbf{Blockchain system.}
Blockchain technology has emerged as a groundbreaking innovation, yet it is not immune to the prevalent threat of denial of service attacks.
Vasek et al.~\cite{vasek2014empirical} have documented a significant number of DDoS attacks targeting the Bitcoin ecosystem, identifying 142 unique instances across 40 services.
Their research indicates that approximately 7\% of service operators have been subjected to such attacks, with currency exchanges and mining pools being the most frequently targeted.

Further exploring this avenue, studies by Johnson et al.~\cite{johnson2014game} and Wu et al.~\cite{wu2020survive} elaborate on the strategies used by malicious entities within the competitive landscape of mining.
Johnson et al.~\cite{johnson2014game} reveal the strategic trade-offs faced by resource-limited attackers: Either to allocate computing resources to their mining efforts or to engage in DDoS attacks to diminish the success rate of rival pools.
Through game-theoretical modeling, they identify optimal DDoS strategies, which include the selection of victim pools and the allocation of resources for the attack.
Wu et al.~\cite{wu2020survive} extend this analysis to a dynamic environment where miners frequently switch pools, leading to evolving pool sizes.
They model this interaction as a general-sum stochastic game and develop a Nash learning algorithm to deduce near-optimal attack strategies, thereby maximizing the attacker's rewards.

Li et al.~\cite{li2021deter} examine a different attack vector within the Ethereum network, focusing on the abuse of transaction handling mechanisms.
They identify how malicious actors can disrupt the network by sending malformed transactions with nonces that exceed the expected value or by initiating transactions that overdraft an account's balance.
These transactions can lead to the eviction of legitimate transactions from a victim's transaction pool and their replacement with invalid ones, effectively preventing the victim from disseminating valid transactions or including them in the blockchain.
Notably, these attacks incur minimal costs, as they consume negligible amounts of Ether.

Other research efforts, such as those by Heilman et al.~\cite{heilman2015eclipse} and Tran et al.~\cite{tran2020stealthier}, focus on connection manipulation attacks aimed at isolating nodes from the blockchain network.
The Eclipse attack~\cite{heilman2015eclipse} exploits Bitcoin's peer selection process, allowing attackers with numerous IP addresses to flood a victim's peer address database with malicious nodes, effectively segregating the victim from legitimate peers.
The EREBUS attack~\cite{tran2020stealthier} leverages an adversary's position as a man-in-the-middle Autonomous System to influence peering decisions over time, eventually replacing all of a victim's peers with spoofed ones, thus isolating them from the network.

Gervais et al.~\cite{gervais2015tampering} and Walck et al.~\cite{walck2019tendrilstaller} have researched methods to introduce delays into blockchain operations.
They exploit the data request protocols of Bitcoin, where nodes avoid redundant data requests from their peers.
Attackers can take advantage of this by advertising transactions to a victim node, causing it to wait indefinitely for data that the attacker never sends~\cite{gervais2015tampering}.
The TendrilStaller attack~\cite{walck2019tendrilstaller} delays block propagation to the victim with fewer attack resources.
The attack exploits a recent block propagation protocol which prescribes a Bitcoin node to select three neighbors that can send unsolicited blocks.
As a result, the attacker can induce the victim to select three attack nodes, which perform the delaying procedure to make the victim node stuck in the waiting process.

Finally, studies by Apostolaki et al.~\cite{apostolaki2017hijacking} and Li et al.~\cite{li2023bijack} highlight vulnerabilities in the underlying network protocols (BGP and TCP) used by blockchain systems.
These works illustrate how attackers can leverage AS-level BGP hijacks to intercept and disrupt Bitcoin traffic to and from victim nodes.
Specifically, Apostolaki et al.~\cite{apostolaki2017hijacking} demonstrate how adversaries with control over an Autonomous System can manipulate routing tables by executing BGP hijacks.
This is done by broadcasting spoofed BGP announcements that claim ownership of a victim node's IP prefix.
As a result, the malicious AS can intercept all traffic intended for the victim, selectively filter out Bitcoin traffic, and drop those packets.
The Bijack attack~\cite{li2023bijack} takes advantage of a specific flaw in the assignment method of the IPID field within the TCP protocol (Section~\ref{subsec:adversarial-ddos-bypassing-detection}).
The vulnerability enables attackers to deduce the active TCP connections of a victim node, including sensitive information such as the TCP sequence number, victim node’s port number, and the IP and port numbers of peers.
Armed with this information, the attacker can forge TCP RST packets to sever these connections, forcing the victim node to disconnect from the blockchain network.

\subsubsection{Computing Infrastructure}
\textbf{Remote direct memory access system (RDMA).}
Remote Direct Memory Access (RDMA) technology has seen a rapid adoption in a variety of settings, spanning from private data centers to multi-tenant cloud environments.
A notable example of its application is in distributed machine learning, where RDMA's ability to facilitate direct memory access from a client to a remote server's memory via an RDMA-enabled network interface card offers significant performance improvements.
This is chiefly because the data transfer operation circumvents the operating system and traditional network stack, leading to a more efficient communication process.

However, the introduction of RDMA has not come without its security implications.
In the work of Wang et al.~\cite{wanglordma}, it is highlighted that the congestion control mechanisms provided by the RDMA API, particularly Priority-based Flow Control (PFC) and Data Center Quantized Congestion Notification (DCQCN), inadvertently create new opportunities for denial of service attacks.
Attackers can exploit these mechanisms by initiating low-rate DDoS attacks with potentially severe consequences.
The adversaries direct bots to intermittently send bursts of traffic to a chosen egress port.
The pulsating nature of this traffic can manipulate the behavior of DCQCN, which employs an Additive-Increase/Multiplicative-Decrease (AIMD) algorithm similar to that used by TCP congestion control.
Consequently, legitimate traffic destined for the targeted port can be unfairly penalized and throttled, mimicking the effects of a TCP slow-rate DDoS attack.

The situation is exacerbated by the behavior of PFC during periods of congestion.
In an effort to prevent packet loss, PFC issues a PAUSE frame that travels in the opposite direction of the congested traffic, instructing all upstream switches to halt forwarding operations.
This has a cascading effect, as not only is the traffic heading towards the targeted egress port affected, but so too is any unrelated traffic that happens to traverse the impacted switches.
Such collateral damage extends the disruptive impact of the attack well beyond its intended target, illustrating the potential for widespread disruption within an RDMA-enabled network infrastructure.

\textbf{Serverless platform.}
The advent of serverless computing has introduced a paradigm where users can deploy and execute web applications through serverless functions without the overhead of managing servers.
This model only requires users to pay for the actual compute resources used during the execution of these functions.
A characteristic of serverless platforms is the assignment of platform-provided IP addresses, known as egress IPs, for outbound connectivity from these functions.
Notably, these egress IPs are shared among multiple serverless functions.

A study by Xiong et al.~\cite{xiong2021warmonger} exposes a vulnerability inherent to this architecture, wherein an attacker can orchestrate a DDoS attack by exploiting these shared egress IPs \cite{xiong2021warmonger}.
The attack is carried out by deploying several malicious serverless functions that generate a high volume of intrusive requests, such as HTTP floods, using the platform's egress IPs.
These requests are directed towards a targeted server, with the intention of overwhelming it.
Given that egress IPs are typically few in number and remain constant over time, a defensive action taken by the targeted server—such as blocking these IPs to mitigate the attack—can inadvertently lead to collateral damage.
Specifically, legitimate users of the serverless platform who share the blocked egress IPs find themselves inadvertently denied access to the targeted server.
This not only disrupts the services offered by the targeted server but also impacts the availability of services reliant on the serverless platform, illustrating a significant security concern within the serverless computing model.

\subsection{Adversarial Attack}\label{subsec:adversarial-ddos-bypassing-detection}
Recent trends in DDoS attacks have seen a shift towards the deployment of adversarial tactics.
In these sophisticated attacks, the perpetrator meticulously designs malicious traffic to mimic legitimate network flows.
This deceptive strategy is intended to evade current detection and mitigation systems, allowing the harmful data to pass unchecked and be received by the targeted victim.
This section delves into the recent advancements in understanding these covert attack methodologies, and we categorize these attacks according to their targeted security enforcement.

\subsubsection{Learning-based Detection System}
In recent years, the deployment of machine learning (ML) and deep learning (DL) techniques for intrusion detection has seen a significant rise, with systems being trained on substantial traffic data to distinguish between normal and malicious flows and filter out the latter in real-time.
However, as these detection systems become more sophisticated, so as the methods employed by attackers to circumvent them.

Fogla et al.~\cite{fogla2006evading} address how rule-based intrusion detection systems, where rules are usually inferred by learning techniques (e.g., decision tree), can be evaded through polymorphic blending attacks (PBAs).
These attacks cleverly disguise malicious packets to appear statistically similar to legitimate traffic, thus evading detection.
The core challenge lies in mutating malicious packets such that they conform to the regular grammar of the detection system (e.g., requirement on the payload size).
The authors demonstrate the computational complexity in finding an optimal PBA, highlighting its NP-complete nature.
To tackle this, they recommend the use of satisfiability (SAT) or integer linear programming (ILP) solvers to discover near-optimal PBAs.

Yan et al.~\cite{yan2023automatic} examine the vulnerabilities in ML-based intrusion detection systems, particularly their susceptibility to evasion attacks using adversarial examples.
In a simulated attack scenario, the adversary has limited feedback, e.g., only the binary result of detection success or failure.
By interacting with the system, the adversary discerns patterns of benign and malicious DDoS traffic.
With this knowledge, substitute models are trained using ensemble learning to approximate the decision boundaries of the target system.
This equips the adversary with a quasi-white-box view, facilitating the creation of adversarial traffic samples that are statistically representative of network traffic while still harboring malicious payloads.
These samples are then used to successfully bypass the actual IDS, exposing significant vulnerabilities in these systems.

Abusnaina et al.\cite{abusnaina2019examining} delve into adversarial learning attacks against deep-learning-based detection systems.
They observe that while general adversarial attacks (e.g., those documented by Papernot et al.\cite{papernot2016limitations} and Moosavi-Dezfooli et al.\cite{moosavi2016deepfool}) can induce misclassification in standard tasks like image recognition, they fail to generate adversarial network flows that require to maintain the characteristics of legitimate traffic.
To overcome this, they introduce a flow-merge technique, which merges attributes of benign flows with a mask flow using operations such as accumulation or averaging, thereby crafting adversarial flows that evade detection.
Similarly, Hashemi et al.\cite{hashemi2019towards} pinpoint manipulations, such as the splitting and injecting of packet payloads, that can modify network features perceived by a detection system without violating network protocol requirements.

Mustapha et al.~\cite{mustapha2023detecting} propose the use of Generative Adversarial Networks (GANs) for creating adversarial flows.
They utilize a Wasserstein GAN (WGAN), which includes a generator that creates malicious flow samples from random noise, aiming to mirror the distribution of benign traffic data.
Meanwhile, the discriminator, acting as a surrogate for the target detection system, aims to differentiate between genuine and synthetic samples.
The closed feedback loop between the generator and discriminator ensures a continuous refinement of the adversarial samples. 
The process iterates until the discriminator's accuracy plummets, at which point the generated flows can effectively evade the target system.

Finally, Matta et al.~\cite{matta2017ddos} conceptualize randomized DDoS attacks, a particularly stealthy form of DDoS.
The process begins with bots monitoring online activity to capture normal traffic patterns, which are then used to compile an emulation dictionary.
Bots mimic legitimate traffic by selecting packets from this dictionary at random.
To balance the trade-off between message innovation (uniqueness) and independence (redundancy), the bots employ randomization in their message selection and transmission rates.
This method generates malicious flows with high innovation rates, effectively fooling detection systems.

\subsubsection{Commercial DDoS protection}
DoS attacks continue to evolve, necessitating sophisticated defensive measures from Internet Service Providers (ISPs) and cloud service providers like CloudFlare, who offer traffic scrubbing services to their customers.
These services typically employ strategies such as IP hiding and address validation to distinguish and filter out malicious traffic.
Despite these measures, research indicates that attackers can still circumvent these defenses.

Jin et al.~\cite{jin2018your} examined the dependence of Denial of Service Protection Services (DPS) on concealing the server's true IP address and the effectiveness of traffic scrubbing techniques.
They found that DPS works by providing a false DNS record to mask the server's actual IP address.
Consequently, both legitimate and malicious traffic are directed to a scrubbing center, where the latter is intended to be filtered out.
However, this defense has vulnerabilities, especially when changes are made to the DPS configuration.
If a user discontinues their DPS service or switches providers, the original DPS may retain records of the server's actual IP address.
Attackers can exploit this by querying the name servers of the former DNS provider, thus unmasking the target server's IP and allowing them to bypass the DPS altogether.

Nosyk et al. \cite{nosyk2023closed} introduced a scanning algorithm designed to detect networks that forego Source Address Validation (SAV), which is a critical defense against amplification attacks.
SAV works by rejecting packets with spoofed source IP addresses at the network's edge.
The study demonstrated that attackers could easily discern whether a network has implemented SAV.
By sending spoofed DNS requests to each host within a target network and observing responses from an attacker-controlled authoritative name server, it can be determined whether SAV is absent.
If the network lacks SAV, it becomes a potential target for amplification attacks.
This scanning method revealed that a significant portion of networks, 49\% of IPv4 and 26\% of IPv6 Autonomous Systems (AS), do not implement SAV, leaving millions of DNS resolvers vulnerable to these attacks.
Similarly, the Spoofer project, maintained by CAIDA~\cite{spoofer}, measures a network's vulnerability to spoofing by sending packets with forged source addresses to a measurement server.
The success or failure of these transmissions reveals whether the network can be exploited for spoofing.
Lone et al.~\cite{lone2017using} propose another active inference technique using routing loops identified in traceroute data.
An attacker dispatches a traceroute packet with a fabricated source IP address to a customer network within an Internet Service Provider (ISP).
If the ISP fails to filter out the spoofed packet, it forwards the packet to the customer.
The absence of accurate routing for the fake address causes the packet to oscillate between the customer and ISP, indicating that the ISP is susceptible to source IP spoofing.

Further research highlighted by Wu et al.~\cite{wu2014software} and Shankesi et al.~\cite{shankesi2010resource} has shown that client puzzle schemes, designed to mitigate DDoS attacks, are not foolproof against determined adversaries.
Attackers can leverage computational resources such as GPUs or integrated CPU-GPU systems to solve puzzles more rapidly.
If the puzzle is parallelizable, an attacker might distribute the task across hundreds of GPU cores, significantly decreasing the time required to solve it.
Alternatively, if the puzzle function is non-parallelizable, the attacker might inundate the server with requests, assigning each GPU core to solve different puzzles independently.
This technique effectively reduces the time needed to solve these challenges, thereby increasing the potency of the attack.

\subsubsection{Protocol Security Enforcement}
Communication protocols are typically engineered with inherent security mechanisms, enabling the parties involved in the communication to validate the messages they receive.
Nonetheless, contemporary research has uncovered that these security provisions are not infallible; indeed, attackers have identified and exploited design flaws within these mechanisms to circumvent security checks..

Cao et al.~\cite{cao2016off} identified a critical vulnerability in TCP, wherein a mechanism designed to protect against DoS attacks inadvertently introduces a new attack surface.
The introduction of a global rate limit, as specified in RFC 5961, was meant to mitigate DDoS attacks.
However, it inadvertently introduced a new attack vector.
By sending spoofed packets, attackers can manipulate the global rate limit counter, a shared resource, and monitor its effects to deduce the existence of a TCP connection and its sequence number.
Consequently, this information enables them to disrupt the connection by transmitting a malicious RST packet with the correct sequence number, impersonating the victim.

Feng et al.~\cite{feng2020off} discovered a similar exploit within the Linux kernel's mixed IPID assignment method, which was originally implemented to counter TCP hijacking.
Attackers can observe changes in the IPID counter, induced by spoofed packets, to infer details about active TCP connections and hijack them using malicious RST packets.
Wang et al.~\cite{wang2024off} extended this attack vector to WiFi networks by demonstrating that the size of encrypted frames can be observed and used to infer TCP connection details, including sequence and acknowledgment numbers.

Further examining the ICMP protocol, Feng et al.~\cite{feng2022off} revealed a disconnection between the legitimacy check mechanism for ICMP redirect messages and a suite of stateless protocols such as UDP, ICMP, GRE, IPIP, and SIT.
This gap allows off-path attackers to craft evasive ICMP error messages that bypass the legitimacy checks, leading to the revival of ICMP redirect attacks.
These attackers can orchestrate stealthy DoS attacks, tricking public servers into redirecting their traffic into black holes with just one forged ICMP redirect message.

Feng et al.~\cite{feng2022pmtud} also investigated the interaction between IP fragmentation and TCP, challenging the assumption that IP is protected from fragmentation attacks by the default implementation of Path Maximum Transmission Unit Discovery (PMTUD).
They found that ICMP error messages could desynchronize the path MTU values between the IP and TCP layers.
This desynchronization can result in IP fragmentation, even when PMTUD is used, allowing an off-path attacker to trigger fragmentation and inject malicious packet fragments, causing legitimate packets to be lost.

\subsection{Summary}
We have summarized the attacks discussed previously in Table~\ref{tab:survey-attack-summary}.
To categorize these attacks, we examine them from six distinct perspectives, including the protocols and systems they target.
Among these perspectives, we emphasize two specific characteristics of malicious traffic: traffic type and traffic pattern.
Regarding traffic types, we distinguish attacks as either direct or indirect. Direct attacks involve sending malicious traffic straight to the victim, whereas indirect attacks route the malicious traffic through intermediaries, such as DNS resolvers.
In terms of traffic patterns, we categorize malicious traffic based on its volume and timing.
In terms of volume, attacks can be volumetric (i.e., large quantities of data), or low-rate (i.e., smaller amounts of data).
In terms of timing, attacks are classified by their duration and frequency, such as continuous flooding or periodic pulsing.
\begin{table*}
\centering
\caption{Summary of surveyed DDoS attacks}
\scalebox{0.75}{
\begin{tblr}{
  hlines,
  vlines,
}
\textbf{Research} & \textbf{Category} & \textbf{Target} & \textbf{Exploited Features} & \textbf{Traffic Type} & \textbf{Traffic Pattern} & \textbf{Impact} \\
       \begin{tabular}[c]{@{}l@{}}\cite{luo2014mathematical,tang2013modeling,schuchard2010losing},\\\cite{wang2002detecting}\end{tabular}        &              Transport-Layer Protocol                    &         TCP        &                  \begin{tabular}[c]{@{}l@{}}Session management,\\congestion control\end{tabular}                  &           Direct               &                 \begin{tabular}[c]{@{}l@{}}Volumetric flooding,\\Low-rate pulsing\end{tabular}                   &       Service down for TCP servers          \\
       \cite{nawrocki2021quicsand}        &              Transport-Layer Protocol                    &         QUIC        &                  \begin{tabular}[c]{@{}l@{}}Session management,\\flexible ID setting\end{tabular}                  &           (In)Direct              &                 Volumetric flooding                   &       Service down for QUIC servers          \\
       \cite{gilad2011fragmentation,atlasis2012attacking}        &              Network-Layer Protocol                    &         IP        &                  \begin{tabular}[c]{@{}l@{}}Header fragmentation\end{tabular}                  &           Direct              &                 Low-Rate flooding                   &       Service down for target servers          \\
       \cite{dantas2014selective}        &             Application-Layer Protocol                     &        HTTP         &              Request fragmentation                      &            Direct              &                 Low-rate pulsing                  &        Server socket exhaustion         \\
        \cite{beckett2017http,praseed2019multiplexed}       &                 Application-Layer Protocol                 &         HTTP/2        &          Multiplexing                          &      Direct                    &         Volumetric     flooding                      &            Server CPU exhaustion                \\
        \cite{sisalem2006denial,tang2014sip}       &                 Application-Layer Protocol                 &         SIP        &            \begin{tabular}[c]{@{}l@{}}Session management,\\flexible ID setting\end{tabular}                        &         Direct                 &         Volumetric    flooding                       &           \begin{tabular}[c]{@{}l@{}}Proxy socket exhaustion,\\stop benign VoIP calls\end{tabular}                \\
        \begin{tabular}[c]{@{}l@{}}\cite{yin2023waterpurifier,kim2017preventing,yazdani2022mirrors},\\\cite{griffioen2021scan}\end{tabular}       &                 Application-Layer Protocol                 &         DNS        &              \begin{tabular}[c]{@{}l@{}}Flexible ID setting,\\recursive resolution\end{tabular}                      &           (In)Direct               &              Low-rate flooding                    &        \begin{tabular}[c]{@{}l@{}}Service down for target servers,\\service down for DNS servers\end{tabular}               \\
        \cite{mohammed2023detection}       &                 IoT Protocol                 &         Modbus        &          Memory allocation                          &         Direct                 &               Low-rate flooding                     &           Service down for control units                \\
        \cite{wang2022zigbee}       &                 IoT Protocol                 &         Zigbee        &          Network rejoin procedure                          &         Direct                 &               Low-rate flooding                     &           Devices unable to join the network                \\
        \cite{studer2009coremelt,kang2013crossfire}       &                 Networking Infrastructure                 &         Routing system        &          Traffic concentration                          &         Direct                 &               Volumetric flooding                     &           Cut off connections to a region                \\
        \cite{shin2013avant,shin2013attacking,cao2019crosspath}       &                 Networking Infrastructure                 &         SDN system        &            \begin{tabular}[c]{@{}l@{}}Control-data plane separation,\\shared path for data and control\end{tabular}                        &         Direct                 &              Volumetric flooding                      &         Service down for switches/controllers                  \\
        \cite{gasti2013and,mannes2019naming}       &                 Networking Infrastructure                 &         NDN system       &              Content caching                      &           Direct               &              Volumetric flooding                    &        Service down for NDN routers               \\
        \begin{tabular}[c]{@{}l@{}}\cite{henrydoss2014critical,silva2020repel,olimid20205g},\\ \cite{sattar2019towards,javadpour2023reinforcement,mirsky2020ddos} \end{tabular}       &                 Networking Infrastructure                 &         Cellular network        &         \begin{tabular}[c]{@{}l@{}}LTE control-data separation,\\5G network slicing,\\unidentified emergency call\end{tabular}                           &           (In)Direct               &             Volumetric flooding                       &        \begin{tabular}[c]{@{}l@{}}Service down for stations,\\user equipment, and 911 center\end{tabular}                   \\
        \cite{tushir2020quantitative,vukovic2014security}       &                 Distributed System                 &         IoT system       &           \begin{tabular}[c]{@{}l@{}}Power support,\\device interaction\end{tabular}                         &   Direct                       &           \begin{tabular}[c]{@{}l@{}}Volumetric flooding\end{tabular}                         &       
        \begin{tabular}[c]{@{}l@{}}Service down for home devices\\and smart grid nodes\end{tabular}                    \\
        \begin{tabular}[c]{@{}l@{}}\cite{vasek2014empirical,johnson2014game,wu2020survive},\\\cite{li2021deter,heilman2015eclipse,tran2020stealthier},\\\cite{gervais2015tampering,walck2019tendrilstaller,apostolaki2017hijacking}\end{tabular}       &                 Distributed System                 &         Blockchain  system      &      \begin{tabular}[c]{@{}l@{}}Mining pool,\\transaction handling,\\peer selection\end{tabular}                           &            Direct              &       Volumetric flooding                             &        \begin{tabular}[c]{@{}l@{}}Service down for blockchain node\end{tabular}                   \\
        \cite{wanglordma}       &                 Computing Infrastructure                 &         RDMA  system      &            Congestion control                        &        (In)Direct                  &            Volumetric flooding                        &       \begin{tabular}[c]{@{}l@{}}Service down for direct and\\indirect RDMA API calls\end{tabular}                    \\

        \cite{xiong2021warmonger}       &                 Computing Infrastructure                 &         Serverless platform        &      Shared egress IP                           &            Indirect              &       Volumetric flooding                             &        Deployed web service down                  \\
        \begin{tabular}[c]{@{}l@{}}\cite{fogla2006evading,yan2023automatic,abusnaina2019examining},\\\cite{papernot2016limitations,moosavi2016deepfool,hashemi2019towards},\\\cite{mustapha2023detecting,matta2017ddos}\end{tabular}       &                 Adversarial DoS                 &         Learning-based IDS        &      Adversarial learning                           &            Direct              &       Volumetric flooding                             &        Bypass IDS detection                \\
        \begin{tabular}[c]{@{}l@{}}\cite{jin2018your,nosyk2023closed,wu2014software},\\\cite{shankesi2010resource}\end{tabular}       &                 Adversarial DDoS                 &         \begin{tabular}[c]{@{}l@{}}Commercial DoS\\protection\end{tabular}        &      \begin{tabular}[c]{@{}l@{}}IP hiding,\\address validation,\\client puzzle\end{tabular}                           &            Direct              &       Low-rate pulsing                             &        \begin{tabular}[c]{@{}l@{}}Bypass commercial DoS protection\end{tabular}                  \\
        \begin{tabular}[c]{@{}l@{}}\cite{cao2016off,feng2020off,wang2024off},\\\cite{feng2022off,feng2022pmtud}\end{tabular}       &                 Adversarial DDoS                 &         \begin{tabular}[c]{@{}l@{}}Protocol security\\enforcement\end{tabular}        &      \begin{tabular}[c]{@{}l@{}}TCP global rate limit,\\ICMP redirect,\\fragmentation protection\end{tabular}                           &            Direct              &       Low-rate pulsing                             &        \begin{tabular}[c]{@{}l@{}}Leverage security enforcement to\\construct DoS attacks\end{tabular}                  \\
\end{tblr}
}
\label{tab:survey-attack-summary}
\end{table*}

Our literature review reveals a concerning evolution: DoS attacks are becoming increasingly diverse, extending their reach to a broader spectrum of network protocols and capitalizing on weaknesses within newly established systems.
These advanced attacks are characterized by their nimbleness and intricacy, often bypassing traditional security measures with alarming facility.
In light of this context, it is essential to thoroughly examine the attributes and evolving patterns of these DoS attacks, enabling security professionals to more effectively identify new attack vectors as new protocols and systems are designed.
In this section, we dissect the current trends in DoS attacks.
In Section~\ref{sec:open-problem}, we offer strategic insights for conducting comprehensive investigations into the attack surfaces of new protocols and systems.

\textit{\textbf{DDoS attacks are exploiting advanced protocol features to increase the attack efficiency, stealthiness, and severity.}}
Traditional DDoS attacks often exploit familiar aspects of network protocols including congestion control, identity spoofing, and packet fragmentation.
However, as network protocols have evolved, they have developed sophisticated features that now present new opportunities for attackers.
Features such as HTTP/2 multiplexing, DNS recursive resolution, Modbus flexible header, and Zigbee network rejoin are now being leveraged to construct more potent and covert DDoS attacks.

Specifically, attackers are drawn to these advanced features for several reasons.
(1) \textit{Traffic amplification}.
Modern protocol features can significantly amplify traffic. This means attackers can use fewer resources to launch larger attacks.
For example, HTTP/2 multiplexing enables attackers to achieve up to 95 times the attack bandwidth compared to HTTP/1.0 traffic, under the same packet transmission rate.
(2) \textit{Increased stealthiness}.
These features often allow attackers to indirectly route malicious traffic towards the target, enhancing the stealthiness of the attack.
A notable tactic involves exploiting DNS recursive resolution, where attackers distribute malicious DNS requests across multiple resolvers.
These resolvers then unwittingly forward the flood of requests to the target authoritative server, complicating the tracing process and obscuring the origins of the attack.
(3) \textit{Resource diversity exploitation}.
By interacting with various system properties of the target host, these protocol features expand the attack surface.
This diversification enables attackers to manipulate different system properties and cause more severe disruptions.
For instance, attackers can use specially crafted Modbus packets to target a controller unit’s memory, or issue malicious Zigbee rejoin requests to overload the routing device's child table.

The foregoing discussion makes it clear that the evolution of network protocols, while enhancing efficiency and introducing new functionalities to modern computing environments, also creates new vulnerabilities in the realm of DDoS attacks.
As these protocols become more complex, they not only broaden the attack surface but also introduce subtle vulnerabilities that can be exploited in unexpected and innovative ways.

\textit{\textbf{DDoS attacks are increasingly targeting advanced systems and exploiting emerging vulnerabilities.}}
The landscape of DDoS attacks is undergoing a transformative shift, mirroring the rapid evolution of technological systems.
The focus of these attacks has broadened, moving beyond traditional web servers to encompass an array of sophisticated and emerging systems, such as SDN, cellular network, IoT, and blockchain systems.
Attackers are not only exploiting known issues including uncontrolled resource consumption (CWE-400) and Network amplification (CWE-406), but are also identifying and leveraging new vulnerabilities in these emerging systems.
In particular, our survey highlights three particularly concerning vulnerabilities.

(1) \textit{Separation of data and control planes}.
One significant change in network management is the separation of data and control planes, a technique used in popular systems such as SDN and cellular networks.
This architecture allows network administrators to manage, configure, and optimize network behavior from a centralized location.
However, this centralization also creates a critical vulnerability.
An attacker can target this central point by flooding the control-plane communication channel with fake control messages, disrupting the normal operations of the control plane.
Additionally, because the centralized controller relies on physical resources, e.g., CPU and memory, to manage the network, an attacker could deploy malicious bots to generate a massive number of suspicious network flows.
This tactic strains the controller's resources, making it difficult to maintain effective network monitoring and management.

(2) \textit{Vulnerable resource sharing mechanism}.
Resource sharing is a common trail for efficient resource utilization.
For example, functions running on a serverless platform may share the same egress IP addresses.
As another example, network slicing in 5G allows multiple virtual networks to operate on the same physical network infrastructure, optimizing CPU, memory, and bandwidth usage.
However, these shared environments can significantly increase the risk of widespread disruptions.
By targeting a single vulnerable service (e.g., a vulnerable serverless function), the attacker can impact the underlying infrastructure and paralyze all services co-located on the same infrastructure.
Even worse, if attackers have access to the infrastructure, they can initiate cyber attacks in representative of all co-located services, triggering others to blacklist them.
For instance, the attacker can rent a serverless platform and deploy malicious serverless functions to initiate cyber attacks on legitimate parties (e.g., DNS resolvers).
Consequently, all legitimate services sharing the same egress IP would also suffer from denial of service, even though they are not directly involved in the attack.

(3) \textit{Exploitable component interaction}.
Modern systems often feature complex inter-dependencies among their components.
For example, in a blockchain network, each node communicates with its nearest neighbors to stay updated about the entire network.
Similarly, in smart grid systems, nodes share their state estimates with each other, and each node bases its control decisions on the data received from its neighbors.
However, these intricate interactions can also introduce vulnerabilities.
Attackers can exploit these relationships to launch DDoS attacks.
For instance, an attacker might manipulate the peer selection process in Bitcoin to isolate a target node from the rest of the network.
Additionally, attackers can disseminate false information, such as incorrect state measurements, to disrupt the decision-making processes of benign nodes.
This can effectively paralyze their operational logic, leading to broader system disruptions.

The discussion underscores a critical development in the realm of cyber threats, highlighting how complexity in advanced systems can be both a driver of innovation and a magnet for vulnerabilities.
This dual nature presents significant challenges as we strive to advance technologically while securing the systems against increasingly sophisticated threats.

\textit{\textbf{The scale of traffic associated with DDoS attacks has significantly diminished: Even a single message can cause denial of service.}}
The evolution of DDoS attacks has seen a marked transition from the traditional volumetric flooding techniques to more insidious low-rate strategies.
These sophisticated attacks exploit specific design or implementation weaknesses inherent in communication protocols and systems, negating the need for attackers to generate large volumes of traffic.
For example, the TCP congestion control mechanism, IP fragmentation process, and DNS amplification can all be manipulated due to their protocol features.
One particularly concerning exploitation method involves the erroneous handling of DNS error messages~\cite{pan2024loopy}.
Attackers can leverage this flaw to send a single message that generates persistent, malicious traffic, thereby efficiently draining the targeted server's resources.
Similar exploitative tactics can be observed in systems with advanced features, such as Named Data Networking (NDN) content caching, interactions within smart grid centers, and the peer selection process in blockchain networks.
These examples underscore the adaptability of attackers in using complex system functionalities to their advantage.

\textit{\textbf{Adversarial DDoS tactics are constantly evolving, targeting various types of detection systems and exhibiting diverse levels of attack costs.}}
Adversarial DDoS tactics employ sophisticated and aggressive methods designed to disrupt the normal operations of targeted services and evade detection systems.
Our research identifies three types of adversarial attacks: Adversarial machine learning, bypassing commercial DDoS protection, and exploiting protocol security enforcement.
To assess these tactics, we here provide a detailed overview of their workflows, and analyze their advantages and disadvantages, focusing on factors such as attack cost and stealthiness.

(1) \textit{Adversarial machine learning}.
This tactic targets machine learning-based detection systems and typically unfolds in two phases:  Mimicking the decision boundary of the target detection system, and generating malicious samples that fall within these boundaries.
To understand the behavior of the target detection system, the adversary generates a large number of test samples and collects feedback (labels) from the system.
Then this feedback is used to refine the sample generator.
This iterative process continues over multiple rounds until the majority of the generated samples can bypass the detection system, effectively inferring its decision boundary.
Once the boundary is understood, the refined generator can then produce malicious samples tailored for subsequent attacks.

This tactic is applicable to a broad range of learning-based detection systems.
Moreover, it does not require extensive prior knowledge of the target system, making it a flexible approach for attackers.
However, the major drawback of this method is its high resource consumption.
The attacker must generate and test a potentially vast number of samples to accurately infer the decision boundary of the detection system.
Specifically, popular techniques like Generative Adversarial Networks (GANs) may require an extensive number of iterations to converge, and it is challenging to train the generator.
Moreover, the frequent submission of test samples can trigger system alerts, making it relatively easy for the detection system to recognize an ongoing attack.
This can lead to countermeasures such as blocking the source of the traffic (e.g., IP blocking), further complicating the attacker's efforts.

(2) \textit{Bypassing commercial DDoS protection}.
Commercial DDoS protection services employ several strategies to safeguard against denial of service attacks, e.g., IP hiding and source address validation.
However, our survey reveals that attackers have developed multiple techniques to bypass these defenses, exploiting inherent vulnerabilities or oversights of commercial protection systems.
By identifying and exploiting weaknesses (e.g., non-implemented SAV and retained DNS records) attackers can tailor their strategies to specific network vulnerabilities.
The cost of an attack and its stealthiness vary significantly based on the targeted vulnerabilities and the techniques employed.
For example, the process of scanning for SAV implementation is costly and lacks stealth due to the detectability of large-scale network scans.
Conversely, exploiting DNS records to uncover hidden IP addresses is generally low-cost and can be highly stealthy.
In this scenario, attackers can utilize publicly available datasets collected from various vantage points, making the method less conspicuous and more accessible.

(3) \textit{Exploiting protocol security enforcement}.
Recent research has revealed a variety of techniques that attackers use to exploit design flaws in protocol security mechanisms for conducting denial-of-service attacks.
These attackers cleverly use security mechanisms (e.g., the TCP global rate limit and the IPID counter) as unintended side channels.
This manipulation allows them to extract critical information about network communications (e.g., TCP session number) or to alter these communications detrimentally.
One of the primary challenges in addressing these attacks is their detection difficulty, as they utilize legitimate functions of communication protocols.
Once attackers gather critical data, such as TCP session numbers, they can initiate highly targeted attacks with minimal traffic (e.g., a single TCP RST packet).
This subtlety means that traditional DDoS metrics, which often focus on large volumes of traffic, fail to identify these attacks.
However, these techniques demand a sophisticated understanding of protocol dynamics and behaviors, and their effectiveness often hinges on the specific implementations of the protocol stack.
Furthermore, because protocol updates and patches can address these vulnerabilities, the lifespan of such attack methods may be limited.
Once a patch is applied, the methods that previously exploited these flaws can become obsolete.

\section{DDoS Detection}\label{ddos-detection}
\begin{figure*}
    \centering
    \includegraphics[scale=.35]{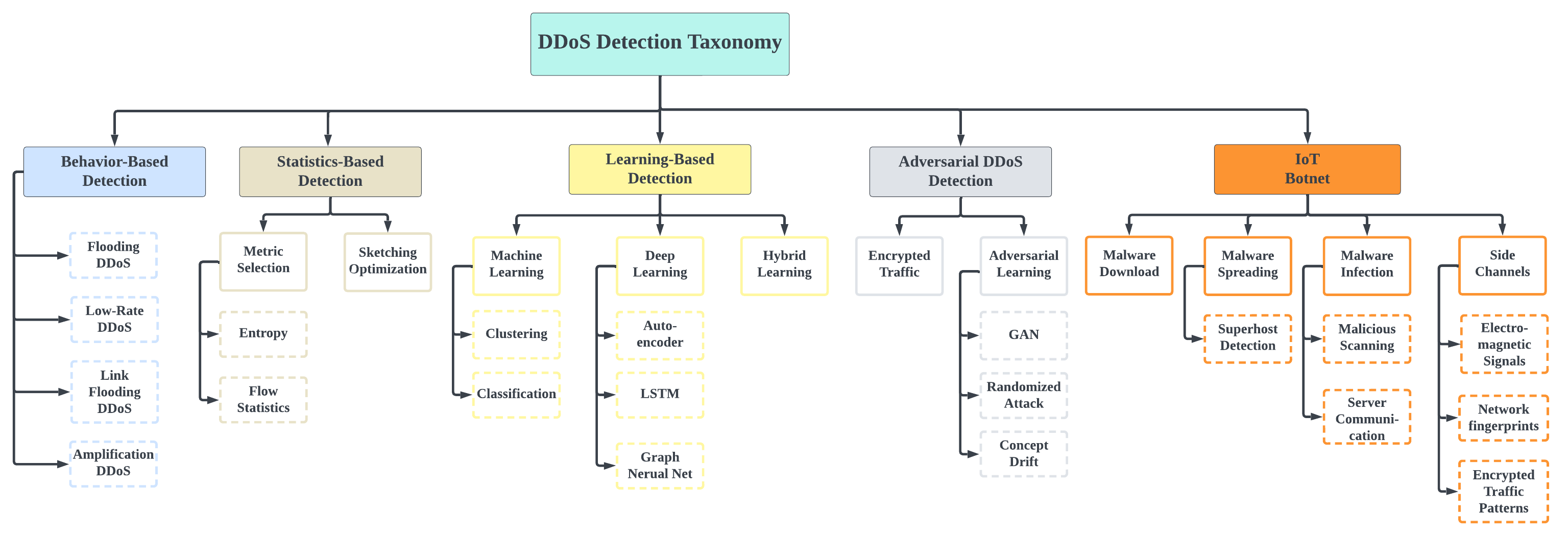}
    \caption{Taxonomy of DDoS detection methods.}
    \label{fig:taxonomy-ddos-detection}
\end{figure*}

In this section, we report existing works which focus on DDoS detection.
DDoS attacks pose significant threats to the stability and security of online services.
Therefore, effective detection of such attacks are paramount.
We classify existing detection methods into five distinct categories, each with its approach to identifying DDoS activities.
Figure~\ref{fig:taxonomy-ddos-detection} shows the overview.
Specifically, compared with existing surveys (Section~\ref{subsec:related-work}), our work proposes a more comprehensive detection taxonomy including five categories.
Besides behavior-based, statistics-based, and learning-based detection methods (Section~\ref{subsec:behavior-based-detection} - Section~\ref{subsec:learning-based-detection}), our work covers the adversarial-based detection methods (Section~\ref{subsec:adversarial-ddos-detection}) and botnet detection methods (Section~\ref{subsec-iot-botnet-detection}) which are rarely discussed in existing works.
\begin{itemize}
    \item \textit{Behavior-based detection.}
    These methods are predicated on the premise that legitimate users and attackers exhibit inherently different interaction patterns with network resources.
    The detection process involves continuous monitoring and analysis of network traffic to capture and assess behavioral signatures that can differentiate between benign and malignant traffics.
    \item \textit{Statistics-based detection.}
    Statistics-based detection techniques employ mathematical models to analyze network traffic.
    These methods utilize various statistical metrics (e.g., entropy) to establish normal traffic profiles.
    When certain metrics exceed predefined thresholds, an alert is triggered, indicating a possible DDoS attack.
    \item \textit{Learning-based detection.}
    Learning-based detection represents a sophisticated category that harnesses the power of machine learning (ML) and deep learning (DL) algorithms.
    These systems are trained on large datasets of network traffic to distinguish between benign and malicious flows.
    \item \textit{Adversarial DDoS detection.}
    Adversarial DDoS detection focuses on identifying traffic that has been manipulated to evade traditional detection systems.
    \item \textit{IoT Botnet detection.}
    Unlike the other categories, IoT Botnet detection methods concentrate on the early stages of DDoS attacks, namely the botnet formation phase.
    These techniques strive to identify compromised IoT devices that are being recruited into a botnet
\end{itemize}

\subsection{Behavior-Based Detection}\label{subsec:behavior-based-detection}
In this section, we delve into detection techniques by analyzing the unique behavioral patterns of attack traffics.
Recognizing that various DDoS attack strategies—such as flooding and reflection—exhibit distinct behavioral signatures, these detection methods are often specialized to effectively target and identify the specific characteristics of a given attack type.
To provide a coherent and methodical overview, we categorize these detection techniques by the attack types they are designed to detect.
\subsubsection{Flooding DDoS}
To tackle the issue of detecting flooding traffics produced by malicious bots, Scherrer et al.~\cite{scherrer2023albus} introduced ALBUS. The fundamental principle behind ALBUS is that malicious flows exhibit a pattern of consistently high data transfer rates within short time frames, in contrast to legitimate flows, which may only sporadically experience bursts and not sustain them.
As a result, ALBUS employs the Leaky Bucket (LB) algorithm to monitor data packet flow, effectively identifying when a flow exceeds a predefined rate, indicative of a burst.
Due to the potential vast number of flows, continuously monitoring all of them would be impractical because of the excessive memory consumption it would require. To mitigate this, ALBUS uses flow sampling, where only a subset of flows is monitored at any given time. This selection is made by mapping flows to specific monitoring points, referred to as checkpoints, using a hashing function. Checkpoints continuously monitor flows that exhibit persistent bursty behavior, thereby increasing the likelihood of accurately identifying abnormal flows. In contrast, flows that do not consistently show burstiness are removed from monitoring and classified as benign.

Tandon et al.~\cite{tandon2021defending} develop a system named FRADE, which employs heuristic rules to distinguish between application-layer DDoS requests and legitimate requests. The core concept is that despite similarities in requests generated by bots and human users, there are discernible differences in the dynamics of their activities—specifically, the frequency and sequence of page visits. To detect malicious requests, FRADE analyzes web server access logs to calculate the rate of client-server interactions and the transition probabilities between page pairs. A request flow is flagged as suspicious if the frequency of page visits is excessively high or if the transition probabilities between pages are unusually low.
Furthermore, FRADE incorporates a honeypot strategy by deploying special web objects, such as hyperlinks that are not typically navigated to by human users. Interaction with these honeypot elements is considered a strong indicator of malicious flows.
Flows that activate these traps are deemed malicious, and their subsequent requests are consequently blocked, thereby impeding their ability to participate in a DDoS attack.

\subsubsection{Low-Rate DDoS}
Low-rate DDoS attacks (e.g., Slowloris) operate by sending low volumes of network traffic to exploit vulnerabilities and monopolize critical, limited resources.
Unlike flooding DDoS, these low-rate attacks are subtle, cost-effective, and increasingly common.
To detect these attacks, Tandon et al.~\cite{tandon2023leader} introduce Leader, a defense mechanism that exploits the resource consumption patterns.
Rather than identifying characteristic patterns of low-rate traffics, Leader employs OS-level tracing to monitor detailed resource usage—like CPU cycles, memory allocation, and sequences of processing function calls—for each network connection.
By using this resource usage data as a basis, Leader constructs a model of normal traffic behavior using one-class SVM and elliptic envelope techniques.
During its operational phase, Leader employs anomaly detection to scrutinize each incoming request against the established model.
Requests that diverge significantly from the model are flagged as potential threats.

An important feature of pulse-wave DDoS attacks, which include varieties such as the Shrew and RoQ attacks, is that these attacks dynamically adjust the attack traffic pattern (e.g., pulsing interval) and exploit the reaction time of the detection system, rending them ineffective.
To identify the malicious flows at line rate and in real-time, Alcoz et al.~\cite{alcoz2022aggregate} present ACC-Turbo, which re-imagines the standard Aggregate-based Congestion Control (ACC) mechanism by integrating a real-time clustering algorithm.
By doing so, it can detect the onset of pulse-wave patterns as they emerge, rather than after they have established a foothold.

\subsubsection{Link Flooding DDoS}
Link Flooding Attacks (LFAs, e.g., Crossfire) target routing systems and have garnered increasing attention from security analysts. These attacks involve bots sending packets to publicly accessible decoy servers, which indirectly flood a node that is not an apparent target. This "under-the-radar" activity cumulatively floods specific target links, impacting the intended target node. Given the severity and stealthiness of LFAs, a range of detection methods have been proposed.

Liaskos et al.~\cite{liaskos2018network,rezapour2021rl} employ traffic engineering techniques, combined with reinforcement learning, to identify malicious flows involved in LFAs.
The underlying principle is that bots will shift their decoy servers and select new critical links to maintain the impact on the target whenever the network topology is altered.
Network operators can exploit this behavior by periodically changing the network topology and monitoring the flows that frequently contribute to the congestion of critical links.
A reinforcement learning strategy is utilized: The probability that a flow is malicious is increased each time it is associated with a congestion event.
In a similar vein, Gkounis et al.\cite{gkounis2016interplay,liaskos2016novel} suggest traffic rerouting to compel malicious flows to relocate to new destinations, thereby hastening their detection.
Kang et al.\cite{kang2016spiffy} propose a rerouting scheme designed to make malevolent flows escalate their traffic volume, which ultimately leads to their detection.
Kang et al.~\cite{kang2016spiffy} also analyze the behavior of a cost-sensitive attacker who employs a consistent and optimal strategy for sending traffic. SPIFFY, their proposed method, increases the bandwidth of a targeted bottleneck link and observes the traffic response. Legitimate traffic sources generally adjust their throughput in response to the added bandwidth, whereas attack flows, likely already at their maximum capacity, do not change their rate as significantly, allowing for the detection of the malicious flows.

Ma et al.~\cite{ma2019randomized} introduce a game-theoretic detection approach to combat LFAs within the constraints of available resources and against adaptive adversaries. Their technique models the conflict as a Stackelberg security game, wherein the defender employs a randomized mixed-detection strategy to optimize detection effectiveness. This strategy unpredictably varies the monitored links, complicating the task for a knowledgeable adversary to evade detection without exhibiting malicious patterns. By integrating models of rational and boundedly rational adversary behavior, the defense adapts dynamically to potential actions of the adversary, with the aim of identifying traffic anomalies indicative of an ongoing LFA.

\subsubsection{Amplification DDoS}
Amplification DDoS attacks pose significant challenges to internet security, often leveraging IP spoofing to magnify their impact. Research in this domain has primarily focused on three aspects: address spoofing detection, request validation, and honeypots.

To counter IP spoofing, Dainotti et al.~\cite{dainotti2013estimating} estimate legitimate address spaces within autonomous systems (AS) by reconstructing bidirectional flows from NetFlow records.
They filter out flows with a high packet volume, assuming these are likely non-spoofed, and use the source IP addresses of these flows to establish a baseline of legitimate address spaces.
Similarly, Lichtblau et al.\cite{lichtblau2017detection} analyze BGP routing data to map relationships among AS and their associated address spaces, facilitating the identification of legitimate source-destination IP pairs
When traffic with source and destination addresses mismatching any legitimate pair is captured, it is flagged for potential spoofing, enabling administrators, such as ISPs, to filter out the suspicious traffic.

Other studies focus on DNS request verification by pairing queries and responses within the same DNS transaction to identify components of a reflection attack~\cite{di2011protecting,dai2024dampadf}.
Di Paola et al.\cite{di2011protecting} employ Bloom filters to efficiently store and lookup request information, while Dai et al.\cite{dai2024dampadf} enhance this approach with DAmpADF.
Utilizing two Bloom filters, DAmpADF alternately records DNS requests, reducing false positives.
Moreover, it identifies popular DNS servers using an exponential-weakening decay method, allowing requests and responses from these servers to bypass Bloom filter recording.

Honeypots are also deployed to detect malicious hosts attempting to initiate amplification DDoS attacks.
By simulating vulnerable application protocols, such as DNS, honeypots act as reflectors for attackers, thus capturing and characterizing attack traffic.
The Cambridge Cybercrime Center (CCC)\cite{thomas20171000}, AmpPot\cite{kramer2015amppot}, and Honeypot Platform for Intrusion (HPI)~\cite{griffioen2021scan} all implement threshold-based detection methods to distinguish attack flows, although these methods may miss multi-protocol attacks or those below local traffic thresholds.
In response to these limitations, Wagner~\cite{wagner2021united} advocates for a collaborative detection approach.
By sharing information through a DDoS Information Exchange Point (DXP), participating mitigation platforms across different vantage points can detect a significantly higher percentage of attack traffic.

Complementing the above methods, Krupp et al.~\cite{krupp2021bgpeek} introduce the BGPEEK-A-BOO technique, leveraging the attackers' reliance on their service provider's BGP routes.
This method involves BGP Poisoning to isolate specific AS, enabling the traceback of spoofed traffic sources through observations of TTL fluctuations or cessation of attack traffic.
By deploying amplification honeypots and systematically analyzing traffic changes, the method traces malicious flows back to their originating AS.
The approach, validated through simulations and real-world experiments, demonstrates its effectiveness in identifying the sources of malicious traffic without requiring prior knowledge of the attacker or external cooperation.

\subsection{Statistics-Based Detection}
This section provides an overview of current statistics-based detection methods. Initially, we explore the foundational concepts behind these methods and the specific metrics employed for detection.
Subsequently, we delve into the sketching technique, a strategic optimization method designed to mitigate memory and storage constraints in scenarios with a large volume of network flows.

\subsubsection{Metric Selection}
Entropy is a widely used statistical metric that measures the randomness present in packet attributes, serving as a key indicator in identifying anomalous traffic patterns indicative of flooding DDoS attacks.
Kalkan et al.~\cite{kalkan2018jess} introduced JESS, an entropy-based detection method that utilizes joint entropy to assess the randomness across combined attributes of network flows, such as destination IP addresses and transport-layer flags.
By calculating the joint entropy for these attribute sets, JESS effectively distinguishes between normal traffic and potential flooding attacks.
A flow with abnormally low joint entropy is flagged as suspicious, as it likely represents a concerted effort to overwhelm a network resource.

Flow statistics (e.g., flow numbers, flow sizes) are also widely adopted, especially for the detection of link flooding attacks.
For instance, the LinkScope system~\cite{xue2014towards} implements a two-tiered measurement strategy.
Initially, it identifies links that serve a significant number of downstream servers and are, therefore, attractive targets for attackers.
Subsequently, LinkScope employs both end-to-end and hop-by-hop measurements, such as packet loss rates and round-trip times, to monitor these links.
The system then applies the Cumulative Sum (CUSUM) algorithm to detect sudden changes in these metrics, indicative of a congested path.
By correlating the two types of measurements, LinkScope accurately localizes the link under attack.

RADAR~\cite{zheng2018realtime} also focuses on the LFA attacks and performs correlation analyses on network flow information to detect them.
RADAR collects traffic statistics from the Software-Defined Networking (SDN) data plane, applying heuristic rules to identify suspicious patterns, such as regular congestion on specific paths.
To discern malicious traffic, RADAR then conducts adaptive traffic analysis within the SDN control plane, checking whether the flow statistics match those associated with attack patterns, such as synchronized bursts of flows and congestion events.
Ripple and Mew~\cite{xing2021ripple,zhou2023mew} employ programmable switches to facilitate in-network measurement and detection of link-flooding attacks.
They maintain a network-wide "defense panorama," which is a synchronized view of attack signals, enabling the system to monitor for and react to flooding attacks.
Their switches periodically assess congestion levels against predefined thresholds and identify suspect hosts by analyzing flow statistics, such as the number of low-rate flows between pairs of source and destination IP addresses.

\subsubsection{Sketching Optimization}
Monitoring a vast number of data flows and managing their statistics can lead to substantial memory consumption.
To address this challenge, sketching techniques have emerged as a resource-efficient solution, employing compact data structures—referred to as "sketches"—to approximate traffic statistics within well-established error margins. 

Poseidon~\cite{zhang2020poseidon} leverages the Count-Min sketch to estimate the sizes of network flows.
The process begins when a flow is identified, at which point a series of hash functions calculate multiple indices corresponding to this flow.
The Count-Min sketch then increments the flow counters in the register arrays at these indices.
To estimate a flow's size, Poseidon retrieves the values from the counters and adopts the smallest value as the estimate.
This method allows users to specify thresholds to identify potentially malicious flows that exhibit abnormal sizes.
While Poseidon is adept at measuring flow size, it does not cater to other network statistics.
To bridge this gap, Jaqen~\cite{liu2021jaqen} introduces the concept of universal sketches.
Utilizing the Count Sketch, Jaqen is capable of estimating a broader range of network statistics, such as source IPs and source port numbers. 
This versatility enables users to set precise thresholds for specific network features, thereby facilitating more granular anomaly detection.

Tailored to the NDN network measurement, Xu et al.~\cite{xu2022towards} propose the LiEffi-FM Sketch.
This approach is predicated on the observation that during a DDoS attack, malicious bots tend to generate a significant increase in Interest packet requests that share the same name prefix but differ in requested data content.
The LiEffi-FM Sketch efficiently monitors these Interest packets at NDN routers, probabilistically counting the unique data requests linked to a common name prefix.
Xu et al. then employ Monte Carlo hypothesis testing to establish a threshold for distinguishing between benign and malicious requests, thereby enabling ongoing protection with minimal resource expenditure.

\subsection{Learning-Based Detection}\label{subsec:learning-based-detection}
Machine learning and deep learning are critical in DDoS detection for their ability to rapidly analyze and respond to complex and evolving threats.
Unlike traditional methods, these technologies adapt and learn, identifying attack patterns and potential vulnerabilities in real-time, thus enhancing the capability to predict, detect, and mitigate DDoS attacks effectively, ensuring greater network security and service uptime.
In this section, we revisit existing learning-based techniques.

\subsubsection{Machine Learning}
Recent research has leveraged both clustering and classification techniques to enhance the detection of malicious network flows.
Ahmed et al.~\cite{ahmed2018statistical} proposed a method where clustering is used to create distinct fingerprints of web applications by collecting packet-level features, such as IP addresses, and stream-level features, like the number of bytes transferred from client to server.
These fingerprints are then grouped using efficient clustering algorithms, e.g., Mean-shift.
The resulting clusters are associated with corresponding applications, such as HTTP or SMTP, by using labeled training data.
Consequently, when analyzing a client request flow, its fingerprint is compared against this repository to identify the application it corresponds to.
Unmatched flows are considered suspicious.

In parallel, Qin et al.\cite{qin2015ddos} adopted entropy-based features, such as packet size and flow duration, employing the K-means algorithm to distinguish between benign and malicious traffic.
By modeling the normal patterns of request behavior, this technique flags flows with entropy vectors that deviate significantly from those of benign clusters as potential threats.
Bhatia et al.\cite{bhatia2021mstream} further extend this concept by clustering flows that exhibit large volumes of similar activities, considering both categorical and numerical attributes to detect suspicious patterns.

On the classification front, MM et al.\cite{mm2022efficient} introduced a novel approach that utilizes Kernel Principal Component Analysis to refine and select the most relevant flow features, followed by training a Support Vector Machine-based classifier to sort the flow samples.
Similarly, Panigrahi et al.\cite{panigrahi2022intrusion} applied Multi-Objective Evolutionary Feature Selection to pinpoint the most informative flow features and employed a combination of Decision Table and Naive Bayes classifiers to categorize network traffic.
Eshete et al.~\cite{eshete2017dynaminer} design a system, DYNAMINER, which abstracts HTTP transactions into Web Conversation Graphs (WCGs) to capture these dynamics.
The temporal changes reflected in the WCGs, such as node degree variations, are used to train an ensemble random forest classifier to distinguish between benign and compromised flows.

\subsubsection{Deep Learning}
Autoencoder is arguably the most popular deep model for DDoS detection.
The Kitsune framework, as introduced by Mirsky et al.~\cite{mirsky2018kitsune}, employs an ensemble of autoencoders that operate online and in an unsupervised manner to differentiate between normal and anomalous traffic patterns.
This is achieved by feeding network traffic instances to the ensemble, where each autoencoder attempts to reconstruct traffic feature subsets.
The reconstruction quality is assessed using the root mean squared error (RMSE) metric, and a collective RMSE from all autoencoders is produced.
An output module then evaluates this aggregate RMSE against a decision threshold to determine the traffic's nature—benign or potentially malicious.
Aktar et al.\cite{aktar2023towards} propose a deep learning model that leverages a contractive autoencoder to detect DDoS anomalies, furthering the application of autoencoders in this field.

Other deep models, e.g., multi-layer perceptron and LSTM, are also widely used.
Diallo et al.\cite{diallo2021adaptive} introduced ACID, which utilizes a neural network with multiple kernels for effective anomaly detection.
De et al.\cite{de2021detection} selectively use three traffic features (packet count, entropy, and average inter-arrival time) to train a Multi-Layer Perceptron (MLP).
Combined with Fuzzy Logic, this MLP is particularly adept at detecting RoQ DDoS attacks due to its high accuracy.
Wang et al.\cite{wang2020dynamic} also explore MLP-based DDoS detection, employing sequential feature selection to reduce redundancy and irrelevance, and they incorporate a dynamic feedback loop to continuously adapt to changing traffic patterns.
Aydin et al.\cite{aydin2022long} developed LSTM-CLOUD, an LSTM-based system for monitoring network traffic in cloud environments, utilizing historical data to pinpoint potential DDoS attacks.
Xu et al.~\cite{xu2022xatu} exploit predictable attacker behavior and auxiliary signals from prior incidents to train an LSTM network, which employs survival analysis for early attack detection while minimizing false positives.

Graph neural networks also gained attention.
Agiollo et al.~\cite{agiollo2023gnn4ifa} address NDN Interest Flooding Attacks by representing the network as a graph and using Graph Neural Networks (GNNs) for both Supervised Attack Detection (SAD) and Unsupervised Attack Detection (UAD).
SAD classifies the network's state with a trained GNN, while UAD relies on the network's ability to reconstruct masked graph segments and detect anomalies through reconstruction errors.
Duan et al.~\cite{duan2022application} argue that existing DL methods for DDoS detection don't adequately capture IP pair interactions and topological data, which are vital for identifying anomalies.
They suggest a novel approach using Dynamic Line Graph Neural Networks (DLGNN) to analyze dynamic spatiotemporal graphs of network traffic, capturing the intricate spatial and temporal dynamics of IP communications.

Some works try to address key issues during neural network training (e.g., data collection) and detection (e.g., high false positive rates).
Wichtlhuber et al.\cite{wichtlhuber2022ixp} suggest collecting data from ISP blackholed traffic for training deep detection models, as it often contains malicious samples.
Fu et al.\cite{fu2023point} address false positives in learning-based flooding detection systems with pVoxel, which discriminates between sparse benign traffic features and dense malicious traffic features in the feature space, refining detection accuracy. Zhao et al.\cite{zhao2023ernn} improve the accuracy of learning-based systems by training a Recurrent Neural Network (RNN) with both normal and noisy traffic that includes common network-induced phenomena, using a Mealy machine to dynamically adjust the training process and enhance the system's robustness to these phenomena.

\subsubsection{Hybrid Learning Approaches}
Recent advancements have explored the integration of machine learning and deep learning techniques to enhance the accuracy and efficiency of anomaly detection systems.
Dong et al.~\cite{dong2023horuseye} introduced a two-tiered framework named HorusEye, which merges the strengths of machine learning and deep learning to initiate a robust detection process.
The framework commences with the deployment of an isolation forest model, specifically chosen for its ability to rapidly sift through and flag suspicious network traffic at high throughput rates.
Subsequent to the initial screening by the isolation model, traffic deemed suspect undergoes a more thorough investigation utilizing an Asymmetric Autoencoder (AAE).
This AAE is designed with a deep-layered encoder to effectively distill complex data representations, while its paired decoder is tasked with reconstructing the input features, maintaining simplicity to avoid unnecessary computational complexity.
The effectiveness of this analysis is quantified through the Root Mean Squared Error Loss (RMSE), which serves as the criterion for determining the abnormality within the suspect data flow.

In parallel, Long et al.~\cite{long2022hybrid} have devised a detection system employing a Stacked Sparse Autoencoder combined with a Support Vector Machine (SSAE-SVM).
Their approach leverages the autoencoder's unsupervised learning capability to distill a refined representation of the data.
The SVM then steps in, employing these refined features to classify network traffic with a heightened level of precision.

Expanding on these hybrid models, Mahadik et al.~\cite{mahadik2023edge} have crafted a sophisticated CNN-LSTM hybrid model, capitalizing on the CNN's innate proficiency in automatic feature extraction and the LSTM's capacity to retain information over extended sequences.
This model is specifically tailored to identify and classify a spectrum of DDoS attacks, ranging from binary to multi-class categories.

\subsection{Adversarial DDoS Detection}\label{subsec:adversarial-ddos-detection}
Adversarial DDoS detection is crucial in maintaining the resilience and reliability of online services in the face of increasingly sophisticated cyber threats.
In this section, we revisit recent detection techniques which targets two adversarial strategies: Encrypted malicious traffic which bypasses deep packet examination, and adversarial learning which bypasses learning-based detection systems.
\subsubsection{Encrypted Malicious Traffic}
As attackers increasingly utilize encryption, traditional detection methods struggle to identify malicious traffic hidden within encrypted data streams.
Fu et al.~\cite{fu2023detecting} tackle this challenge by recognizing that without the need to examine the encrypted packet payload, the interaction patterns between multiple attack flows exhibit distinctive characteristics compared to legitimate traffic.
They introduce HyperVision, a novel system designed to construct interaction graphs from network flows.
To reduce the complexity of these graphs, HyperVision aggregates brief flows, thereby decreasing the overall graph density.
The system subsequently divides the graph into separate connected components and employs clustering techniques based on high-level statistical indicators, such as flow count and size.
By scrutinizing the deviation of components from the cluster centroids, HyperVision effectively flags anomalies. Within these outliers, the system further clusters the edges to accurately isolate and identify the malicious flows.
This methodology provides a robust framework for the detection of encrypted malicious activities.

Complementing this, Cui et al.~\cite{cui2023cbseq} acknowledge that despite the continual evolution of malware, the fundamental objectives, such as executing DDoS attacks, can be detected through consistent network behavior patterns.
Their approach, CBSeq, strategically disregards the encrypted content of the traffic.
Instead, it focuses on the behavioral attributes of network traffic, capturing essential features like the duration and number of flows.
By clustering similar traffic patterns, CBSeq is able to outline behavior sequences that are indicative of malicious intent.
The core of CBSeq's effectiveness lies in its application of a Transformer-based model, MSFormer.
This model is adept at discerning the subtleties within these behavior sequences, thereby empowering CBSeq to distinguish between benign and malicious network traffic with high accuracy.

\subsubsection{Adversarial Learning}
Adversarial attacks pose a significant threat to learning-based detection systems.
Mustapha et al.~\cite{mustapha2023detecting} initially presented a Long Short-Term Memory (LSTM) method tailored for the detection of adversarial Distributed Denial of Service (DDoS) attacks.
However, they noted its inadequacy when confronted with a range of adversarial DDoS attack types, particularly those synthesized by Generative Adversarial Networks (GANs).
To address this limitation, they refined the LSTM detection framework by incorporating adversarial samples produced by a GAN.
This enhancement significantly bolstered the LSTM model's prowess in recognizing these sophisticated attacks.
Wang et al.~\cite{wang2023bars} address the vulnerability of deep learning-based detection systems to adversarial samples.
They introduce BARS, a robustness certification framework that enhances system resilience by applying customized noise distributions to various features according to their susceptibility to adversarial attacks.
BARS generates adversarial examples to certify and improve the detection system's robustness against such evasion tactics.
Catillo et al.~\cite{catillo2023case} investigate the robustness of both machine learning (e.g., autoencoders) and non-ML-based (e.g., decision trees) intrusion detection systems against adversarial DDoS attacks. Utilizing the CICIDS2017 dataset, their findings suggest that autoencoder-based models are more robust to adversarial samples, while decision trees are significantly more vulnerable.

Fu et al.\cite{fu2021realtime} tackled the randomized attacks, where adversaries blend benign packets with malicious ones to elude detection systems.
Their research found that frequency domain features of network traffic offer greater resilience against such evasion attempts.
By accurately representing traffic patterns with minimal information loss, these features improve detection accuracy and throughput.
Fu et al. employed Discrete Fourier Transformation (DFT) to translate time domain traffic features into the frequency domain, revealing the repetition patterns of traffic.
They then trained a classifier with these features using clustering algorithms to differentiate between benign and malicious traffic in real-time.
Complementing this approach, Fouladi et al.~\cite{fouladi2022novel} observed that specific traffic statistics, such as the count of unique source IP addresses (USIP) and the normalized number of unique destination IP addresses (NUDIP) relative to the total packet count, exhibit notable changes in both time and frequency domains during DDoS attacks.
They applied the continuous wavelet transform (CWT) to transform USIP and NUDIP statistics into two-dimensional time-frequency domain features.
These features were then fed into a Convolutional Neural Network (CNN) classifier that was trained to discriminate between normal and malicious traffic efficiently.

In parallel, Matta et al.\cite{matta2017ddos} and Cirillo et al.\cite{cirillo2021botnet} examine the challenge of detecting randomized DDoS attacks by analyzing user behavior within a botnet.
They posit that members of a botnet are likely to show less message innovation than independent users due to the coordinated nature of their activity.
To quantify this, Matta et al. introduce the Message Innovation Rate (MIR), which assesses the diversity of messages sent over time by user groups.
They also develop an algorithm, BotBuster, that identifies potential botnets by clustering users with low MIR scores.
Cirillo et al. build on this by considering scenarios where different botnet groups use distinct emulation dictionaries, and they validate that BotBuster remains effective even when multiple bot groups are present.
Feng et al.~\cite{feng2020application} propose a system to counter randomized DDoS attacks using a Markov decision process that evaluates traffic legitimacy in context, including server resource usage and client-server interaction history.
Their system employs a reinforcement learning agent to differentiate between legitimate and malicious traffic, dynamically adjusting its responses to minimize disruption to legitimate users and respond effectively to attacks.

Finally, Yang et al.~\cite{yang2021cade} address concept drift in DDoS detection models that occurs when attackers alter their behavior, causing the testing data distribution to deviate from the training data.
To combat this, they propose CADE, which refines the training process by mapping high-dimensional traffic features to a lower-dimensional latent space for clustering similar flows.
CADE then employs contrastive learning to enhance the separation between these clusters.
This method allows for the categorization of malicious samples into fine-grained sub-classes, unveiling diverse attack strategies and improving the model's training robustness against evolving threats.

\subsection{IoT Botnet Detection}\label{subsec-iot-botnet-detection}
The large and growing number of IoT devices, coupled with multiple security vulnerabilities, brings an increasing concern for launching DDoS.
As a result, instead of pinpointing malicious traffics and flows as shown in previous sections, fruitful research works focus on the detection of (infected) IoT devices and malicious device behaviors.

\subsubsection{Malware Download Activity}
Recent research in cybersecurity has targeted the detection of botnets by scrutinizing malware download patterns.
Invernizzi et al.~\cite{invernizzi2014nazca} introduced Nazca, a system that analyzes the web traffic graph to distinguish malware downloads.
The fundamental concept behind Nazca is that the collective analysis of malware downloads reveals distinct and identifiable patterns which are not evident when these downloads are viewed individually, thus exposing their malicious intent.
Nazca tracks HTTP requests, collecting metadata such as endpoints, URIs, and the presence of executable downloads.
It then identifies suspicious downloads based on anomalous traits not typical of legitimate software, like evasion techniques or connections to dubious servers.
By clustering these events, Nazca improves the accuracy of bot detection and reduces false positives.

Furthering this line of research, Kwon et al.~\cite{kwon2015dropper} studied complex malware, like trojans, that trigger subsequent downloads.
They developed a downloader graph to represent the relationships between downloaded executables on infected hosts. By examining the structural nuances of these graphs, e.g., the rate of new executable downloads and overall structure, they trained a random forest machine learning classifier to differentiate between benign and malicious download activities, enhancing the detection of compromised systems.

\subsubsection{Malware Spreading Activity}
Superspreaders are unique hosts that are characterized by their extensive network of distinct connections.
Within the realm of DDoS attacks, these superspreaders commonly represent infected machines that reach out to numerous other systems to propagate DDoS malware.
Consequently, their detection is pivotal for bolstering network security.

A fundamental approach to identifying superspreaders is by monitoring all unique destination contacts made by each host through the use of a hash table, as demonstrated by Flowscan~\cite{plonka2000flowscan}.
However, this technique, which relies on maintaining the state of each network flow, can be prohibitively memory-intensive, especially on high-speed networks where it becomes impractical.
To address the limitations of monitoring large volumes of hosts, Kamiyama et al.~\cite{kamiyama2007simple} introduced a method that employs hash-based flow sampling for pinpointing potential superspreaders.
This technique begins by sampling packets according to the hash of the flow key.
Subsequently, a Bloom filter is applied to ascertain whether the sampled packet corresponds to a new flow.
If identified as new, the associated host's counter in the host table is incremented.
Hosts whose counters exceed a predetermined threshold are then classified as superspreaders.
Although this sampling strategy is capable of keeping up with high-speed network links, its accuracy is compromised, which can be problematic.

To facilitate more efficient storage of flow information, some researchers have turned to sketching techniques.
Guan et al.~\cite{guan2009new}, for instance, developed reversible sketches that are capable of estimating the in/out degrees of hashed hosts.
However, the hash functions utilized are computationally intensive, owing to their reliance on complex arithmetic operations, particularly when processing IP addresses.
Liu et al. \cite{liu2015detection} sought to alleviate these computational demands by introducing a novel sketch called the Vector Bloom Filter (VBF).
This filter eschews the maintenance of explicit host identifiers, yet it remains capable of reconstructing the identities of superpoints and assessing their cardinalities.
Building upon these developments, Tang et al.~\cite{tang2022high} crafted the SpreadSketch, a pragmatic sketch data structure tailored for the real-time detection of superspreaders.
SpreadSketch assigns a binary hash string to each connection, which serves as an estimate of the source's fan-out.
Furthermore, by combining multiple instances of SpreadSketch, it is possible to achieve a comprehensive network-wide perspective, which is instrumental in the reconstruction and identification of all superspreaders.

\subsubsection{Malware Infection Activity}
Several research works have developed techniques to detect compromised IoT devices by examining their behavioral patterns.
Antonakakis et al.~\cite{antonakakis2017understanding} identify Mirai-infected devices by monitoring for Mirai-style scanning activities.
Herwig et al.~\cite{herwig2019measurement} detect devices infected with the Hajime malware by analyzing the public Distributed Hash Table (DHT) utilized by Hajime bots for command-and-control (C\&C) communication.
Tegeler et al.~\cite{tegeler2012botfinder} observe that botnets from the same family often exhibit consistent patterns in their C\&C communications, including specific data upload formats and timing patterns for connections to C\&C servers.
Based on these observations, they introduce BOTFINDER, a system that uses five distinct features from bot traffic, such as the average connection duration, and applies clustering to model these bot families.
This clustering helps to determine the infection status of a host and identify the type of malware present.

It is important to note that these detection methods focus on identifying devices compromised by particular malware, such as Mirai or Hajime.
Conversely, Guo et al.~\cite{guo2020detecting} propose two algorithms designed to profile network activities of devices and detect IoT devices irrespective of the specific malware.
Their methods rely on knowledge of the servers with which these devices communicate, typically operated by IoT manufacturers.
The first algorithm inspects the destination IP addresses and DNS queries in client-generated traffic; if these are associated with an IoT manufacturer, the client is likely an IoT device.
The second algorithm uses active scanning to detect IoT devices by identifying TLS certificates containing IoT manufacturer names.
These approaches enable real-time classification of devices as benign or infected by examining their network communications.

Sikder et al.~\cite{sikder20176thsense} demonstrate that benign user activities on IoT devices typically trigger a specific set of sensors, whereas infected devices often disrupt these patterns. By learning the normal sensor data patterns associated with user activities, their system, 6thSense, employs machine learning models, such as Naive Bayes, to detect anomalous sensor activity. Meanwhile, AEGIS [7] profiles the context of user activities and sensor-device interactions. It incorporates smartphone app context, like user interactions with device controls, and employs a Markov Chain-based machine learning technique to identify abnormal behaviors in smart home environments.

\subsubsection{Side Channels}
Recent research has explored the use of side channels to detect compromised IoT devices, employing signals such as electromagnetic (EM) emanations, network traffic fingerprints, and even encrypted traffic patterns.
Khan et al.~\cite{khan2019idea} demonstrate that EM signals emitted by IoT devices exhibit distinct patterns when the devices execute benign applications versus when they participate in DDoS attacks.
Their proposed system, IDEA, leverages EM signals as a side channel to discern DDoS activities on IoT devices.
IDEA operates by first establishing a baseline of EM patterns from a secure device.
Subsequent monitoring of a target device's EM signals allows for the detection of deviations from this baseline.
When EM signal reconstruction errors occur, IDEA interprets these as indicative of anomalous or possibly malicious activities.

Beyond EM signals, researchers have shown that network traffic fingerprints can act as effective side channels.
Shodan, a search engine detailed by the authors~\cite{shodan}, gathers information on devices connected to the internet by scanning IPv4 addresses across select ports.
It identifies devices by correlating textual matches, like "IP camera," with service banners and specific device information.
CAIDA extends this approach to pinpoint compromised IoT devices that initiate communication with allocated but unassigned IP addresses~\cite{torabi2018inferring}.
Censys, another tool comparable to Shodan, allows for community-contributed rules that facilitate the identification of device manufacturers and models via textual patterns in device banners~\cite{durumeric2015search}.
Furthermore, Mirian et al.~\cite{mirian2016internet} focus on industrial control systems (ICS), scanning the IPv4 space with ICS-specific protocols.
Their findings unearth over 60,000 publicly accessible ICS devices, which could potentially be targets for DDoS exploits.

The work by Acar et al.~\cite{acar2020peek} indicates that even when IoT communication is encrypted, valuable insights can be gleaned from the analysis of metadata such as packet lengths and traffic rates.
For example, the pattern of larger packets is often associated with a smart camera transmitting video data, whereas smaller packets might indicate a temperature sensor's data transmission.
Through the application of machine learning techniques, such as k-Nearest Neighbors (kNN), on the sniffed encrypted traffic and its metadata, classifiers can determine device types, states, and user behaviors.
Anomalies in these classifications can signal abnormal device operations, such as involvement in flooding attacks.
Additional studies, like Meidan et al.\cite{meidan2017profiliot}, apply machine learning models to LAN-side measurements to identify IoT devices based on their traffic flow statistics.
They utilize an array of features from network, transport, and application layers, e.g., the number of bytes and HTTP GET requests.
Le et al.~\cite{le2019unearthing} take a unique approach by converting DNS names into embeddings—numeric representations that encapsulate the semantic content of the DNS names.
A deep learning model, specifically a multi-layer perceptron, then classifies devices as IoT or non-IoT based on these embeddings derived from their DNS queries.

\subsection{Summary}
In this section, we outline the essential features required for effective DDoS detection in emerging technologies.
We also address the challenges involved in implementing these features and offer practical guidelines for developing advanced detection systems.

\textit{\textbf{Accurate differentiation between normal clients and adversaries enhances DDoS detection.}}
    Identifying deviations in behavior is an effective strategy for delineating legitimate user activity from DDoS activities.
    Such deviations can be discerned through various indicators.
    For example, anomalies in network traffic bursts, patterns of page visits, resource demands triggered by incoming requests, and instances of address spoofing serve as reliable markers to distinguish between ordinary users and malicious adversaries.
    Incorporating these behavioral signatures into the framework of statistical and machine learning algorithms holds the potential to advance the development of a sophisticated and adaptive DDoS detection system.
    This fusion of behavioral analysis with computational techniques can enhance the system's ability to respond to evolving attack vectors in real-time.
    
\textit{\textbf{Attack-agnostic detection methods are preferred, but it is important to minimize the occurrence of false positives.}}
    DDoS attacks manifest in multiple forms, such as volumetric flooding and low-rate attacks.
    These variations underscore the necessity for a robust and versatile DDoS detection mechanism capable of identifying a broad spectrum of attack patterns.
    Traditional detection methods, tailored to specific attack types, offer high accuracy by leveraging characteristics unique to each attack.
    However, the practical deployment of such specialized detection methods in diverse network environments is impractical for several reasons.

    First, the deployment of attack-specific detection systems often entails substantial costs, which are exacerbated in networks with limited resources, such as those incorporating programmable switches.
    The financial and computational overheads make these methods less viable in resource-constrained settings.
    Second, the dynamic nature of attack strategies poses a significant challenge
    Attackers frequently modify their tactics, necessitating an equally dynamic and responsive detection policy.
    Designing such a flexible policy enforcement that can quickly adapt to changing attack patterns is a complex task.
    Third, the rapid emergence of new DDoS attack vectors complicates the maintenance of up-to-date detection strategies.
    Real-time updates to supervised detection models, such as retraining with new data and relabeling, are hindered by high deployment costs and inevitable delays.
    This lag leaves networks vulnerable to novel attacks.

    In light of these challenges, unsupervised learning emerges as a promising avenue for achieving generalized DDoS detection.
    This approach involves modeling the normal traffic patterns of various application scenarios.
    With a baseline of expected behavior established, the detection system can proactively identify anomalies, regardless of the specific attack technique employed.
    Moreover, unsupervised learning systems can adapt over time, adjusting their baseline models to reflect evolving legitimate user behaviors.
    This continuous adaptation is critical for maintaining detection accuracy in dynamic network environments.

    Despite its potential, unsupervised learning is not without its pitfalls.
    Notably, the tendency for high false-positive rates must be addressed to prevent the erroneous classification of legitimate traffic as malicious.
    Systems like Nazca and DAmpADF represent significant strides in overcoming these challenges, offering refined algorithms to mitigate the issue of false positives while maintaining robust detection capabilities.

\textit{\textbf{Efficient detection is crucial for managing the substantial volumes of modern network traffic.}}
    To effectively manage the escalating volume of network traffic, DDoS detection systems must ensure efficient flow processing speeds and optimized memory utilization.
    The burgeoning traffic presents complex challenges in terms of both processing and storage of flow data.

    In response to these challenges, researchers have proposed various solutions, which can be broadly categorized as software-based or hardware-based approaches.
    Among the software-based strategies, sketching stands out.
    This method employs a streamlined data structure that trades a marginal loss of accuracy for substantial reductions in memory requirements.
    Unlike conventional data structures, such as hash tables, sketches are far more scalable in the face of increasing data volumes.
    Furthermore, their compact design enables faster processing speeds, which is critical for real-time DDoS detection and mitigation.
    
    On the hardware front, the advent of programmable switches has been a game-changer.
    These devices come equipped with advanced hardware primitives, e.g., in-network packet parsing, which facilitate packet parsing and processing at line rates.
    Additionally, programmable switches offer remarkable flexibility; they can be updated to comprehend newly emerging protocols and to address novel attack vectors.
    This adaptability provides a significant advantage over traditional fixed-function hardware, which lacks the capability to evolve in tandem with the dynamic nature of network threats.

\textit{\textbf{Cross-domain data sharing can increase detection accuracy.}}
    DDoS attacks are often characterized by their distributed nature, which can span multiple jurisdictions and involve countless infected devices.
    The complexity and scale of these attacks necessitate a coordinated response from various entities within the internet infrastructure.
    Effective collaboration among Autonomous Systems (ASes), Internet Service Providers (ISPs), and routers is crucial for timely detection and mitigation of these threats.

    Programmable switches, like those highlighted in the work of Jaqen and Mew, demonstrate the potential of such collaboration.
    These switches enable the sharing of monitoring metrics across different ASes and ISPs, facilitating a comprehensive overview of network conditions.
    This shared intelligence can be pivotal for several reasons:
    (1) Proactive Threat Intelligence Sharing.
    By exchanging traffic information, ASes and ISPs can identify emerging attack patterns early and disseminate warnings to preempt potential attacks.
    (2) Resource Optimization: Collaboration allows for pooling resources and expertise, which can lead to more efficient use of available bandwidth and computational power.

    While sharing information is vital for defense, it also raises concerns about user privacy and data protection.
    ASes and ISPs must establish trust relationships and agree on frameworks that respect privacy laws and user consent.
\section{Detection System Deployment}\label{sec:ddos-detection-system-deployment}



\subsection{SDN}
The advent of SDN represents a substantial advancement in network management and security.
SDN introduces innovative core concepts, such as the separation of the control plane from the data plane.
This separation provides a flexible framework for implementing DDoS defense strategies.
In this section, we will explore the benefits of SDN and examine contemporary research that leverages SDN capabilities to enhance DDoS defense mechanisms.

\subsubsection{Programmability}
SDN's programmability facilitates the rapid creation and deployment of software-based features to detect and counteract DoS attacks, eliminating the need for hardware upgrades.
Specifically, it enables rapid deployment of software-based detection targeting various types of DDoS attacks.
Tang et al.'s Performance and Features (P\&F) framework~\cite{tang2021performance} utilizes SDN's programmability to combat Low-rate Denial of Service (LDoS) attacks by incorporating detection and mitigation directly within the SDN controller, showcasing how software solutions can swiftly adapt to security challenges.
Similarly, the LFADefender system~\cite{wang2018detecting} leverages SDN to identify and thwart Link Flood Attacks (LFA) by analyzing network flows and adjusting switch rules in real-time, demonstrating SDN's capability to quickly respond to threats.
SDNShield~\cite{chen2021sdnshield} employs a three-stage defense against SYN flooding attacks, using SDN's programmable features for statistical analysis, authentication of TCP handshakes, and a recovery mechanism for legitimate traffic, emphasizing SDN's role in maintaining network security with agility.

SDN enables quick deployment of security measures and statistical-based detection methods through software. The Joint Entropy Statistical Scheme (JESS)~\cite{kalkan2018jess} leverages SDN's programmability for real-time entropy-based analysis, dynamically creating security rules to defend against DoS attacks, bypassing hardware modifications.
Additionally, the DOCUS framework~\cite{shalini2022docus} utilizes SDN's programmable controller to implement a Cumulative Sum (CUSUM) algorithm, which monitors connection patterns for early DDoS detection.
Finally, Long et al.~\cite{long2022hybrid} use SDN's flexibility to incorporate entropy-based anomaly detection and a hybrid machine learning model for real-time DoS attack classification. Through software updates to the SDN controller, the network autonomously adjusts flow tables to counter threats, showcasing SDN's ability to facilitate immediate, software-centric security responses without hardware upgrades.

SDN's programmability allows for the deployment of advanced learning-based detection systems without hardware upgrades.
Najar's study~\cite{najar2024cyber} demonstrates SDN's capacity to swiftly integrate deep learning techniques such as CNNs for traffic analysis, with preprocessing strategies like Balanced Random Sampling adapting to new attack patterns. This showcases SDN's adaptability in cyber defense.
Hnamte et al.~\cite{hnamte2024ddos} further illustrate SDN's capabilities by developing a DNN model for DDoS detection that can be quickly updated and redeployed in response to evolving threats, emphasizing the value of SDN's flexible infrastructure.
Ribeiro et al.~\cite{ribeiro2023detecting} present an SDN-based architecture that incorporates ML models for real-time traffic monitoring and malicious flow detection. The architecture also employs Moving Target Defense (MTD) to redirect attacks, leveraging SDN's dynamic reconfiguration to mitigate damage, thus exemplifying SDN's role in facilitating the swift implementation of effective software-defined security measures against DoS attacks.

Finally, SDN affords privacy and efficiency enhancement in DDoS attack detection. Zhu et al.~\cite{zhu2018privacy} leverages SDN's programmability to integrate an optimized k-Nearest Neighbors (kNN) algorithm into the SDN controller, enabling encrypted traffic analysis for privacy-preserving DDoS detection.
Rashidi et al.~\cite{rashidi2017collaborative} utilizes a game-theoretic model within the SDN control plane for dynamic resource allocation, illustrating SDN's capacity for efficient detection supports.

\subsubsection{Separation of Control and Data Planes}
Unlike traditional network architecture where the control logic is embedded within each network device (e.g., routers and switches), SDN decouples the control plane from the data plane enables centralized management of the network, allowing for more coordinated and efficient control decisions with a comprehensive overview.
The separation mechanism between control and data planes in SDN offers versatile deployment options for detection methods.
Rezapour et al.~\cite{rezapour2021rl} demonstrate the use of reinforcement learning within the context of SDN to combat link-flooding attacks.
The applied algorithms, ERA and DRA, benefit from the SDN's control-data plane separation by dynamically adapting routing decisions.
Yue et al.~\cite{yue2024ccs} leverage SDN's architecture for efficient LDoS attack defense, employing data plane anomaly detection to offload the controller and facilitate swift centralized responses.
Upon detecting irregularities, switches prompt a controller-led global analysis with Bayesian voting. Confirmed attacks trigger an immediate, optimized rerouting response, showcasing SDN's potential for coordinated, dynamic network defense.

SDN's centralized control management and unified network view enable rapid detection and coordinated mitigation.
Cao et al.~\cite{cao2021detecting} capitalize on the centralized view of the network provided by the SDN controller to maintain an accurate and up-to-date map of the network topology and monitor the state of switches.
This comprehensive visibility is crucial for early detection of link-flooding attacks, as the system can quickly notice and react to changes in traffic flow that may signal an attack.
Similarly, Najar et al.~\cite{najar2024cyber} show how SDN's unified network view helps the CNN models for traffic analysis, pinpointing anomalies across all switches.
Moreover, this centralized intelligence allows the SDN controller to swiftly command rate-limiting and flow removal responses.
Li et al.~\cite{li2023path} demonstrate that SDN's centralized control model is vital for strategic network management. Utilizing the CNNQ algorithm, the framework capitalizes on SDN's global network view to optimize NSF deployment paths, thereby enhancing the network's defense against DDoS attacks.


\subsection{Programmable Switch}\label{subsec:programmable-switch}
The cybersecurity landscape is currently undergoing significant changes due to the emergence of programmable switches.
These switches offer impressive data-plane programmability that surpasses that of SDN, enabling the customization of packet parsing and processing at full network speed.
Consequently, there is a growing body of research dedicated to implementing DDoS defense techniques directly on programmable switches to take advantage of their rapid in-network processing capabilities.
However, the limited resources available on these switches, such as registers and memory, pose substantial challenges for the deployment of such techniques. As a result, recent studies have focused on developing sophisticated DDoS defense mechanisms that are not only effective but also resource-efficient. This section will explore the latest advancements in leveraging the potential of programmable switches to enhance DDoS defense mechanisms.

\subsubsection{Data-Plane Traffic Statistics Analysis}
Programmable switches have become a promising tool for traffic analysis, enabling the rapid collection of network statistics and performing statistic-based DDoS defense.
Given the inherent constraints of programmable switches, such as limited register and memory capacity, data sketching techniques are extensively employed.

Liu et al.~\cite{liu2016one} propose a traffic analysis method using universal sketching to optimize data collection within programmable switches.
The approach starts with packet sampling via flowkeys hashed by predefined functions.
These packets update sketch counters in the on-chip SRAM through the Count Sketch algorithm, while a dedicated list of heavy flowkeys is managed in a fixed-size TCAM for rapid updates.
A P4 program controls packet processing, allowing packets to sequentially move through sampling, sketching, and storage stages.
For DDoS detection, the technique tracks the number of unique flows to a host against a predefined threshold.
Programmable switches maintain counters for these flows using the data plane's universal sketch primitive, enabling real-time detection and mitigation of DDoS attacks with the current switch infrastructure.

Similarly, Jaqen~\cite{liu2021jaqen} applies universal sketching to programmable switches for DDoS attack detection.
The system architecture bifurcates into data and control planes.
The data plane integrates universal sketches with a signature detector to gauge attack-related metrics.
Efficiency is achieved by condensing short hashes into long ones to decrease hash calculations and by updating a single Count Sketch instance per packet to minimize memory accesses.
The control plane utilizes an API, featuring a Query function for metric retrieval and thresholds to sift through traffic for anomalies, thus identifying potential DDoS events.

Ding et al.~\cite{ding2021tracking} introduced two sketch-based methods, P4LogLog and P4NEntropy, which are designed to function within the constraints of programmable switches.
P4LogLog is designed to estimate the flow cardinality, which is the number of unique network flows. It operates by updating a register for each incoming packet. The update is based on the hash of packet attributes, allowing the system to maintain a count of distinct flows without storing each flow identifier.
P4NEntropy aims to estimate the normalized entropy of network traffic.  P4NEntropy leverages the flow cardinality estimated by P4LogLog and combines it with a sketch data structure, such as the Count Sketch, to estimate the counts of packets per flow. The switch then calculates the entropy using operations supported by the P4 language.

\subsubsection{Data-Plane Machine Learning}
There is a growing interest in the application of learning-based methods on programmable switches.
However, the practical implementation of machine learning models on such switches is challenging.
The primary obstacle is the limited set of operations that programmable switches support. For example, they typically lack the ability to perform multiplication or division, which are fundamental operations in many machine learning algorithms.
Additionally, the hardware's architecture, which revolves around match-action tables, is not naturally conducive to the complex computations required by these models.
Therefore, adapting machine learning models to function within these constraints is an active area of research.
The goal is to develop methods that can translate the sophisticated processes of learning-based models into operations compatible with the streamlined, efficient environment of programmable switches.

Barradas et al.~\cite{barradas2021flowlens} focus on using programmable switches for classification-based DDoS mitigation through the Flow Marker Accumulator (FMA), a data structure for efficient flow classification within the switch's data plane. The FMA captures flow markers (i.e., simplified encodings of packet distributions) using quantization to bin continuous features and truncation to fit the switch's memory constraints. Tailored for simplicity to match the switch's computational limitations, the FMA, written in P4 language, balances memory use and classification accuracy and is distributed across the switch pipeline using match-action units.

Alcoz et al.~\cite{alcoz2022aggregate} have implemented a DDoS detection system using online clustering techniques within programmable switches, functioning at line rate.
This system integrates a data plane-based online-clustering module for attack detection with a programmable-scheduling module that operates across the control and data planes to mitigate attacks.
The control plane routinely analyzes clusters to determine scheduling policies for the data plane, which processes incoming packets and manages traffic based on these policies.
The system accommodates the limitations of programmable switches by employing registers for ordinal feature clusters and bloom filters for nominal features, with a resubmission mechanism to update clusters for new traffic patterns.

Zhou et al.~\cite{zhou2023efficient} have developed a method to deploy decision tree-based ML models, specifically Random Forest (RF) and XGBoost (XGB), on programmable switches.
Their system, NetBeacon, features an innovative IDP (In-band Network Telemetry Data Plane) design that employs a sequential multi-phase model architecture.
It is tailored to process per-packet and flow-level features at line speed, including the computation of aggregate and summary statistics such as maximum, minimum, mean, and variance.
To circumvent the issue of match-action table bloat, NetBeacon introduces "range marking", a technique that maps numerical ranges to unique bit strings, enabling efficient representation within match/action tables.
Recognizing the hardware's limitations, NetBeacon applies alternative methods such as approximate calculations and bit shifting to estimate statistical measurements like variance and mean, avoiding operations like multiplication and division that are unsupported on programmable switches.

Similarly, Dong et al.~\cite{dong2023horuseye} introduce HorusEye, a system that implements an ensemble of Isolation Trees (iTrees) on programmable switches for anomaly detection.
HorusEye transforms the iForest model into a collection of whitelist matching rules, enabling the deployment on the switches' match-action tables.
It operates by first parsing incoming packets to attribute them to their respective flows, employing bi-hash algorithms for accurate flow identification.
Subsequently, HorusEye captures burst-level traffic features using a defined segmentation threshold and leverages bi-directional flow matching.
With the traffic features extracted, the system applies the derived iForest whitelist rules to detect anomalies.

Recent research has been focusing on how to efficiently compute network features, which are vital for ML detection, on programmable switches.
Romeiras et al.~\cite{romeiras2023poster} introduce Peregrine, which equips the data plane to calculate flow features in real-time as the switch processes packets, utilizing the PISA pipeline's capabilities for basic arithmetic and stateful memory operations.
It uses the switch's stateful memory to maintain counters for metrics such as packets and bytes per flow, updating these with each passing packet.
Peregrine computes a variety of statistics, including simple unidirectional metrics and more complex bidirectional ones, like mean, variance, standard deviation, magnitude, radius, and an approximate covariance. 
Similarly, Doriguzzi et al.~\cite{doriguzzi2024introducing} demonstrate how a programmable switch can incrementally calculate statistics, such as the average packet size per flow.

\subsubsection{Data-Plane Security Primitive}
In addition to offering direct defense solutions against specific DDoS attacks, several research efforts are dedicated to the creation of security primitives for programmable switches.
These primitives serve as fundamental building blocks that enable users to develop their own tailored defense strategies to counter DDoS threats effectively.

Xing et al.~\cite{xing2021ripple} introduce Ripple, a framework that incorporates a set of seven security primitives tailored for link-flooding DDoS (e.g., Crossfire and Coremelt attacks).
Ripple's standout feature is its ability to construct a defense panorama.
This is a comprehensive view that captures network-wide threat signals derived from local traffic patterns on each switch.
With the panorama, users can specify their detection and mitigation policies using Ripple's primitives, and these policies will be translated into a coordinated collection of P4 programs by the Ripple compiler.
By doing so, users can concentrate on designing their defense strategies at a high level, while Ripple takes care of the underlying complexity, generating the necessary P4 code to implement these strategies across the network's switches.

Mew~\cite{zhou2023mew} introduces a framework that enables programmable switches to synchronize information based on specific criteria.
It achieves multi-level cooperation through four APIs: \textit{Monitor} for state storage configuration, \textit{Sync} for state dissemination within a range, \textit{Request} for defining state request modes and intervals, and \textit{Trigger} for executing actions upon certain conditions.
These APIs simplify the design of detection and mitigation mechanisms.
To manage the limited resources on programmable switches, Mew implements a lightweight distribution protocol that balances storage by selecting the least-utilized switch with a greedy algorithm, adjusting the distribution of states over time.
Additionally, Mew incorporates a memory resizing mechanism that facilitates memory sharing and reallocation among defense functions, allowing for efficient utilization of switch memory and enabling more functions to run concurrently.

\subsection{Summary}
In this section, we summarize the advanced features brought by these advanced network hardware for efficient and effective DDoS detection.

\textbf{\textit{Emerging network hardware facilitates line-speed DDoS defense.}}
    The escalation of network traffic volumes presents a significant challenge to traditional defenses against DDoS attacks (e.g., middleware at edge routers and scrubbing centers).
    In this context, the advent of data-plane programmability, as seen in modern network hardware, offers a transformative approach to mitigating DDoS risks.
    By leveraging the flexibility of control plane decision-making along with the efficient execution capabilities of the data plane, networks can deploy more sophisticated and responsive defense mechanisms.
    
\textbf{\textit{Emerging network hardware facilitates the deployment of coordinated DDoS defenses.}}
    DDoS attacks (e.g., link flooding attacks) are being more sophisticated and difficult to detect and mitigate.
    Addressing these threats requires a dynamic and coordinated defense systems.
    SDN and programmable switches offer a beacon of hope in this cyber arms race.
    Two frameworks that epitomize this potential are Ripple and Mew.
    By constructing a network-wide defense panorama and enabling efficient information synchronization and memory management, these frameworks provide the tools necessary to counteract the stealth and complexity of modern DDoS threats.

\textbf{\textit{Techniques are advancing to develop sophisticated detection models that can operate on hardware with limited capabilities in a flexible manner.}}
    The advancements in DDoS detection techniques for hardware with limited capabilities, like programmable switches, showcase the ongoing innovation within the field of network security.
    One example of this is how complex machine learning models, such as decision trees, can be converted into formats like match-action rules that are compatible with the limited processing capabilities of programmable switches.
    This conversion process enables the implementation of advanced DDoS detection models on these powerful yet constrained devices.
    The development of security primitives also simplifies the deployment process of defense strategies.
    Additionally, the inherent programmability of these switches offers security analysts the flexibility to swiftly change and implement customized defense strategies without significant downtime.
    This adaptability is crucial for maintaining robust defense mechanisms against DDoS attacks in a dynamic threat landscape.

\textbf{\textit{Efficient resource management on these hardware is of paramount importance.}}
    The efficient management of advanced network hardware resources is a cornerstone in the battle against DDoS attacks.
    Techniques such as packet transformation~\cite{holland2021new}, data sketching~\cite{liu2016one}, and dynamic resource allocation~\cite{rashidi2017collaborative}, coupled with the programmability and cooperation enabled by frameworks like Mew, provide a robust foundation for network defense.
    These methods ensure that despite their limited resources, advanced network devices can effectively detect and mitigate DDoS attacks, preserving network reliability and service availability.



\section{Open Problems and Future Work}\label{sec:open-problem}

\subsection{Uncovering DDoS Vulnerabilities}
As discussed in Section~\ref{sec:ddos-attack}, DDoS attacks have emerged as a formidable weapon, constantly evolving to exploit the vulnerabilities of a myriad of protocols and systems.
With the dawn of new technological eras, we witness the birth of innovative protocols and intricate systems at a blistering pace~\cite{li2024survey,chen2024frequency}.
Yet, this rapid march of progress casts a shadow—a vast, unexplored territory where potential weaknesses against DDoS onslaughts lie hidden.
As a result, there is a critical need to forge a comprehensive analysis framework, one that is meticulously designed to dissect and scrutinize the DDoS attack vectors and vulnerabilities inherent within these nascent innovations.

Towards this end, we aim to summarize the critical features which are helpful to study the vulnerability of emerging protocols and features.
Specifically, we surprisingly find that despite the evolution of target protocols, certain patterns of vulnerability (e.g., session management) recur, suggesting a failure to learn from past mistakes.
This recurrence indicates a systemic issue within protocol design processes, wherein lessons from historical vulnerabilities are not adequately integrated into new developments.
In response to this challenge, we propose a comprehensive analysis of common protocol components that have historically and nowadays introduced vulnerabilities.
This analysis is intended to serve as a resource for security professionals and protocol designers, enabling them to more effectively evaluate the security posture of new protocols.
By identifying and understanding these commonalities in vulnerability patterns, it is possible to anticipate potential attack vectors, ensuring a higher degree of security evaluation in the early stages of protocol development.

\textbf{\textit{Improper session management.}}
    Improper session management is a central vulnerability that can lead to DDoS conditions in various protocols.
    The attacks on TCP, QUIC, and SIP protocols illustrate how exploitation of session management can overwhelm and incapacitate servers~\cite{nawrocki2021quicsand,sisalem2006denial,tang2014sip}.
    For instance, in TCP, the SYN flooding attack preys on the limit of the server’s backlog queue for half-open connections~\cite{luo2014mathematical,tang2013modeling,schuchard2010losing,wang2002detecting}.
    When examining new protocols, it is crucial to consider the robustness of session management mechanisms.
    Researchers and developers should validate the following aspects.
    First, the mechanisms for managing sessions should be scalable and resilient to unexpected surges in connection attempts.
    Second, protocols must incorporate aging-out strategies to quickly clean up sessions created by malicious session requests.

\textbf{\textit{Identity spoofing.}}
    Identity spoofing serves as a linchpin for DDoS attacks, including protocol-specific amplification and flooding attacks, e.g., QUIC, SIP, and DNS~\cite{yin2023waterpurifier,kim2017preventing,yazdani2022mirrors}.
    For example, attackers can exploit QUIC/DNS's UDP foundation by sending Initial packets with a spoofed source IP, which forces the server to respond with disproportionately large packets to the victim's address~\cite{nawrocki2021quicsand}.
    In VoIP environments, attackers can flood SIP routers with spoofed BYE messages, disrupting active calls~\cite{tang2014sip}.
    When evaluating new protocols, it is crucial to analyze their resilience to identity spoofing. 
    This involves a thorough examination of how the protocol handles source verification.
    Best practices for such evaluations should include stress testing under spoofed conditions, validating whether spoofed requests are filtered or rate-limiting measures are in place and effective.

\textbf{\textit{Packet fragmentation.}}
    Packet fragmentation has emerged as a salient vulnerability pattern that adversaries exploit to orchestrate DoS attacks, which affects IP and HTTP protocols~\cite{gilad2011fragmentation,atlasis2012attacking,dantas2014selective}.
    This pattern exploits the fundamental design of protocols where large packets must be broken down into smaller fragments for transmission and then reassembled at the destination.
    Attackers take advantage of the complexity in this reassembly process by injecting malicious, overlapping IP fragments and culminating in service disruptions or crashes.
    In examining new protocols for similar vulnerability patterns, it is imperative to scrutinize their packet handling and reassembly mechanisms under various edge cases and adversarial conditions.
    For instance, understanding how a protocol deals with fragmented packets and ensuring that it has robust checks against overlapping, missing, or redundant fragments is crucial~\cite{feng2022pmtud}.
    Moreover, mining for vulnerabilities should involve stress-testing the protocol with deliberately malformed packets to observe how it handles exceptional cases.

At the same time, it is of paramount importance to discern recurring vulnerability patterns embedded within emerging systems.
Such vulnerabilities are the Achilles' heel that could precipitate DoS threats. Understanding and addressing these weak links proactively is not just an exercise in threat mitigation but a fundamental strategy to fortify our digital ecosystem against the ever-evolving menace of service disruption.
Herein, we enumerate critical system components and features that are susceptible to DoS threats, offering guidance for scrutinizing new systems for potential vulnerabilities.

\textbf{\textit{System architecture.}}
    Analyzing system architecture is an essential step in identifying potential vulnerabilities to DoS attacks.
    A clear illustration of this can be found in the architecture of LTE networks, where the distinction between control and data planes can be a critical weakness~\cite{javadpour2023reinforcement,mirsky2020ddos,silva2020repel}.
    Similarly, SDN are susceptible to DDoS attacks due to the same principle, and this separation can be exploited by attackers who generate malicious traffic aimed at overwhelming the specific system plane~\cite{shin2013avant,shin2013attacking,cao2019crosspath}.
    
    To uncover comparable DoS vulnerabilities in new systems, security researchers may adopt a systematic approach that includes the following steps.
    (1) Develop a deep understanding of the system's architecture with a keen focus on the separation of planes and their interdependencies.
    It is critical to identify how communication between the control and data planes is managed and to determine potential choke points where traffic could be maliciously concentrated, e.g., the SDN path from the data plane to the control plane.
    (2) Identify and evaluate centralized components within the architecture—such as SDN controllers or management servers—that represent single points of failure.
    If these components are compromised or overwhelmed, it could result in a systemic failure, effectively crippling the network.
    (3)  Conduct controlled and ethical stress testing, simulating targeted attacks on independent planes.
    By doing so, it is possible to assess the system's resilience and response to high volumes of malicious traffic directed at specific planes.
    
\textbf{\textit{Resource sharing.}}
    Resource sharing is a vital component of contemporary networked systems, aimed at enhancing both efficiency and flexibility.
    However, the very act of sharing resources opens the door to potential security vulnerabilities, especially in the face of DDoS attacks.
    At the heart of the issue lies the principle of interdependence within shared resources, which can be exploited to carry out such attacks. Notable examples include network slicing in 5G technologies, the utilization of a common egress IP in serverless computing platforms~\cite{silva2020repel}, and the sharing of a single Software-Defined Networking (SDN) link for both data and control signal transmission~\cite{cao2019crosspath}.

    To identify and understand the possible DoS threats that this resource sharing might incur, it is crucial for security researchers to gain a comprehensive overview of the entire spectrum of shared resources in a given system. This encompasses not only the physical hardware but also the software components that might be utilized concurrently by different users or services.
    Equally important is the need to examine the patterns of resource consumption and to establish the network of interdependence among the system's clients. By creating an appropriate threat model — one that assumes the presence of malicious clients intent on manipulating shared resources — researchers can then investigate the impact of resource consumption on legitimate users, with a particular focus on assessing its effects on the reliability of the system.

\textbf{\textit{Component dependency.}}
    Modern systems often involve complex inter-dependencies among components.
    These dependencies can create unforeseen DoS vulnerabilities, particularly when one component's performance is contingent upon availability or reliability of other components.
    For example, in a routing system, a large number of important server nodes usually rely on a few common critical links.
    Identification of these links can help to reveal potential vulnerabilities for launching link-flooding attacks~\cite{kang2013crossfire}.
    As another example, figuring out the state dependency among control centers in a smart grid system can help to pinpoint the critical position for injecting false measurement signals and paralyzing the whole grid with little cost~\cite{vukovic2014security}.
    Finally, considering that systems usually rely on underlying network protocols (e.g., the TCP and BGP protocol for blockchain system), the evaluation of underlying protocols' attack surface can bring more insight towards revealing DoS potentials of systems which are built on~\cite{heilman2015eclipse,tran2020stealthier}.

    For security researchers and practitioners looking to unearth similar vulnerabilities in new systems, the following strategies may apply.
    (1) Start by creating a detailed map of the system architecture, highlighting the dependencies between components. Tools like dependency graphs can be invaluable in visualizing and understanding complex interconnections.
    (2) Identify potential checkpoints where traffic or data converges, and assess the impact of their failure. This includes not only physical links but also critical software processes.
    (3) Study the underlying network protocols for known vulnerabilities and consider how they might be exploited in the context of the current system.

\subsection{Construction of Adversarial Attack and Detection Strategies}
As the arms race between cybersecurity defenses and adversarial attacks continues, the complexities of both are escalating.
The emergence of advanced adversarial learning techniques, the proliferation of commercial DDoS protection services, and the strengthening of protocol security mechanisms compel attackers to constantly innovate their strategies to circumvent state-of-the-art DDoS defense solutions.
This dynamic landscape opens up new research avenues in the study of adversarial DDoS attack and detection, which are shown as follows.

\textbf{\textit{Adversarial machine learning for malicious traffic generation.}}
    Attackers are leveraging machine learning (ML) and deep learning (DL) to generate malicious traffic that can elude detection systems~\cite{abusnaina2019examining,mustapha2023detecting,matta2017ddos}.
    By training models that can anticipate the behavior of learning-based detection systems, adversaries can craft traffic that blends with legitimate network activity, thus increasing the difficulty of detection.
    Techniques such as flow-merge and Generative Adversarial Networks (GANs) are particularly effective in refining malicious traffic to mimic benign characteristics, challenging the reliability of current detection methods.
    To counter the adversarial learning and detect the generated traffics, a promising direction is to enhance adversarial training for the detection model.
    Specifically, incorporating a wider array of adversarial examples during the training phase can prepare detection systems to handle unexpected attack vectors.
    Moreover, applying robustness tests across different types of models (e.g., decision trees, autoencoders) can help identify adversarial traffics specific to certain algorithms, leading to more resilient hybrid systems.

\textbf{\textit{Probing and circumventing commercial DoS protection services.}}
    With an increasing reliance on commercial DDoS protection services, assessing their efficacy has become vital~\cite{jin2018your,nosyk2023closed,wu2014software,shankesi2010resource}.
    The first stage in this process involves designing probing techniques to identify whether the victim is under protection.
    Active probing, which uses carefully crafted requests and subsequent response analysis, is the favored approach due to its precision.
    Once a protection service is identified, understanding its life-cycle (e.g., how it manages changes in client behavior and residual DNS records~\cite{jin2018your}) and dissecting its foundational protection mechanisms (e.g., client puzzles) are crucial to reveal vulnerabilities.
    This knowledge is instrumental in developing strategies to investigate the vulnerability of commercial services.

\textbf{\textit{Examination of protocol security mechanism.}}
    The protocol security mechanism is paramount for ensuring the authenticity, confidentiality, and integrity of communications.
    However, these mechanisms often interact with security credentials (e.g., TCP sequence numbers), which can be exploited to glean sensitive information~\cite{cao2016off,wang2024off,feng2022pmtud}.
    This potential side-channel can be weaponized to conduct DDoS attacks, for example, by sending malicious RST packets to prematurely close legitimate connections~\cite{feng2020off,feng2022off}.
    Therefore, rigorously evaluating the resilience of protocol security mechanism against such exploitation is necessary to reveal potential attacks.

\subsection{Privacy-Preserving DDoS Detection}
Various detection techniques necessitate differing amounts of data to discern the distinction between legitimate users and potential attackers.
For example, detection methods that rely on machine learning algorithms utilize traffic feature extraction from both user and attacker data streams to train their models~\cite{ahmed2018statistical,mirsky2018kitsune}.
However, certain techniques raise greater privacy concerns.
Behavior-based detection methods, in particular, often require access to detailed data such as users' browsing history or server resource utilization logs~\cite{tandon2021defending,tandon2023leader}.
These data are used to identify nuanced patterns of behavior, which can then aid in better differentiating between legitimate users and attackers.
Privacy considerations become increasingly complex when there is a need for cross-domain data sharing, where user information is exchanged between different autonomous systems (ASes).

As a result, it is critical to consider the equilibrium between fortifying security measures and upholding user privacy.
Towards this end, techniques like federated learning and data encryption shed lights on addressing the privacy issue.
For instance, Dimolianis et al.~\cite{dimolianis2022ddos} propose to use federated learning techniques, which collaborate multiple Autonomous Systems to train a shared model using their private data.
Each participant trains the model on their local data and computes an update to the model parameters.
As another example, Zhu et al.~\cite{zhu2018privacy} enforce perturbation encryption to encrypt the network traffic.
This encrypted data is then sent to the Computing Server (CS), which is responsible for performing traffic examination and DDoS detection.
This perturbation ensures that the actual traffic data cannot be directly read by the CS, thus preserving the privacy of the data.

However, existing privacy-preserving DDoS detection methods exhibit limitations that warrant further exploration.
The first issue is efficiency. The operations intended for privacy preservation, such as data encryption and federated communication, often introduce additional processing time, which can be significant when dealing with real-time network traffic
It is crucial to develop more advanced methods that can maintain privacy without a substantial impact on detection efficiency.
The second limitation involves the robustness of existing privacy-preserving detection methods, often tailored to defend against specific types of DDoS attacks.
To enhance the robustness of these systems, it is necessary to extend the models to be capable of recognizing a variety of attack vectors concurrently, thereby improving adaptability to different threats.
Lastly, certain methods, particularly those based on federated learning, are vulnerable to data poisoning. Adversaries can manipulate the model by injecting malicious traffic data, potentially compromising the model's integrity. Addressing this vulnerability requires the creation of more resilient models that can detect and mitigate the effects of data poisoning to safeguard the detection process.

\subsection{DDoS Detection Without Control Planes}
Current solutions (e.g., SDN) impose substantial burdens on the control plane.
It is tasked with aggregating traffic statistics, detecting anomalies, and deploying mitigation strategies onto the data plane.
However, the data plane is relegated to basic functions such as rudimentary traffic statistics computations (e.g., packet counts) and implementing detection rules (e.g., flow rules).
This configuration introduces several critical limitations.

\noindent\textit{(1) New attack vectors.} The control plane becomes a target for attackers.
A notable example is the SDN control plane saturation attack (Section \ref{sec:ddos-attack}), which can induce a single point of failure, effectively crippling the entire network.
    As a result, the system intended to safeguard against DDoS attacks can be exploited to disrupt network operations.
    
\noindent\textit{(2) Performance degradation.} Communication between the control and data planes significantly hampers performance.
The necessity for the data plane to send traffic statistics to the control plane, coupled with the control plane's periodic flow rule updates on the data plane, results in increased latency for standard traffic routing due to the additional round-trip communications.
    
\noindent\textit{(3) Resource constraints.} Developing frameworks for control and data plane interaction demands resources, such as memory, which are often scarce and valuable on the data plane.
    This resource allocation can strain the data plane's capabilities, leading to suboptimal performance.

To address the above limitation, a promising strategy is to focus defense mechanisms within the data plane.
This approach narrows potential attack surfaces while leveraging the inherent advantage of high-speed traffic processing, detection, and routing capabilities.
Programmable switches, equipped with advanced hardware primitives, are at the forefront of enabling this shift, with pioneering efforts already underway.
As elaborated in Section \ref{subsec:programmable-switch}, current research is converging on empowering programmable switches with sophisticated statistical computations (e.g., entropy measures) and complex detection methodologies (e.g., machine learning algorithms).
In pursuit of these advancements, certain areas merit further investigation:
\textit{Efficient data structures.}
    The necessity for compact and efficient data structures becomes paramount when dealing with high traffic volumes, particularly given the limited registers and memory available in programmable switches.
    Data sketching emerges as a potent technique to efficiently approximate traffic features under such constraints.
\textit{Complex algorithm deployment.}
    It is also intriguing to consider the deployment of more intricate detection algorithms within these switches.
    A compelling research question is how to effectively translate a fully trained neural network into a set of match-action rules that are compatible with the operational paradigms of switches.

\section{Conclusion}
This paper begins by examining the evolution of DDoS attacks.
We highlight that these attacks are increasingly exploiting new network protocols and systems, and are employing advanced adversarial techniques to circumvent existing detection mechanisms.
We proceed by categorizing current detection methods, classifying them based on the heuristics and techniques they employ.
This analysis underscores the urgent need for modern detection systems that are not only general and efficient but also capable of distinguishing between the behaviors of legitimate users and attackers.
Additionally, we explore the potential of leveraging emerging hardware technologies to achieve line-speed packet processing and detection, offering a promising direction for enhancing response capabilities against DDoS attacks.
In conclusion, the paper provides recommendations for future research aimed at identifying DDoS vulnerabilities in new network protocols and systems.
We also discuss strategies for developing state-of-the-art DDoS detection systems that can effectively respond to the evolving landscape of network threats.

\bibliographystyle{plain}
\bibliography{reference}

\begin{thebibliography}{100}

\bibitem{securelist-amplification-attack}
Ddos attacks in q1 2021.
\newblock \url{https://securelist.com/ddos-attacks-in-q1-2021/102166/}, 2020.

\bibitem{cloudflare-ddos-report}
Ddos threat report for 2023 q4.
\newblock \url{https://blog.cloudflare.com/ddos-threat-report-2023-q4}, 2020.

\bibitem{fbi-amplification-attack}
Fbi warns of new ddos attack vectors: Coap, ws-dd, arms, and jenkins.
\newblock \url{https://www.zdnet.com}, 2020.

\bibitem{shodan}
Shodan.
\newblock \url{https://www.shodan.io/}, 2020.

\bibitem{abhishta2019measuring}
Abhishta Abhishta, Roland van Rijswijk-Deij, and Lambert~JM Nieuwenhuis.
\newblock Measuring the impact of a successful ddos attack on the customer behaviour of managed dns service providers.
\newblock {\em ACM SIGCOMM Computer Communication Review}, 48(5):70--76, 2019.

\bibitem{abusnaina2019examining}
Ahmed Abusnaina, Aminollah Khormali, DaeHun Nyang, Murat Yuksel, and Aziz Mohaisen.
\newblock Examining the robustness of learning-based ddos detection in software defined networks.
\newblock In {\em 2019 IEEE Conference on Dependable and Secure Computing (DSC)}, pages 1--8. IEEE, 2019.

\bibitem{acar2020peek}
Abbas Acar, Hossein Fereidooni, Tigist Abera, Amit~Kumar Sikder, Markus Miettinen, Hidayet Aksu, Mauro Conti, Ahmad-Reza Sadeghi, and Selcuk Uluagac.
\newblock Peek-a-boo: I see your smart home activities, even encrypted!
\newblock In {\em Proceedings of the 13th ACM Conference on Security and Privacy in Wireless and Mobile Networks}, pages 207--218, 2020.

\bibitem{agiollo2023gnn4ifa}
Andrea Agiollo, Enkeleda Bardhi, Mauro Conti, Riccardo Lazzeretti, Eleonora Losiouk, and Andrea Omicini.
\newblock Gnn4ifa: Interest flooding attack detection with graph neural networks.
\newblock In {\em 2023 IEEE 8th European Symposium on Security and Privacy (EuroS\&P)}, pages 615--630. IEEE, 2023.

\bibitem{agrawal2019defense}
Neha Agrawal and Shashikala Tapaswi.
\newblock Defense mechanisms against ddos attacks in a cloud computing environment: State-of-the-art and research challenges.
\newblock {\em IEEE Communications Surveys \& Tutorials}, 21(4):3769--3795, 2019.

\bibitem{ahmed2018statistical}
Muhammad~Ejaz Ahmed, Saeed Ullah, and Hyoungshick Kim.
\newblock Statistical application fingerprinting for ddos attack mitigation.
\newblock {\em IEEE Transactions on Information Forensics and Security}, 14(6):1471--1484, 2018.

\bibitem{aktar2023towards}
Sharmin Aktar and Abdullah~Yasin Nur.
\newblock Towards ddos attack detection using deep learning approach.
\newblock {\em Computers \& Security}, 129:103251, 2023.

\bibitem{al2023bin}
Arwa~Abdulkarim Al~Alsadi, Kaichi Sameshima, Katsunari Yoshioka, Michel Van~Eeten, and Carlos~Hernandez Ga{\~n}{\'a}n.
\newblock Bin there, target that: Analyzing the target selection of iot vulnerabilities in malware binaries.
\newblock In {\em Proceedings of the 26th International Symposium on Research in Attacks, Intrusions and Defenses}, pages 513--526, 2023.

\bibitem{alcoz2022aggregate}
Albert~Gran Alcoz, Martin Strohmeier, Vincent Lenders, and Laurent Vanbever.
\newblock Aggregate-based congestion control for pulse-wave ddos defense.
\newblock In {\em Proceedings of the ACM SIGCOMM 2022 Conference}, pages 693--706, 2022.

\bibitem{antonakakis2017understanding}
Manos Antonakakis, Tim April, Michael Bailey, Matt Bernhard, Elie Bursztein, Jaime Cochran, Zakir Durumeric, J~Alex Halderman, Luca Invernizzi, Michalis Kallitsis, et~al.
\newblock Understanding the mirai botnet.
\newblock In {\em 26th USENIX security symposium (USENIX Security 17)}, pages 1093--1110, 2017.

\bibitem{apostolaki2017hijacking}
Maria Apostolaki, Aviv Zohar, and Laurent Vanbever.
\newblock Hijacking bitcoin: Routing attacks on cryptocurrencies.
\newblock In {\em 2017 IEEE symposium on security and privacy (SP)}, pages 375--392. IEEE, 2017.

\bibitem{atlasis2012attacking}
Antonios Atlasis.
\newblock Attacking ipv6 implementation using fragmentation.
\newblock {\em Blackhat europe}, pages 14--16, 2012.

\bibitem{aydin2022long}
Hakan Ayd{\i}n, Zeynep Orman, and Muhammed~Ali Ayd{\i}n.
\newblock A long short-term memory (lstm)-based distributed denial of service (ddos) detection and defense system design in public cloud network environment.
\newblock {\em Computers \& Security}, 118:102725, 2022.

\bibitem{barradas2021flowlens}
Diogo Barradas, Nuno Santos, Lu{\'\i}s Rodrigues, Salvatore Signorello, Fernando~MV Ramos, and Andr{\'e} Madeira.
\newblock Flowlens: Enabling efficient flow classification for ml-based network security applications.
\newblock In {\em NDSS}, 2021.

\bibitem{beckett2017http}
David Beckett and Sakir Sezer.
\newblock Http/2 cannon: Experimental analysis on http/1 and http/2 request flood ddos attacks.
\newblock In {\em 2017 Seventh International Conference on Emerging Security Technologies (EST)}, pages 108--113. IEEE, 2017.

\bibitem{bhatia2021mstream}
Siddharth Bhatia, Arjit Jain, Pan Li, Ritesh Kumar, and Bryan Hooi.
\newblock Mstream: Fast anomaly detection in multi-aspect streams.
\newblock In {\em Proceedings of the Web Conference 2021}, pages 3371--3382, 2021.

\bibitem{spoofer}
CAIDA.
\newblock Spoofer project.
\newblock \url{https://www.caida.org/projects/spoofer/}, 2014.

\bibitem{cao2019crosspath}
Jiahao Cao, Qi~Li, Renjie Xie, Kun Sun, Guofei Gu, Mingwei Xu, and Yuan Yang.
\newblock The $\{$CrossPath$\}$ attack: Disrupting the $\{$SDN$\}$ control channel via shared links.
\newblock In {\em 28th USENIX Security Symposium (USENIX Security 19)}, pages 19--36, 2019.

\bibitem{cao2021detecting}
Yongyi Cao, Hao Jiang, Yuchuan Deng, Jing Wu, Pan Zhou, and Wei Luo.
\newblock Detecting and mitigating ddos attacks in sdn using spatial-temporal graph convolutional network.
\newblock {\em IEEE Transactions on Dependable and Secure Computing}, 19(6):3855--3872, 2021.

\bibitem{cao2016off}
Yue Cao, Zhiyun Qian, Zhongjie Wang, Tuan Dao, Srikanth~V Krishnamurthy, and Lisa~M Marvel.
\newblock $\{$Off-Path$\}$$\{$TCP$\}$ exploits: Global rate limit considered dangerous.
\newblock In {\em 25th USENIX Security Symposium (USENIX Security 16)}, pages 209--225, 2016.

\bibitem{catillo2023case}
Marta Catillo, Andrea Del~Vecchio, Antonio Pecchia, and Umberto Villano.
\newblock A case study with cicids2017 on the robustness of machine learning against adversarial attacks in intrusion detection.
\newblock In {\em Proceedings of the 18th International Conference on Availability, Reliability and Security}, pages 1--8, 2023.

\bibitem{chadd2018ddos}
Anthony Chadd.
\newblock Ddos attacks: past, present and future.
\newblock {\em Network Security}, 2018(7):13--15, 2018.

\bibitem{chaudhary2023ddos}
Shubhankar Chaudhary and Pramod~Kumar Mishra.
\newblock Ddos attacks in industrial iot: A survey.
\newblock {\em Computer Networks}, page 110015, 2023.

\bibitem{chen2021sdnshield}
Kuan-Yin Chen, Sen Liu, Yang Xu, Ishant~Kumar Siddhrau, Siyu Zhou, Zehua Guo, and H~Jonathan Chao.
\newblock Sdnshield: nfv-based defense framework against ddos attacks on sdn control plane.
\newblock {\em IEEE/ACM Transactions on Networking}, 30(1):1--17, 2021.

\bibitem{chen2008feasibility}
Wei Chen, Yingzhou Zhang, and Yuanchun Wei.
\newblock The feasibility of launching reduction of quality (roq) attacks in 802.11 wireless networks.
\newblock In {\em 2008 14th IEEE International Conference on Parallel and Distributed Systems}, pages 517--524. IEEE, 2008.

\bibitem{chen2024frequency}
Xuehan Chen, Jingjing Tan, Litian Kang, Fengxiao Tang, Ming Zhao, and Nei Kato.
\newblock Frequency selective surface towards 6g communication systems: A contemporary survey.
\newblock {\em IEEE Communications Surveys \& Tutorials}, 2024.

\bibitem{choi2022understanding}
Jinchun Choi, Afsah Anwar, Abdulrahman Alabduljabbar, Hisham Alasmary, Jeffrey Spaulding, An~Wang, Songqing Chen, DaeHun Nyang, Amro Awad, and David Mohaisen.
\newblock Understanding internet of things malware by analyzing endpoints in their static artifacts.
\newblock {\em Computer Networks}, 206:108768, 2022.

\bibitem{cirillo2021botnet}
Michele Cirillo, Mario Di~Mauro, Vincenzo Matta, and Marco Tambasco.
\newblock Botnet identification in ddos attacks with multiple emulation dictionaries.
\newblock {\em IEEE Transactions on Information Forensics and Security}, 16:3554--3569, 2021.

\bibitem{cui2023cbseq}
Susu Cui, Cong Dong, Meng Shen, Yuling Liu, Bo~Jiang, and Zhigang Lu.
\newblock Cbseq: A channel-level behavior sequence for encrypted malware traffic detection.
\newblock {\em IEEE Transactions on Information Forensics and Security}, 2023.

\bibitem{dai2024dampadf}
Yunwei Dai, Tao Huang, and Shuo Wang.
\newblock Dampadf: A framework for dns amplification attack defense based on bloom filters and nampkeeper.
\newblock {\em Computers \& Security}, 139:103718, 2024.

\bibitem{dainotti2013estimating}
Alberto Dainotti, Karyn Benson, Alistair King, KC~Claffy, Michael Kallitsis, Eduard Glatz, and Xenofontas Dimitropoulos.
\newblock Estimating internet address space usage through passive measurements.
\newblock {\em ACM SIGCOMM Computer Communication Review}, 44(1):42--49, 2013.

\bibitem{dantas2014selective}
Yuri~Gil Dantas, Vivek Nigam, and Iguatemi~E Fonseca.
\newblock A selective defense for application layer ddos attacks.
\newblock In {\em 2014 IEEE Joint Intelligence and Security Informatics Conference}, pages 75--82. IEEE, 2014.

\bibitem{de2018ddos}
Michele De~Donno, Nicola Dragoni, Alberto Giaretta, and Angelo Spognardi.
\newblock Ddos-capable iot malwares: comparative analysis and mirai investigation.
\newblock {\em Security and Communication Networks}, 2018(1):7178164, 2018.

\bibitem{de2021detection}
Vin{\'\i}cius de~Miranda~Rios, Pedro~RM In{\'a}cio, Damien Magoni, and M{\'a}rio~M Freire.
\newblock Detection of reduction-of-quality ddos attacks using fuzzy logic and machine learning algorithms.
\newblock {\em Computer Networks}, 186:107792, 2021.

\bibitem{di2011protecting}
Sebastiano Di~Paola and Dario Lombardo.
\newblock Protecting against dns reflection attacks with bloom filters.
\newblock In {\em International Conference on Detection of Intrusions and Malware, and Vulnerability Assessment}, pages 1--16. Springer, 2011.

\bibitem{diallo2021adaptive}
Alec~F Diallo and Paul Patras.
\newblock Adaptive clustering-based malicious traffic classification at the network edge.
\newblock In {\em IEEE INFOCOM 2021-IEEE Conference on Computer Communications}, pages 1--10. IEEE, 2021.

\bibitem{dimolianis2022ddos}
Marinos Dimolianis, Dimitrios~K Kalogeras, Nikos Kostopoulos, and Vasilis Maglaris.
\newblock Ddos attack detection via privacy-aware federated learning and collaborative mitigation in multi-domain cyber infrastructures.
\newblock In {\em 2022 IEEE 11th International Conference on Cloud Networking (CloudNet)}, pages 118--125. IEEE, 2022.

\bibitem{ding2021tracking}
Damu Ding, Marco Savi, and Domenico Siracusa.
\newblock Tracking normalized network traffic entropy to detect ddos attacks in p4.
\newblock {\em IEEE Transactions on Dependable and Secure Computing}, 19(6):4019--4031, 2021.

\bibitem{dong2023horuseye}
Yutao Dong, Qing Li, Kaidong Wu, Ruoyu Li, Dan Zhao, Gareth Tyson, Junkun Peng, Yong Jiang, Shutao Xia, and Mingwei Xu.
\newblock $\{$HorusEye$\}$: A realtime $\{$IoT$\}$ malicious traffic detection framework using programmable switches.
\newblock In {\em 32nd USENIX Security Symposium (USENIX Security 23)}, pages 571--588, 2023.

\bibitem{doriguzzi2024introducing}
Roberto Doriguzzi-Corin, Luis Augusto~Dias Knob, Luca Mendozzi, Domenico Siracusa, and Marco Savi.
\newblock Introducing packet-level analysis in programmable data planes to advance network intrusion detection.
\newblock {\em Computer Networks}, 239:110162, 2024.

\bibitem{duan2022application}
Guanghan Duan, Hongwu Lv, Huiqiang Wang, and Guangsheng Feng.
\newblock Application of a dynamic line graph neural network for intrusion detection with semisupervised learning.
\newblock {\em IEEE Transactions on Information Forensics and Security}, 18:699--714, 2022.

\bibitem{durumeric2015search}
Zakir Durumeric, David Adrian, Ariana Mirian, Michael Bailey, and J~Alex Halderman.
\newblock A search engine backed by internet-wide scanning.
\newblock In {\em Proceedings of the 22nd ACM SIGSAC conference on computer and communications security}, pages 542--553, 2015.

\bibitem{eshete2017dynaminer}
Birhanu Eshete and VN~Venkatakrishnan.
\newblock Dynaminer: Leveraging offline infection analytics for on-the-wire malware detection.
\newblock In {\em 2017 47th Annual IEEE/IFIP International Conference on Dependable Systems and Networks (DSN)}, pages 463--474. IEEE, 2017.

\bibitem{feng2020off}
Xuewei Feng, Chuanpu Fu, Qi~Li, Kun Sun, and Ke~Xu.
\newblock Off-path tcp exploits of the mixed ipid assignment.
\newblock In {\em Proceedings of the 2020 ACM SIGSAC Conference on Computer and Communications Security}, pages 1323--1335, 2020.

\bibitem{feng2022off}
Xuewei Feng, Qi~Li, Kun Sun, Zhiyun Qian, Gang Zhao, Xiaohui Kuang, Chuanpu Fu, and Ke~Xu.
\newblock $\{$Off-Path$\}$ network traffic manipulation via revitalized $\{$ICMP$\}$ redirect attacks.
\newblock In {\em 31st USENIX Security Symposium (USENIX Security 22)}, pages 2619--2636, 2022.

\bibitem{feng2022pmtud}
Xuewei Feng, Qi~Li, Kun Sun, Ke~Xu, Baojun Liu, Xiaofeng Zheng, Qiushi Yang, Haixin Duan, and Zhiyun Qian.
\newblock Pmtud is not panacea: Revisiting ip fragmentation attacks against tcp.
\newblock In {\em NDSS}, 2022.

\bibitem{feng2020application}
Yebo Feng, Jun Li, and Thanh Nguyen.
\newblock Application-layer ddos defense with reinforcement learning.
\newblock In {\em 2020 IEEE/ACM 28th International Symposium on Quality of Service (IWQoS)}, pages 1--10. IEEE, 2020.

\bibitem{fogla2006evading}
Prahlad Fogla and Wenke Lee.
\newblock Evading network anomaly detection systems: formal reasoning and practical techniques.
\newblock In {\em Proceedings of the 13th ACM conference on Computer and communications security}, pages 59--68, 2006.

\bibitem{fouladi2022novel}
Ramin~Fadaei Fouladi, Orhan Ermi{\c{s}}, and Emin Anarim.
\newblock A novel approach for distributed denial of service defense using continuous wavelet transform and convolutional neural network for software-defined network.
\newblock {\em Computers \& Security}, 112:102524, 2022.

\bibitem{fu2021realtime}
Chuanpu Fu, Qi~Li, Meng Shen, and Ke~Xu.
\newblock Realtime robust malicious traffic detection via frequency domain analysis.
\newblock In {\em Proceedings of the 2021 ACM SIGSAC Conference on Computer and Communications Security}, pages 3431--3446, 2021.

\bibitem{fu2023detecting}
Chuanpu Fu, Qi~Li, and Ke~Xu.
\newblock Detecting unknown encrypted malicious traffic in real time via flow interaction graph analysis.
\newblock {\em arXiv preprint arXiv:2301.13686}, 2023.

\bibitem{fu2023point}
Chuanpu Fu, Qi~Li, Ke~Xu, and Jianping Wu.
\newblock Point cloud analysis for ml-based malicious traffic detection: Reducing majorities of false positive alarms.
\newblock In {\em Proceedings of the 2023 ACM SIGSAC Conference on Computer and Communications Security}, pages 1005--1019, 2023.

\bibitem{gasti2013and}
Paolo Gasti, Gene Tsudik, Ersin Uzun, and Lixia Zhang.
\newblock Dos and ddos in named data networking.
\newblock In {\em 2013 22nd International Conference on Computer Communication and Networks (ICCCN)}, pages 1--7. IEEE, 2013.

\bibitem{gervais2015tampering}
Arthur Gervais, Hubert Ritzdorf, Ghassan~O Karame, and Srdjan Capkun.
\newblock Tampering with the delivery of blocks and transactions in bitcoin.
\newblock In {\em Proceedings of the 22nd ACM SIGSAC Conference on Computer and Communications Security}, pages 692--705, 2015.

\bibitem{gilad2011fragmentation}
Yossi Gilad and Amir Herzberg.
\newblock Fragmentation considered vulnerable: blindly intercepting and discarding fragments.
\newblock In {\em 5th USENIX Workshop on Offensive Technologies (WOOT 11)}, 2011.

\bibitem{gkounis2016interplay}
Dimitrios Gkounis, Vasileios Kotronis, Christos Liaskos, and Xenofontas Dimitropoulos.
\newblock On the interplay of link-flooding attacks and traffic engineering.
\newblock {\em ACM SIGCOMM Computer Communication Review}, 46(2):5--11, 2016.

\bibitem{griffioen2021scan}
Harm Griffioen, Kris Oosthoek, Paul van~der Knaap, and Christian Doerr.
\newblock Scan, test, execute: Adversarial tactics in amplification ddos attacks.
\newblock In {\em Proceedings of the 2021 ACM SIGSAC Conference on Computer and Communications Security}, pages 940--954, 2021.

\bibitem{guan2009new}
Xiaohong Guan, Pinghui Wang, and Tao Qin.
\newblock A new data streaming method for locating hosts with large connection degree.
\newblock In {\em GLOBECOM 2009-2009 IEEE Global Telecommunications Conference}, pages 1--6. IEEE, 2009.

\bibitem{guo2020detecting}
Hang Guo and John Heidemann.
\newblock Detecting iot devices in the internet.
\newblock {\em IEEE/ACM Transactions on Networking}, 28(5):2323--2336, 2020.

\bibitem{hashemi2019towards}
Mohammad~J Hashemi, Greg Cusack, and Eric Keller.
\newblock Towards evaluation of nidss in adversarial setting.
\newblock In {\em Proceedings of the 3rd ACM CoNEXT Workshop on Big DAta, Machine Learning and Artificial Intelligence for Data Communication Networks}, pages 14--21, 2019.

\bibitem{he2009reduction}
Yanxiang He, Qiang Cao, Yi~Han, Libing Wu, and Tao Liu.
\newblock Reduction of quality (roq) attacks on structured peer-to-peer networks.
\newblock In {\em 2009 IEEE International Symposium on Parallel \& Distributed Processing}, pages 1--9. IEEE, 2009.

\bibitem{heilman2015eclipse}
Ethan Heilman, Alison Kendler, Aviv Zohar, and Sharon Goldberg.
\newblock Eclipse attacks on $\{$Bitcoin’s$\}$$\{$peer-to-peer$\}$ network.
\newblock In {\em 24th USENIX security symposium (USENIX security 15)}, pages 129--144, 2015.

\bibitem{henrydoss2014critical}
James Henrydoss and Terry Boult.
\newblock Critical security review and study of ddos attacks on lte mobile network.
\newblock In {\em 2014 IEEE Asia Pacific Conference on Wireless and Mobile}, pages 194--200. IEEE, 2014.

\bibitem{herwig2019measurement}
Stephen Herwig, Katura Harvey, George Hughey, Richard Roberts, and Dave Levin.
\newblock Measurement and analysis of hajime, a peer-to-peer iot botnet.
\newblock In {\em Network and Distributed Systems Security (NDSS) Symposium}, 2019.

\bibitem{hnamte2024ddos}
Vanlalruata Hnamte, Ashfaq~Ahmad Najar, Hong Nhung-Nguyen, Jamal Hussain, and Manohar~Naik Sugali.
\newblock Ddos attack detection and mitigation using deep neural network in sdn environment.
\newblock {\em Computers \& Security}, 138:103661, 2024.

\bibitem{holland2021new}
Jordan Holland, Paul Schmitt, Nick Feamster, and Prateek Mittal.
\newblock New directions in automated traffic analysis.
\newblock In {\em Proceedings of the 2021 ACM SIGSAC Conference on Computer and Communications Security}, pages 3366--3383, 2021.

\bibitem{invernizzi2014nazca}
Luca Invernizzi, Stanislav Miskovic, Ruben Torres, Christopher Kruegel, Sabyasachi Saha, Giovanni Vigna, Sung-Ju Lee, and Marco Mellia.
\newblock Nazca: Detecting malware distribution in large-scale networks.
\newblock In {\em NDSS}, volume~14, pages 23--26, 2014.

\bibitem{javadpour2023reinforcement}
Amir Javadpour, Forough Ja’fari, Tarik Taleb, and Chafika Benza{\"\i}d.
\newblock Reinforcement learning-based slice isolation against ddos attacks in beyond 5g networks.
\newblock {\em IEEE Transactions on Network and Service Management}, 2023.

\bibitem{jin2018your}
Lin Jin, Shuai Hao, Haining Wang, and Chase Cotton.
\newblock Your remnant tells secret: Residual resolution in ddos protection services.
\newblock In {\em 2018 48th Annual IEEE/IFIP International Conference on Dependable Systems and Networks (DSN)}, pages 362--373. IEEE, 2018.

\bibitem{johnson2014game}
Benjamin Johnson, Aron Laszka, Jens Grossklags, Marie Vasek, and Tyler Moore.
\newblock Game-theoretic analysis of ddos attacks against bitcoin mining pools.
\newblock In {\em Financial Cryptography and Data Security: FC 2014 Workshops, BITCOIN and WAHC 2014, Christ Church, Barbados, March 7, 2014, Revised Selected Papers 18}, pages 72--86. Springer, 2014.

\bibitem{kalkan2018jess}
K{\"u}bra Kalkan, Levent Altay, G{\"u}rkan G{\"u}r, and Fatih Alag{\"o}z.
\newblock Jess: Joint entropy-based ddos defense scheme in sdn.
\newblock {\em IEEE Journal on Selected Areas in Communications}, 36(10):2358--2372, 2018.

\bibitem{kamiyama2007simple}
Noriaki Kamiyama, Tatsuya Mori, and Ryoichi Kawahara.
\newblock Simple and adaptive identification of superspreaders by flow sampling.
\newblock In {\em IEEE INFOCOM 2007-26th IEEE International Conference on Computer Communications}, pages 2481--2485. IEEE, 2007.

\bibitem{kang2016spiffy}
Min~Suk Kang, Virgil~D Gligor, Vyas Sekar, et~al.
\newblock Spiffy: Inducing cost-detectability tradeoffs for persistent link-flooding attacks.
\newblock In {\em NDSS}, volume~1, pages 53--55, 2016.

\bibitem{kang2013crossfire}
Min~Suk Kang, Soo~Bum Lee, and Virgil~D Gligor.
\newblock The crossfire attack.
\newblock In {\em 2013 IEEE symposium on security and privacy}, pages 127--141. IEEE, 2013.

\bibitem{khan2019idea}
Haider~Adnan Khan, Nader Sehatbakhsh, Luong~N Nguyen, Robert~L Callan, Arie Yeredor, Milos Prvulovic, and Alenka Zaji{\'c}.
\newblock Idea: Intrusion detection through electromagnetic-signal analysis for critical embedded and cyber-physical systems.
\newblock {\em IEEE Transactions on Dependable and Secure Computing}, 18(3):1150--1163, 2019.

\bibitem{kim2017preventing}
Soyoung Kim, Sora Lee, Geumhwan Cho, Muhammad~Ejaz Ahmed, Jaehoon Jeong, and Hyoungshick Kim.
\newblock Preventing dns amplification attacks using the history of dns queries with sdn.
\newblock In {\em Computer Security--ESORICS 2017: 22nd European Symposium on Research in Computer Security, Oslo, Norway, September 11-15, 2017, Proceedings, Part II 22}, pages 135--152. Springer, 2017.

\bibitem{kramer2015amppot}
Lukas Kr{\"a}mer, Johannes Krupp, Daisuke Makita, Tomomi Nishizoe, Takashi Koide, Katsunari Yoshioka, and Christian Rossow.
\newblock Amppot: Monitoring and defending against amplification ddos attacks.
\newblock In {\em Research in Attacks, Intrusions, and Defenses: 18th International Symposium, RAID 2015, Kyoto, Japan, November 2-4, 2015. Proceedings 18}, pages 615--636. Springer, 2015.

\bibitem{krupp2021bgpeek}
Johannes Krupp and Christian Rossow.
\newblock Bgpeek-a-boo: Active bgp-based traceback for amplification ddos attacks.
\newblock In {\em 2021 IEEE European Symposium on Security and Privacy (EuroS\&P)}, pages 423--439. IEEE, 2021.

\bibitem{kumari2023comprehensive}
Pooja Kumari and Ankit~Kumar Jain.
\newblock A comprehensive study of ddos attacks over iot network and their countermeasures.
\newblock {\em Computers \& Security}, page 103096, 2023.

\bibitem{kwon2015dropper}
Bum~Jun Kwon, Jayanta Mondal, Jiyong Jang, Leyla Bilge, and Tudor Dumitra{\c{s}}.
\newblock The dropper effect: Insights into malware distribution with downloader graph analytics.
\newblock In {\em Proceedings of the 22nd ACM SIGSAC Conference on Computer and Communications Security}, pages 1118--1129, 2015.

\bibitem{le2019unearthing}
Franck Le, Mudhakar Srivatsa, and Dinesh Verma.
\newblock Unearthing and exploiting latent semantics behind dns domains for deep network traffic analysis.
\newblock In {\em Proc. Workshop AI for Internet of Things}, pages 1--6, 2019.

\bibitem{li2021deter}
Kai Li, Yibo Wang, and Yuzhe Tang.
\newblock Deter: Denial of ethereum txpool services.
\newblock In {\em Proceedings of the 2021 ACM SIGSAC Conference on Computer and Communications Security}, pages 1645--1667, 2021.

\bibitem{li2023path}
Man Li, Shuangxing Deng, Huachun Zhou, and Yajuan Qin.
\newblock A path selection scheme for detecting malicious behavior based on deep reinforcement learning in sdn/nfv-enabled network.
\newblock {\em Computer Networks}, 236:110034, 2023.

\bibitem{li2023comprehensive}
Qing Li, He~Huang, Ruoyu Li, Jianhui Lv, Zhenhui Yuan, Lianbo Ma, Yi~Han, and Yong Jiang.
\newblock A comprehensive survey on ddos defense systems: New trends and challenges.
\newblock {\em Computer Networks}, page 109895, 2023.

\bibitem{li2023bijack}
Shaoyu Li, Shanghao Shi, Yang Xiao, Chaoyu Zhang, Y~Thomas Hou, and Wenjing Lou.
\newblock Bijack: Breaking bitcoin network with tcp vulnerabilities.
\newblock In {\em European Symposium on Research in Computer Security}, pages 306--326. Springer, 2023.

\bibitem{li2024survey}
Yuan Li, Hao Zhang, Chen Zhang, Tao Huang, and F~Richard Yu.
\newblock A survey of quantum internet protocols from a layered perspective.
\newblock {\em IEEE Communications Surveys \& Tutorials}, 2024.

\bibitem{liaskos2018network}
Christos Liaskos and Sotiris Ioannidis.
\newblock Network topology effects on the detectability of crossfire attacks.
\newblock {\em IEEE Transactions on Information Forensics and Security}, 13(7):1682--1695, 2018.

\bibitem{liaskos2016novel}
Christos Liaskos, Vasileios Kotronis, and Xenofontas Dimitropoulos.
\newblock A novel framework for modeling and mitigating distributed link flooding attacks.
\newblock In {\em IEEE INFOCOM 2016-The 35th Annual IEEE International Conference on Computer Communications}, pages 1--9. IEEE, 2016.

\bibitem{lichtblau2017detection}
Franziska Lichtblau, Florian Streibelt, Thorben Kr{\"u}ger, Philipp Richter, and Anja Feldmann.
\newblock Detection, classification, and analysis of inter-domain traffic with spoofed source ip addresses.
\newblock In {\em Proceedings of the 2017 Internet Measurement Conference}, pages 86--99, 2017.

\bibitem{liu2015detection}
Weijiang Liu, Wenyu Qu, Jian Gong, and Keqiu Li.
\newblock Detection of superpoints using a vector bloom filter.
\newblock {\em IEEE Transactions on Information Forensics and Security}, 11(3):514--527, 2015.

\bibitem{liu2016one}
Zaoxing Liu, Antonis Manousis, Gregory Vorsanger, Vyas Sekar, and Vladimir Braverman.
\newblock One sketch to rule them all: Rethinking network flow monitoring with univmon.
\newblock In {\em Proceedings of the 2016 ACM SIGCOMM Conference}, pages 101--114, 2016.

\bibitem{liu2021jaqen}
Zaoxing Liu, Hun Namkung, Georgios Nikolaidis, Jeongkeun Lee, Changhoon Kim, Xin Jin, Vladimir Braverman, Minlan Yu, and Vyas Sekar.
\newblock Jaqen: A $\{$High-Performance$\}$$\{$Switch-Native$\}$ approach for detecting and mitigating volumetric $\{$DDoS$\}$ attacks with programmable switches.
\newblock In {\em 30th USENIX Security Symposium (USENIX Security 21)}, pages 3829--3846, 2021.

\bibitem{lone2017using}
Qasim Lone, Matthew Luckie, Maciej Korczy{\'n}ski, and Michel Van~Eeten.
\newblock Using loops observed in traceroute to infer the ability to spoof.
\newblock In {\em Passive and Active Measurement: 18th International Conference, PAM 2017, Sydney, NSW, Australia, March 30-31, 2017, Proceedings 18}, pages 229--241. Springer, 2017.

\bibitem{long2022hybrid}
Zhang Long and Wang Jinsong.
\newblock A hybrid method of entropy and ssae-svm based ddos detection and mitigation mechanism in sdn.
\newblock {\em Computers \& Security}, 115:102604, 2022.

\bibitem{luo2014mathematical}
Jingtang Luo, Xiaolong Yang, Jin Wang, Jie Xu, Jian Sun, and Keping Long.
\newblock On a mathematical model for low-rate shrew ddos.
\newblock {\em IEEE Transactions on Information Forensics and Security}, 9(7):1069--1083, 2014.

\bibitem{ma2019randomized}
Xiaobo Ma, Bo~An, Mengchen Zhao, Xiapu Luo, Lei Xue, Zhenhua Li, Tony~TN Miu, and Xiaohong Guan.
\newblock Randomized security patrolling for link flooding attack detection.
\newblock {\em IEEE Transactions on Dependable and Secure Computing}, 17(4):795--812, 2019.

\bibitem{mahadik2023edge}
Shalaka~S Mahadik, Pranav~M Pawar, and Raja Muthalagu.
\newblock Edge-hetiot defense against ddos attack using learning techniques.
\newblock {\em Computers \& Security}, 132:103347, 2023.

\bibitem{mannes2019naming}
Elisa Mannes and Carlos Maziero.
\newblock Naming content on the network layer: A security analysis of the information-centric network model.
\newblock {\em ACM Computing Surveys (CSUR)}, 52(3):1--28, 2019.

\bibitem{matta2017ddos}
Vincenzo Matta, Mario Di~Mauro, and Maurizio Longo.
\newblock Ddos attacks with randomized traffic innovation: Botnet identification challenges and strategies.
\newblock {\em IEEE Transactions on Information Forensics and Security}, 12(8):1844--1859, 2017.

\bibitem{meidan2017profiliot}
Yair Meidan, Michael Bohadana, Asaf Shabtai, Juan~David Guarnizo, Mart{\'\i}n Ochoa, Nils~Ole Tippenhauer, and Yuval Elovici.
\newblock Profiliot: A machine learning approach for iot device identification based on network traffic analysis.
\newblock In {\em Proceedings of the symposium on applied computing}, pages 506--509, 2017.

\bibitem{mirian2016internet}
Ariana Mirian, Zane Ma, David Adrian, Matthew Tischer, Thasphon Chuenchujit, Tim Yardley, Robin Berthier, Joshua Mason, Zakir Durumeric, J~Alex Halderman, et~al.
\newblock An internet-wide view of ics devices.
\newblock In {\em 2016 14th Annual Conference on Privacy, Security and Trust (PST)}, pages 96--103. IEEE, 2016.

\bibitem{mirkovic2004taxonomy}
Jelena Mirkovic and Peter Reiher.
\newblock A taxonomy of ddos attack and ddos defense mechanisms.
\newblock {\em ACM SIGCOMM Computer Communication Review}, 34(2):39--53, 2004.

\bibitem{mirsky2018kitsune}
Yisroel Mirsky, Tomer Doitshman, Yuval Elovici, and Asaf Shabtai.
\newblock Kitsune: An ensemble of autoencoders for online network intrusion detection.
\newblock In {\em Network and Distributed Systems Security (NDSS) Symposium}, 2018.

\bibitem{mirsky2020ddos}
Yisroel Mirsky and Mordechai Guri.
\newblock Ddos attacks on 9-1-1 emergency services.
\newblock {\em IEEE Transactions on Dependable and Secure Computing}, 18(6):2767--2786, 2020.

\bibitem{mm2022efficient}
Gowthul~Alam MM, Michael~Raj TF, et~al.
\newblock An efficient svm based deho classifier to detect ddos attack in cloud computing environment.
\newblock {\em Computer Networks}, 215:109138, 2022.

\bibitem{mohammed2023detection}
Abubakar~Sadiq Mohammed, Eirini Anthi, Omer Rana, Neetesh Saxena, and Pete Burnap.
\newblock Detection and mitigation of field flooding attacks on oil and gas critical infrastructure communication.
\newblock {\em Computers \& Security}, 124:103007, 2023.

\bibitem{moosavi2016deepfool}
Seyed-Mohsen Moosavi-Dezfooli, Alhussein Fawzi, and Pascal Frossard.
\newblock Deepfool: a simple and accurate method to fool deep neural networks.
\newblock In {\em Proceedings of the IEEE conference on computer vision and pattern recognition}, pages 2574--2582, 2016.

\bibitem{mustapha2023detecting}
Ali Mustapha, Rida Khatoun, Sherali Zeadally, Fadlallah Chbib, Ahmad Fadlallah, Walid Fahs, and Ali El~Attar.
\newblock Detecting ddos attacks using adversarial neural network.
\newblock {\em Computers \& Security}, 127:103117, 2023.

\bibitem{najar2024cyber}
Ashfaq~Ahmad Najar and S~Manohar Naik.
\newblock Cyber-secure sdn: A cnn-based approach for efficient detection and mitigation of ddos attacks.
\newblock {\em Computers \& Security}, 139:103716, 2024.

\bibitem{nawrocki2021quicsand}
Marcin Nawrocki, Raphael Hiesgen, Thomas~C Schmidt, and Matthias W{\"a}hlisch.
\newblock Quicsand: quantifying quic reconnaissance scans and dos flooding events.
\newblock In {\em Proceedings of the 21st ACM internet measurement conference}, pages 283--291, 2021.

\bibitem{nosyk2023closed}
Yevheniya Nosyk, Maciej Korczy{\'n}ski, Qasim Lone, Marcin Skwarek, Baptiste Jonglez, and Andrzej Duda.
\newblock The closed resolver project: Measuring the deployment of inbound source address validation.
\newblock {\em IEEE/ACM Transactions on Networking}, 2023.

\bibitem{obaidat2023creating}
Islam Obaidat, Bennett Kahn, Fatemeh Tavakoli, and Meera Sridhar.
\newblock Creating a large-scale memory error iot botnet using ns3dockeremulator.
\newblock In {\em 2023 53rd Annual IEEE/IFIP International Conference on Dependable Systems and Networks (DSN)}, pages 470--479. IEEE, 2023.

\bibitem{olimid20205g}
Ruxandra~F Olimid and Gianfranco Nencioni.
\newblock 5g network slicing: A security overview.
\newblock {\em IEEE Access}, 8:99999--100009, 2020.

\bibitem{pa2016iotpot}
Yin Minn~Pa Pa, Shogo Suzuki, Katsunari Yoshioka, Tsutomu Matsumoto, Takahiro Kasama, and Christian Rossow.
\newblock Iotpot: A novel honeypot for revealing current iot threats.
\newblock {\em Journal of Information Processing}, 24(3):522--533, 2016.

\bibitem{pan2024loopy}
Yepeng Pan, Anna Ascheman, and Christian Rossow.
\newblock Loopy hell(ow): Infinite traffic loops at the application layer.
\newblock In {\em 33rd USENIX security symposium (USENIX security 24)}, pages 111--125, 2024.

\bibitem{panigrahi2022intrusion}
Ranjit Panigrahi, Samarjeet Borah, Moumita Pramanik, Akash~Kumar Bhoi, Paolo Barsocchi, Soumya~Ranjan Nayak, and Waleed Alnumay.
\newblock Intrusion detection in cyber--physical environment using hybrid na{\"\i}ve bayes—decision table and multi-objective evolutionary feature selection.
\newblock {\em Computer Communications}, 188:133--144, 2022.

\bibitem{papernot2016limitations}
Nicolas Papernot, Patrick McDaniel, Somesh Jha, Matt Fredrikson, Z~Berkay Celik, and Ananthram Swami.
\newblock The limitations of deep learning in adversarial settings.
\newblock In {\em 2016 IEEE European symposium on security and privacy (EuroS\&P)}, pages 372--387. IEEE, 2016.

\bibitem{plonka2000flowscan}
Dave Plonka.
\newblock $\{$FlowScan$\}$: A network traffic flow reporting and visualization tool.
\newblock In {\em 14th Systems Administration Conference (LISA 2000)}, 2000.

\bibitem{praseed2018ddos}
Amit Praseed and P~Santhi Thilagam.
\newblock Ddos attacks at the application layer: Challenges and research perspectives for safeguarding web applications.
\newblock {\em IEEE Communications Surveys \& Tutorials}, 21(1):661--685, 2018.

\bibitem{praseed2019multiplexed}
Amit Praseed and P~Santhi Thilagam.
\newblock Multiplexed asymmetric attacks: Next-generation ddos on http/2 servers.
\newblock {\em IEEE Transactions on Information Forensics and Security}, 15:1790--1800, 2019.

\bibitem{qin2015ddos}
Xi~Qin, Tongge Xu, and Chao Wang.
\newblock Ddos attack detection using flow entropy and clustering technique.
\newblock In {\em 2015 11th International Conference on Computational Intelligence and Security (CIS)}, pages 412--415. IEEE, 2015.

\bibitem{rashidi2017collaborative}
Bahman Rashidi, Carol Fung, and Elisa Bertino.
\newblock A collaborative ddos defence framework using network function virtualization.
\newblock {\em IEEE Transactions on Information Forensics and Security}, 12(10):2483--2497, 2017.

\bibitem{rezapour2021rl}
Amir Rezapour and Wen-Guey Tzeng.
\newblock Rl-shield: mitigating target link-flooding attacks using sdn and deep reinforcement learning routing algorithm.
\newblock {\em IEEE Transactions on Dependable and Secure Computing}, 19(6):4052--4067, 2021.

\bibitem{ribeiro2023detecting}
Marcos~Aur{\'e}lio Ribeiro, Mauro Sergio~Pereira Fonseca, and Juliana de~Santi.
\newblock Detecting and mitigating ddos attacks with moving target defense approach based on automated flow classification in sdn networks.
\newblock {\em Computers \& Security}, 134:103462, 2023.

\bibitem{rodriguez2013survey}
Rafael~A Rodr{\'\i}guez-G{\'o}mez, Gabriel Maci{\'a}-Fern{\'a}ndez, and Pedro Garc{\'\i}a-Teodoro.
\newblock Survey and taxonomy of botnet research through life-cycle.
\newblock {\em ACM Computing Surveys (CSUR)}, 45(4):1--33, 2013.

\bibitem{romeiras2023poster}
Jo{\~a}o Romeiras~Amado, Francisco~Chami{\c{c}}a Pereira, Salvatore Signorello, Miguel Correia, and Fernando Ramos.
\newblock Poster: In-network ml feature computation for malicious traffic detection.
\newblock In {\em Proceedings of the ACM SIGCOMM 2023 Conference}, pages 1105--1107, 2023.

\bibitem{sachdeva2010ddos}
Monika Sachdeva, Gurvinder Singh, Krishan Kumar, and Kuldip Singh.
\newblock Ddos incidents and their impact: A review.
\newblock {\em Int. Arab J. Inf. Technol.}, 7(1):14--20, 2010.

\bibitem{sattar2019towards}
Danish Sattar and Ashraf Matrawy.
\newblock Towards secure slicing: Using slice isolation to mitigate ddos attacks on 5g core network slices.
\newblock In {\em 2019 IEEE Conference on Communications and Network Security (CNS)}, pages 82--90. IEEE, 2019.

\bibitem{scherrer2023albus}
Simon Scherrer, Jo~Vliegen, Arish Sateesan, Hsu-Chun Hsiao, Nele Mentens, and Adrian Perrig.
\newblock Albus: a probabilistic monitoring algorithm to counter burst-flood attacks.
\newblock In {\em 2023 42nd International Symposium on Reliable Distributed Systems (SRDS)}, pages 162--172. IEEE, 2023.

\bibitem{schuchard2010losing}
Max Schuchard, Abedelaziz Mohaisen, Denis Foo~Kune, Nicholas Hopper, Yongdae Kim, and Eugene~Y Vasserman.
\newblock Losing control of the internet: using the data plane to attack the control plane.
\newblock In {\em Proceedings of the 17th ACM conference on Computer and communications security}, pages 726--728, 2010.

\bibitem{shalini2022docus}
PV~Shalini, V~Radha, and Sriram~G Sanjeevi.
\newblock Docus-ddos detection in sdn using modified cusum with flash traffic discrimination and mitigation.
\newblock {\em Computer Networks}, 217:109361, 2022.

\bibitem{shankesi2010resource}
Ravinder Shankesi, Omid Fatemieh, and Carl~A Gunter.
\newblock Resource inflation threats to denial of service countermeasures.
\newblock {\em Dept. Comput. Sci., UIUC, Champaign, IL, USA, Tech. Rep}, 2010.

\bibitem{shin2013attacking}
Seungwon Shin and Guofei Gu.
\newblock Attacking software-defined networks: A first feasibility study.
\newblock In {\em Proceedings of the second ACM SIGCOMM workshop on Hot topics in software defined networking}, pages 165--166, 2013.

\bibitem{shin2013avant}
Seungwon Shin, Vinod Yegneswaran, Phillip Porras, and Guofei Gu.
\newblock Avant-guard: Scalable and vigilant switch flow management in software-defined networks.
\newblock In {\em Proceedings of the 2013 ACM SIGSAC conference on Computer \& communications security}, pages 413--424, 2013.

\bibitem{sikder20176thsense}
Amit~Kumar Sikder, Hidayet Aksu, and A~Selcuk Uluagac.
\newblock $\{$6thSense$\}$: A context-aware sensor-based attack detector for smart devices.
\newblock In {\em 26th USENIX Security Symposium (USENIX Security 17)}, pages 397--414, 2017.

\bibitem{silva2020repel}
Renato~S Silva, Carlos~Colman Meixner, Rafael~S Guimar{\~a}es, Thierno Diallo, Borja~O Garcia, Lu{\'\i}s~FM de~Moraes, and Magnos Martinello.
\newblock Repel: a strategic approach for defending 5g control plane from ddos signalling attacks.
\newblock {\em IEEE Transactions on Network and Service Management}, 18(3):3231--3243, 2020.

\bibitem{sisalem2006denial}
Dorgham Sisalem, Jiri Kuthan, and Sven Ehlert.
\newblock Denial of service attacks targeting a sip voip infrastructure: attack scenarios and prevention mechanisms.
\newblock {\em IEEE Network}, 20(5):26--31, 2006.

\bibitem{sommese2022investigating}
Raffaele Sommese, KC~Claffy, Roland van Rijswijk-Deij, Arnab Chattopadhyay, Alberto Dainotti, Anna Sperotto, and Mattijs Jonker.
\newblock Investigating the impact of ddos attacks on dns infrastructure.
\newblock In {\em proceedings of the 22nd ACM Internet Measurement Conference}, pages 51--64, 2022.

\bibitem{srinivasa2022bad}
Shreyas Srinivasa, Dimitrios Georgoulias, Jens~Myrup Pedersen, and Emmanouil Vasilomanolakis.
\newblock A bad idea: Weaponizing uncontrolled online-ides in availability attacks.
\newblock In {\em 2022 IEEE European Symposium on Security and Privacy Workshops (EuroS\&PW)}, pages 82--92. IEEE, 2022.

\bibitem{studer2009coremelt}
Ahren Studer and Adrian Perrig.
\newblock The coremelt attack.
\newblock In {\em European Symposium on Research in Computer Security}, pages 37--52. Springer, 2009.

\bibitem{tandon2021defending}
Rajat Tandon, Abhinav Palia, Jaydeep Ramani, Brandon Paulsen, Genevieve Bartlett, and Jelena Mirkovic.
\newblock Defending web servers against flash crowd attacks.
\newblock In {\em International Conference on Applied Cryptography and Network Security}, pages 338--361. Springer, 2021.

\bibitem{tandon2023leader}
Rajat Tandon, Haoda Wang, Nicolaas Weideman, Shushan Arakelyan, Genevieve Bartlett, Christophe Hauser, and Jelena Mirkovic.
\newblock Leader: Defense against exploit-based denial-of-service attacks on web applications.
\newblock In {\em Proceedings of the 26th International Symposium on Research in Attacks, Intrusions and Defenses}, pages 744--758, 2023.

\bibitem{tang2021performance}
Dan Tang, Yudong Yan, Siqi Zhang, Jingwen Chen, and Zheng Qin.
\newblock Performance and features: Mitigating the low-rate tcp-targeted dos attack via sdn.
\newblock {\em IEEE Journal on Selected Areas in Communications}, 40(1):428--444, 2021.

\bibitem{tang2014sip}
Jin Tang, Yu~Cheng, Yong Hao, and Wei Song.
\newblock Sip flooding attack detection with a multi-dimensional sketch design.
\newblock {\em IEEE Transactions on Dependable and Secure Computing}, 11(6):582--595, 2014.

\bibitem{tang2022high}
Lu~Tang, Yao Xiao, Qun Huang, and Patrick~PC Lee.
\newblock A high-performance invertible sketch for network-wide superspreader detection.
\newblock {\em IEEE/ACM Transactions on Networking}, 2022.

\bibitem{tang2013modeling}
Yajuan Tang, Xiapu Luo, Qing Hui, and Rocky~KC Chang.
\newblock Modeling the vulnerability of feedback-control based internet services to low-rate dos attacks.
\newblock {\em IEEE transactions on information forensics and security}, 9(3):339--353, 2013.

\bibitem{tegeler2012botfinder}
Florian Tegeler, Xiaoming Fu, Giovanni Vigna, and Christopher Kruegel.
\newblock Botfinder: Finding bots in network traffic without deep packet inspection.
\newblock In {\em Proceedings of the 8th international conference on Emerging networking experiments and technologies}, pages 349--360, 2012.

\bibitem{thomas20171000}
Daniel~R Thomas, Richard Clayton, and Alastair~R Beresford.
\newblock 1000 days of udp amplification ddos attacks.
\newblock In {\em 2017 APWG Symposium on Electronic Crime Research (eCrime)}, pages 79--84. IEEE, 2017.

\bibitem{torabi2018inferring}
Sadegh Torabi, Elias Bou-Harb, Chadi Assi, Mario Galluscio, Amine Boukhtouta, and Mourad Debbabi.
\newblock Inferring, characterizing, and investigating internet-scale malicious iot device activities: A network telescope perspective.
\newblock In {\em 2018 48th Annual IEEE/IFIP International Conference on Dependable Systems and Networks (DSN)}, pages 562--573. IEEE, 2018.

\bibitem{tran2020stealthier}
Muoi Tran, Inho Choi, Gi~Jun Moon, Anh~V Vu, and Min~Suk Kang.
\newblock A stealthier partitioning attack against bitcoin peer-to-peer network.
\newblock In {\em 2020 IEEE Symposium on Security and Privacy (SP)}, pages 894--909. IEEE, 2020.

\bibitem{tushir2020quantitative}
Bhagyashri Tushir, Yogesh Dalal, Behnam Dezfouli, and Yuhong Liu.
\newblock A quantitative study of ddos and e-ddos attacks on wifi smart home devices.
\newblock {\em IEEE Internet of Things Journal}, 8(8):6282--6292, 2020.

\bibitem{vasek2014empirical}
Marie Vasek, Micah Thornton, and Tyler Moore.
\newblock Empirical analysis of denial-of-service attacks in the bitcoin ecosystem.
\newblock In {\em Financial Cryptography and Data Security: FC 2014 Workshops, BITCOIN and WAHC 2014, Christ Church, Barbados, March 7, 2014, Revised Selected Papers 18}, pages 57--71. Springer, 2014.

\bibitem{vukovic2014security}
Ognjen Vukovi{\'c} and Gy{\"o}rgy D{\'a}n.
\newblock Security of fully distributed power system state estimation: Detection and mitigation of data integrity attacks.
\newblock {\em IEEE Journal on Selected Areas in Communications}, 32(7):1500--1508, 2014.

\bibitem{wagner2021united}
Daniel Wagner, Daniel Kopp, Matthias Wichtlhuber, Christoph Dietzel, Oliver Hohlfeld, Georgios Smaragdakis, and Anja Feldmann.
\newblock United we stand: Collaborative detection and mitigation of amplification ddos attacks at scale.
\newblock In {\em Proceedings of the 2021 ACM SIGSAC conference on computer and communications security}, pages 970--987, 2021.

\bibitem{walck2019tendrilstaller}
Matthew Walck, Ke~Wang, and Hyong~S Kim.
\newblock Tendrilstaller: Block delay attack in bitcoin.
\newblock In {\em 2019 IEEE international conference on Blockchain (Blockchain)}, pages 1--9. IEEE, 2019.

\bibitem{wang2018data}
An~Wang, Wentao Chang, Songqing Chen, and Aziz Mohaisen.
\newblock A data-driven study of ddos attacks and their dynamics.
\newblock {\em IEEE Transactions on Dependable and Secure Computing}, 17(3):648--661, 2018.

\bibitem{wang2018delving}
An~Wang, Wentao Chang, Songqing Chen, and Aziz Mohaisen.
\newblock Delving into internet ddos attacks by botnets: characterization and analysis.
\newblock {\em IEEE/ACM Transactions on Networking}, 26(6):2843--2855, 2018.

\bibitem{wang2002detecting}
Haining Wang, Danlu Zhang, and Kang~G Shin.
\newblock Detecting syn flooding attacks.
\newblock In {\em Proceedings. Twenty-first annual joint conference of the IEEE computer and communications societies}, volume~3, pages 1530--1539. IEEE, 2002.

\bibitem{wang2022zigbee}
Jincheng Wang, Zhuohua Li, Mingshen Sun, and John~CS Lui.
\newblock Zigbee’s network rejoin procedure for iot systems: vulnerabilities and implications.
\newblock In {\em Proceedings of the 25th International Symposium on Research in Attacks, Intrusions and Defenses}, pages 292--307, 2022.

\bibitem{wang2018detecting}
Juan Wang, Ru~Wen, Jiangqi Li, Fei Yan, Bo~Zhao, and Fajiang Yu.
\newblock Detecting and mitigating target link-flooding attacks using sdn.
\newblock {\em IEEE Transactions on dependable and secure computing}, 16(6):944--956, 2018.

\bibitem{wang2023bars}
Kai Wang, Zhiliang Wang, Dongqi Han, Wenqi Chen, Jiahai Yang, Xingang Shi, and Xia Yin.
\newblock Bars: Local robustness certification for deep learning based traffic analysis systems.
\newblock In {\em NDSS}, 2023.

\bibitem{wang2020dynamic}
Meng Wang, Yiqin Lu, and Jiancheng Qin.
\newblock A dynamic mlp-based ddos attack detection method using feature selection and feedback.
\newblock {\em Computers \& Security}, 88:101645, 2020.

\bibitem{wanglordma}
Shicheng Wang, Menghao Zhang, Yuying Du, Ziteng Chen, Zhiliang Wang, Mingwei Xu, Renjie Xie, and Jiahai Yang.
\newblock Lordma: A new low-rate dos attack in rdma networks.

\bibitem{wang2024off}
Ziqiang Wang, Xuewei Feng, Qi~Li, Kun Sun, Yuxiang Yang, Mengyuan Li, and Ke~Xu.
\newblock Off-path tcp hijacking in wi-fi networks: A packet-size side channel attack.
\newblock {\em arXiv preprint arXiv:2402.12716}, 2024.

\bibitem{wichtlhuber2022ixp}
Matthias Wichtlhuber, Eric Strehle, Daniel Kopp, Lars Prepens, Stefan Stegmueller, Alina Rubina, Christoph Dietzel, and Oliver Hohlfeld.
\newblock Ixp scrubber: learning from blackholing traffic for ml-driven ddos detection at scale.
\newblock In {\em Proceedings of the ACM SIGCOMM 2022 Conference}, pages 707--722, 2022.

\bibitem{wu2020survive}
Shuangke Wu, Yanjiao Chen, Minghui Li, Xiangyang Luo, Zhe Liu, and Lan Liu.
\newblock Survive and thrive: A stochastic game for ddos attacks in bitcoin mining pools.
\newblock {\em IEEE/ACM Transactions on Networking}, 28(2):874--887, 2020.

\bibitem{wu2014software}
Yongdong Wu, Zhigang Zhao, Feng Bao, and Robert~H Deng.
\newblock Software puzzle: A countermeasure to resource-inflated denial-of-service attacks.
\newblock {\em IEEE Transactions on Information Forensics and security}, 10(1):168--177, 2014.

\bibitem{xing2021ripple}
Jiarong Xing, Wenqing Wu, and Ang Chen.
\newblock Ripple: A programmable, decentralized $\{$Link-Flooding$\}$ defense against adaptive adversaries.
\newblock In {\em 30th USENIX Security Symposium (USENIX Security 21)}, pages 3865--3881, 2021.

\bibitem{xiong2021warmonger}
Junjie Xiong, Mingkui Wei, Zhuo Lu, and Yao Liu.
\newblock Warmonger: inflicting denial-of-service via serverless functions in the cloud.
\newblock In {\em Proceedings of the 2021 ACM SIGSAC Conference on Computer and Communications Security}, pages 955--969, 2021.

\bibitem{xu2022towards}
Zhiwei Xu, Xin Wang, and Yujun Zhang.
\newblock Towards persistent detection of ddos attacks in ndn: A sketch-based approach.
\newblock {\em IEEE Transactions on Dependable and Secure Computing}, 2022.

\bibitem{xu2022xatu}
Zhiying Xu, Sivaramakrishnan Ramanathan, Alexander Rush, Jelena Mirkovic, and Minlan Yu.
\newblock Xatu: Boosting existing ddos detection systems using auxiliary signals.
\newblock In {\em Proceedings of the 18th International Conference on emerging Networking EXperiments and Technologies}, pages 1--17, 2022.

\bibitem{xue2014towards}
Lei Xue, Xiapu Luo, Edmond~WW Chan, and Xian Zhan.
\newblock Towards detecting target link flooding attack.
\newblock In {\em 28th Large Installation System Administration Conference (LISA14)}, pages 90--105, 2014.

\bibitem{yan2023automatic}
Haonan Yan, Xiaoguang Li, Wenjing Zhang, Rui Wang, Hui Li, Xingwen Zhao, Fenghua Li, and Xiaodong Lin.
\newblock Automatic evasion of machine learning-based network intrusion detection systems.
\newblock {\em IEEE Transactions on Dependable and Secure Computing}, 2023.

\bibitem{yan2015software}
Qiao Yan, F~Richard Yu, Qingxiang Gong, and Jianqiang Li.
\newblock Software-defined networking (sdn) and distributed denial of service (ddos) attacks in cloud computing environments: A survey, some research issues, and challenges.
\newblock {\em IEEE communications surveys \& tutorials}, 18(1):602--622, 2015.

\bibitem{yang2021cade}
Limin Yang, Wenbo Guo, Qingying Hao, Arridhana Ciptadi, Ali Ahmadzadeh, Xinyu Xing, and Gang Wang.
\newblock $\{$CADE$\}$: Detecting and explaining concept drift samples for security applications.
\newblock In {\em 30th USENIX Security Symposium (USENIX Security 21)}, pages 2327--2344, 2021.

\bibitem{yazdani2022mirrors}
Ramin Yazdani, Alden Hilton, Jeroen van~der Ham, Roland van Rijswijk-Deij, Casey Deccio, Anna Sperotto, and Mattijs Jonker.
\newblock Mirrors in the sky: on the potential of clouds in dns reflection-based denial-of-service attacks.
\newblock In {\em Proceedings of the 25th International Symposium on Research in Attacks, Intrusions and Defenses}, pages 263--275, 2022.

\bibitem{yin2023waterpurifier}
Lihua Yin, Muyijie Zhu, Wenxin Liu, Xi~Luo, Chonghua Wang, and Yangyang Li.
\newblock Waterpurifier: A scalable system to prevent the dns water torture attack in 5g-enabled siot network.
\newblock {\em Computer Communications}, 199:186--195, 2023.

\bibitem{yue2024ccs}
Meng Yue, Qingxin Yan, Zichao Lu, and Zhijun Wu.
\newblock Ccs: A cross-plane collaboration strategy to defend against ldos attacks in sdn.
\newblock {\em IEEE Transactions on Network and Service Management}, 2024.

\bibitem{zhang2020poseidon}
Menghao Zhang, Guanyu Li, Shicheng Wang, Chang Liu, Ang Chen, Hongxin Hu, Guofei Gu, Qianqian Li, Mingwei Xu, and Jianping Wu.
\newblock Poseidon: Mitigating volumetric ddos attacks with programmable switches.
\newblock In {\em the 27th Network and Distributed System Security Symposium (NDSS 2020)}, 2020.

\bibitem{zhang2024revealing}
Zhiyi Zhang, Guorui Xiao, Sichen Song, R~Can Aygun, Angelos Stavrou, Lixia Zhang, and Eric Osterweil.
\newblock Revealing protocol architecture’s design patterns in the volumetric ddos defense design space.
\newblock {\em IEEE Communications Surveys \& Tutorials}, 2024.

\bibitem{zhao2023ernn}
Ziming Zhao, Zhaoxuan Li, Jialun Jiang, Fengyuan Yu, Fan Zhang, Congyuan Xu, Xinjie Zhao, Rui Zhang, and Shize Guo.
\newblock Ernn: Error-resilient rnn for encrypted traffic detection towards network-induced phenomena.
\newblock {\em IEEE Transactions on Dependable and Secure Computing}, 2023.

\bibitem{zheng2018realtime}
Jing Zheng, Qi~Li, Guofei Gu, Jiahao Cao, David~KY Yau, and Jianping Wu.
\newblock Realtime ddos defense using cots sdn switches via adaptive correlation analysis.
\newblock {\em IEEE Transactions on Information Forensics and Security}, 13(7):1838--1853, 2018.

\bibitem{zhou2023efficient}
Guangmeng Zhou, Zhuotao Liu, Chuanpu Fu, Qi~Li, and Ke~Xu.
\newblock An efficient design of intelligent network data plane.
\newblock In {\em 32nd USENIX Security Symposium (USENIX Security 23)}, pages 6203--6220, 2023.

\bibitem{zhou2023mew}
Huancheng Zhou, Sungmin Hong, Yangyang Liu, Xiapu Luo, Weichao Li, and Guofei Gu.
\newblock Mew: Enabling large-scale and dynamic link-flooding defenses on programmable switches.
\newblock In {\em 2023 IEEE Symposium on Security and Privacy (SP)}, pages 3178--3192. IEEE, 2023.

\bibitem{zhu2018privacy}
Liehuang Zhu, Xiangyun Tang, Meng Shen, Xiaojiang Du, and Mohsen Guizani.
\newblock Privacy-preserving ddos attack detection using cross-domain traffic in software defined networks.
\newblock {\em IEEE Journal on Selected Areas in Communications}, 36(3):628--643, 2018.

\end{thebibliography}

\end{document}